\documentclass[useAMS,usenatbib,a4paper]{mn2e}

\usepackage{graphicx, epsfig}
\usepackage{float}
\usepackage{subfigure}
\usepackage{natbib}
\usepackage{color}
\usepackage{cancel}
\usepackage{bm}
\usepackage{amsbsy,amssymb,amsmath}
\usepackage[draft]{hyperref}
\usepackage{enumitem}
\usepackage{txfonts}
\usepackage{gensymb}
\usepackage{ulem}

\def\simlt{\lower.5ex\hbox{$\; \buildrel < \over \sim \;$}}
\def\simgt{\lower.5ex\hbox{$\; \buildrel > \over \sim \;$}}

\newcommand*\Diff[1]{\mbox{d}^#1}

\makeatletter
\renewcommand{\p@subfigure}{}
\makeatother

\date{\today}

\author[Romeo A., Metcalf R.B. \& Pourtsidou A.]{Alessandro Romeo$^{1}$, R. Benton Metcalf$^{1,2}$, Alkistis Pourtsidou$^{3,4}$ \\ 
$^1$ Dipartimento di Fisica e Astronomia, Universit\'a di Bologna, via Gobetti 93/2, 40129 Bologna, Italy \\
$^2$ INAF-Osservatorio Astronomico di Bologna, via Ranzani 1, 40127 Bologna, Italy \\
$^3$ School of Physics and Astronomy, Queen Mary University of London, Mile End Road, London E1 4NS, United Kingdom \\
$^4$ Institute of Cosmology \& Gravitation, University of Portsmouth, Burnaby Road, Portsmouth, PO1 3FX, United Kingdom}

\begin{document}

\title{Simulations for 21\,cm radiation lensing at EoR redshifts}
\maketitle

\begin{abstract}
We introduce simulations aimed at assessing how well weak gravitational lensing of 21cm radiation from the Epoch of Reionization ($z \sim 8$) can be measured by an SKA-like radio telescope. A simulation pipeline has been implemented to study the performance of lensing reconstruction techniques. We show how well the lensing signal can be reconstructed using the three-dimensional quadratic lensing estimator in Fourier space assuming different survey strategies. The numerical code introduced in this work is capable of dealing with issues that can not be treated analytically such as the discreteness of visibility measurements and the inclusion of a realistic model for the antennae distribution. This paves the way for future numerical studies implementing more realistic reionization models, foreground subtraction schemes, and testing the performance of lensing estimators that take into account the non Gaussian distribution of HI after reionization. If multiple frequency channels covering $z \sim 7-11.6$ are combined, Phase 1 of SKA-Low should be able to obtain good quality images of the lensing potential with a total resolution of $\sim 1.6$ arcmin. The SKA-Low Phase 2 should be capable of providing images with high-fidelity even using data from $z\sim 7.7 - 8.3$. We perform tests aimed at evaluating the numerical implementation of the mapping reconstruction. We also discuss the possibility of measuring an accurate lensing power spectrum. Combining data from $z \sim 7-11.6$ using the SKA2-Low telescope model, we find constraints comparable to sample variance in the range $L<1000$, even for survey areas as small as $25\mbox{ deg}^2$.
\end{abstract}


\section{Introduction}
21~cm cosmology opens a unique observational window to previously unexplored cosmological epochs such as the Epoch of Reionization (EoR), the Cosmic Dawn and the Dark Ages \citep{FurlanettoOhBriggs06} using powerful radio interferometers such as the planned Square Kilometer Array (SKA)\footnote{http://www.skatelescope.org/} \citep{PritchardIchiki15}. Furthermore, 21~cm radiation emitted from sources at lower, post-reionization redshifts can be used to measure the Baryonic Acoustic Oscillations (BAO) with the intensity mapping technique \citep{Battye:2004re, ChangPen08,Peterson:2009ka,AnsariCampagne12,BattyeBrowne13,SmootDebono14}. In this paper we will concentrate on another possible application of this radiation, measuring weak gravitational lensing. 

21~cm radiation is generated by the hyperfine, spin flip, transition of neutral hydrogen. When the Cosmic Microwave Background (CMB) photons and the neutral hydrogen spin temperature become thermally decoupled the radiation is potentially observable in absorption or emission depending on whether the spin temperature is lower or higher than CMB temperature. In principle, the 21~cm line can give us access to a huge volume of the currently unobserved Universe covering the redshift range $z \sim 6-200$ during which the neutral fraction of hydrogen is high, as well as more recent, post-reionization, epochs where neutral hydrogen (HI) is found only within galaxies. The redshifted 21~cm line allows us to obtain a 3D map of the Universe, across the sky and along cosmic time by observing in a range of frequencies.

Current and planned experiments like the SKA, LOFAR\footnote{http://www.lofar.org/} (Low Frequency Array), PAPER\footnote{http://eor.berkeley.edu/} (Precision Array for Probing the Epoch of Reionization), and MWA\footnote{http://www.haystack.mit.edu/ast/arrays/mwa/} (Murchinson Widefield Array) have investigations of the high-redshift Universe through HI as their primary or one of their primary goals.
Ly-$\alpha$ photons from the first generation of stars and quasars efficiently raise the spin temperature above the CMB temperature and make the 21~cm brightness temperature effectively proportional to the hydrogen density and neutral fraction. This enables these observations to potentially map out the distribution of HI in three dimensions during the EoR or obtain its power spectrum \citep{BarkanaLoeb05,McQuinnZahnZaldarriaga06,MaoTegmarkMcQuinnZaldarriaga08}. 

At lower redshifts and higher frequencies, the technique of HI intensity mapping, which treats the 21~cm temperature field as a continuous, unresolved background and thus does not rely on detecting individual galaxies, can be used to measure the Baryonic Acoustic Oscillations (BAO), measure redshift space distortions, perform weak lensing studies, test Einstein's theory of general relativity and constrain primordial non-Gaussianity (see, for example,  \citep{Santos:2015gra} and \citep{HallBonvin13}). 

The main focus of this paper is the possibility of doing weak gravitational lensing studies using the 21~cm emission from the EoR. Early works \citep{ZahnZaldarriaga06, MetcalfWhite09} showed that if the EoR is at redshift $z \sim 8$ or later, an SKA-like instrument could map the distribution of matter with high fidelity and if a large enough fraction of the sky could be observed a high precision measurement of the convergence power spectrum could be obtained. Weak lensing is measurable because the 21~cm source can be divided up into multiple, statistically independent maps that are nearly identically lensed by the foreground mass distribution.
The lensing signal can be extracted from the data using a Fourier space quadratic lensing estimator, which was originally developed for the CMB case by \citet{HuOkamoto02} and then extended in 3D for the 21~cm case by \citet{ZahnZaldarriaga06}. Here we generalise this estimator to explicitly take into account the beam of the telescope and the gridding of the visibility measurements.

Observing the 21~cm signal from the EoR is challenging. At such low frequencies foreground contamination (mainly synchrotron emission) poses a particular problem. Foregrounds dominate over the cosmological signal by about four orders of magnitude, but studies indicate that they can be successfully removed by taking advantage of their relative coherence in frequency in comparison to the 21~cm signal from structure in the HI distribution \citep{LiuTegmark12, ChapmanAbdalla12, DillonLiuTegmark13}.

In order to assess how well gravitational lensing could actually be measured in realistic observations it is crucial to perform numerical simulations. Previous assessments have been based on simplifying assumptions that make predicting the noise analytically tractable. In particular, the 21~cm emission has been treated as a Gaussian random field and it has been assumed that foreground subtraction is done perfectly with no residual effects that might affect the lensing results. These are both important factors that cannot be handled analytically \citep{Furlanetto16}. Incomplete and uneven $u$-$v$ coverage is another issue that is best treated numerically. Here we introduce a numerical tool that can be used to perform more realistic studies and investigate the aforementioned problems.

This numerical tool can also be used to simulate measurements of gravitational lensing through the Ly$\alpha$ forest 
observed in high redshift quasar and galaxy spectra at redshifts of $z\sim 2.5$ \citep{CroftRomeoMetcalf17}. As with 21cm data, the forest has the advantage of spectral information, so that one could use independent information taken from across a significant redshift range. A realistic estimate of the noise in a Ly$\alpha$ lensing reconstruction has been presented in \citep{MetcalfCroftRomeo17} using the same techniques used in this work, but with different noise characteristics and taking into account the discreteness of the measured Ly$\alpha$ absorption in each pixel.

Currently, the planned SKA telescopes are the only radio telescopes with enough collecting area and sufficient resolution to observe 21\,cm lensing. SKA will be built in two phases (SKA1 and SKA2) and will have arrays for low frequencies (SKA-Low, $50 - 350 \, {\rm MHz}$) and medium frequencies (SKA-Mid, $350-13800 \, {\rm MHz}$). The primary science objective of SKA-Low is to observe the reionization of the universe at high redshift through its signatures in the 21~cm radiation \citep{PritchardIchiki15}. SKA-Mid will be able to measure 21~cm emission from HI in galaxies at lower redshifts ($z \simlt 3$). A possible HI intensity mapping survey performed by SKA-Mid could have important science outputs for cosmology \citep{Bull:2014rha,Santos:2015gra}. It has also been suggested that SKA-Mid should be able to detect weak lensing of 21cm emission from post-reionization redshifts using the intensity mapping method \citep{PourtsidouMetcalf15,PourtsidouMetcalf14}. In this paper, we present simulations for the case of 21cm lensing from EoR redshifts ({\it i.e.} using SKA-Low) and we plan to address lensing from lower redshift sources with SKA-Mid in a future paper.
In addition to developing a simulation technique and code that can later be used for general 21cm lensing studies, the aim of this work is to investigate how well the lensing signal from 21~cm sources at typical EoR redshifts ($z \sim 8$) can be reconstructed using the quadratic estimator technique and the current SKA design. 

The paper is organised as follows: in Section~\ref{sec:TheorBackgr} we describe the formalism and present the lensing reconstruction formulae that will be implemented in our numerical simulation; in Section~\ref{sec:SimDet} we discuss the set up of the simulation, including the generation of the 21~cm source, the simulation of the lensing signal, the modeling of the instrument, and the beaming effects; in Section~\ref{sec:Results} we present our results by discussing their dependencies on different telescope parameters and on the assumptions we have made. We also comment on various numerical aspects. We conclude in Section~\ref{sec:Conclusions}.

\section{21~cm Radiation Lensing Background}
\label{sec:TheorBackgr}

In order to simulate the 21\,cm temperature field and its lensing on a discrete grid, we will employ the discrete estimator formalism described in \citep{PourtsidouMetcalf15}, based on the work by \citep{ZahnZaldarriaga06}. Using the weak lensing approximation and assuming that the source field is Gaussian, an unbiased and optimal ({\it i.e.} minimum variance) quadratic estimator for reconstructing the lensing potential can be derived. In this Section we describe this formalism and generalise it to include the telescope beam.

\subsection{Lensing Preliminaries}
The 21\,cm radiation emitted from sources at a redshift $z_s$ is lensed by the matter distribution lying between us and the emission. Gravitational lensing will shift the observed position of a point on the sky without changing the surface brightness. If the lensing is weak compared to structure in the source, the observed temperature can be expressed as a Taylor expansion of the unlensed temperature:
\begin{equation}
\label{eqn: LensField}
\tilde{T}\left(\bm{\theta},\nu\right) = T\left(\bm{\theta} - \bm{\alpha}(\bm{\theta}),\nu\right) \simeq T\left(\boldsymbol{\theta},\nu\right)  - \bm{\alpha}(\bm{\theta})\cdot\boldsymbol{\nabla}_{\bm{\theta}} T\left(\bm{\theta},\nu\right) + \dots
\end{equation}
where $\bm{\alpha}(\bm{\theta})$ is the deflection caused by lensing (with $\bm{\theta}$ the true position of the source) and dots denote higher-order terms in the expansion. The approximation used in Eq.~(\ref{eqn: LensField}) is  valid in the CMB case because of the smallness of the temperature gradients on medium scales and Silk damping on smaller scales. This expansion is also valid in the 21~cm case, where temperature gradients can be large, but the deflections (or deflection gradients) are small compared to them on all scales of interest. The deflection field $\bm{\alpha}(\bm{\theta})$ is related to the 2D projected lensing potential via $\bm{\nabla{\Phi}} = - \bm{\alpha}(\bm{\theta}) $, in the weak lensing limit. The lensing potential comes from the integration over the redshift direction of the full 3D gravitational potential \citep{BartelmannSchneider01}
\begin{equation}
\Phi = \frac{2}{c^2}\int_0^{z_s}{dz \frac{\mathcal{D}\left(z\right)\mathcal{D}\left(z_s-z\right)}{\mathcal{D}\left(z_s\right)}\phi\left[\mathcal{D}\left(z\right)\bm{\theta}\left(z\right),z\right]},
\end{equation}
where $\mathcal{D}(z)$ is the comoving angular diameter distance at redshift $z$. Taking the observed lensed position to be $\bm{\theta}$ and the unlensed one to be $\bm{\xi}$, the shear $\gamma_{1,2}$ and the convergence $\kappa$ are related to the gravitational potential by the Jacobian matrix
\begin{eqnarray}
\bm{\mathcal{J}}\left(\bm{\theta},z_s\right) = \frac{\partial\bm{\xi}}{\partial\bm{\theta}} &=&  \left(\begin{matrix} 
1 - \kappa - \gamma_1 && -\gamma_2 \\
-\gamma_2 && 1-\kappa +\gamma_1 
\end{matrix}\right) \nonumber \\
&=& \left(\begin{matrix} 
1 - \Phi_{,11}  && \Phi_{,12} \\
\Phi_{,12} && 1 - \Phi_{,22}
\end{matrix}\right),
\end{eqnarray}
where we have neglected any rotational variable in off-diagonal elements and the subscripts 1 and 2 stand for the derivative operation with respect to the two transverse coordinates of the lensing potential. 

The convergence field is related to the potential - or, equivalently, the deflection field - via the Poisson equation $\kappa = - \nabla^2\Phi /2 = \boldsymbol{\nabla}\cdot\bm\alpha /2$ 
Using the Limber approximation \citep{Limber54} for small scales, we can define the power spectrum of the deflection (or convergence) field. This will be related to the 3D density fluctuations power spectrum through 
\begin{equation}
\label{eqn:DefField}
C_L^{\alpha\alpha} = \frac{9\Omega_m^2H_0^3}{L(L+1)c^3}\int_0^{z_s}{dz\, \frac{\mathcal{W}^2(z)}{a^2(z)E(z)}} P_\delta \left( k =\frac{L}{\mathcal{D}(z)},z\right)
\end{equation}
\citep{Kaiser92}, where $E(z) = H(z)/H_0$ and $\mathcal{W}(z) = 1 -\left[ \mathcal{D}\left(z\right)/\mathcal{D}\left(z_s\right)\right]$. $H_0$ is the Hubble parameter today and $\Omega_m$ is the density of the matter in the Universe relative to the critical density. Throughout this work we adopt a standard $\Lambda$CDM cosmology with the Planck parameters set \citep{Ade:2015xua}.

\subsection{21~cm Brightness Temperature Fluctuation Field}
\label{sec:21cmfield}
The brightness temperature for the 21~cm line is given by
\begin{equation} 
\label{eq:T21cm}
\bar{T}(z) \simeq 26 (1+\delta_{\rm b}) x_{\rm H} \left(1 - \frac{T_{\rm{CMB}}}{T_{\rm S}}\right)\left(\frac{\Omega_{\rm b}h^2}{0.022}\right)\left[\left(\frac{0.15}{\Omega_{\rm m}h^2}\right)\left(\frac{1+z}{10}\right)\right]^{1/2} \mbox{mK},
\end{equation}
\citep{FurlanettoOhBriggs06, ZahnZaldarriaga06} 
where $x_{\rm H}$ is the neutral hydrogen fraction, $T_{\rm S}$ is the 21 cm spin transition temperature, $T_{\rm{CMB}} = 2.73(1+z) \mbox{ K}$ is the CMB temperature at redshift $z$, $\delta_{\rm b} = (\rho_{\rm b}-\bar\rho_{\rm b})/\bar\rho_{\rm b}$ is the baryon density contrast measured in redshift space, and $\Omega_{\rm b}$ is the average density of baryons today relative to the critical density. In the regimes of interest here, \textit{i.e.} $z<15$, $T_{\rm S} \gg T_{CMB}$ so that there is no dependence on the CMB temperature. The ionization fraction and the density of HI will depend on the considered epoch, the ionization history and structure formation history.  

The brightness temperature will be represented in the simulation within a rectangular volume centered at a redshift $z$. The comoving, radial length of this volume is  $\mathcal{L}(z,\Delta\nu)$ with $\Delta\nu$ the bandwidth of the observation. We will make the approximation that the angular distance to the simulation box $\mathcal{D}(z)$ is very large compared to $\mathcal{L}(z,\Delta\nu)$ so that the angular sizes of the front and the back of the box are the same. With this and the flat-sky approximation for small patches of the sky, the 3D temperature field is represented in Fourier space by defining the wave vectors $\bm{k}_\perp = \bm{l}/\mathcal{D}(z)$ and $k_\parallel = 2\pi k_p/\mathcal{L}(z,\Delta\nu)$, where $\bm{l} $ is the multipole vector, the Fourier space dual of the angle coordinate, and $k_p$ is an integer which discretises the $k_\parallel$ direction. The frequency band is broken up into many channels which can be interpreted as tangential slices. The Fourier dual of the radial distance is then the discrete values of $k_\parallel$ or $k_p$. Homogeneity dictates that there will be no correlations between modes with different $k_p$. 

We will take the simulation box to be square in the angular dimensions with the obvious extension to rectangular geometry. The angular area of the survey and box will be $\Omega_s$. The number of grid points in each dimension on the sky will be $N_\perp$ so that the total number of grid positions in each frequency channel is $N_s = N_\perp^2$. The angular resolution is $\Delta\theta$ and $(m,n)$ are the pixel indices.

The conversion between radial distance and frequency is given by
\begin{equation}
dr = \frac{c}{H_0}  \frac{dz}{ \sqrt{\Omega_{\rm m} (1+z)^3 + \Omega_{\rm K} (1+z)^2 + \Omega_\Lambda} }  \simeq \frac{c (1+z)^{1/2}}{\nu_{21}H_0\Omega_{\rm m}^{1/2}}\ d\nu,
\end{equation}
where $\Omega_{\rm K}$ is the energy density parameter for curvature, $\Omega_\Lambda$ is the one for a cosmological constant, and the approximation made in the last step  holds at high redshifts when the Universe is matter dominated. Since we are interested in the Epoch of Reionization, this approximation is valid for our purposes. The rest frame frequency is $\nu_{21} = 1420.4$ MHz. With this the total depth of the box can be calculated,
\begin{equation}
\mathcal{L}(z,\Delta\nu) \simeq \frac{c (1+z)^{1/2}}{\nu_{21}H_0\Omega_{\rm m}^{1/2}}\Delta\nu,
\end{equation}
and the frequency of each channel, $\bm{\nu}$, can be converted into radial distances $r_\nu$ within the box.

The Discrete Fourier Transform (DFT) of the temperature intensity field is then
\begin{equation}
\label{eqn:Tfield}
T_{\bm{l},k_p} = \frac{\Omega_s}{N_sN_\nu}\sum_{\bm{m,n,r}}  \exp\left[- 2\pi{\rm i} \left\{ \frac{1}{N_s}\,  \bm{ l} \cdot (m,n) + \frac{1}{\mathcal{L}}\, r_\nu \,k_p \right\} \right] T_{m,n,\nu},
\end{equation}
where $N_\nu$ is the number of channels within the band that is used.

From Eq.(\ref{eqn:Tfield}), we can derive the angular power spectrum of the 21 cm temperature field, $C_{l,k_p}$, defined by
\begin{equation}
\bigg\langle T_{\bm{l},k_p} T^\star_{\bm{l}',k'_p}\bigg\rangle = \Omega_s C_{l,k_p} \delta^{\rm K}_{l,l'}\delta^{\rm K}_{k_\parallel,k'_\parallel}.
\end{equation}
Throughout this paper the averaging operation denoted as $\langle \dots\rangle$ is performed over 21\,cm intensity field realisations. The angular power spectrum is related to the discrete temperature field power spectrum $P_k$ via 
\begin{equation}
\label{eqn:Pk}
P_k = \frac{P_{\Delta T}(k)}{V_s} = \frac{P_{\Delta T}\left[\sqrt{(l/\mathcal{D})^2+(2\pi k_p/\mathcal{L})^2}\right]}{\Omega_s\mathcal{D}^2\mathcal{L}} = \frac{C_{l,k_p}}{\Omega_s}.
\end{equation}
\citep{ZahnZaldarriaga06}.

For our first set of simulations we will adopt a simple model for the brightness temperature distribution which has been used before and can be directly compared to analytic results. We will consider a time before ionization when hydrogen is completely neutral ($x_{\rm H}=1$ in expression~(\ref{eq:T21cm})). The brightness temperature is then only dependent on the density distribution of hydrogen. To model this we will make the assumption that the baryons are not yet significantly biased with respect to the mass so that their power spectrum in redshift space is given by
\begin{equation}
\label{eqn:21cmPS}
P_{\Delta T}(k) = \bar{T}^2(z)\left(1+ f\mu_k^2\right)^2P_\delta(k),
\end{equation}
where $P_\delta(k)$ is the dark matter power spectrum. 

\begin{figure}
\centering
\includegraphics[scale=0.36]{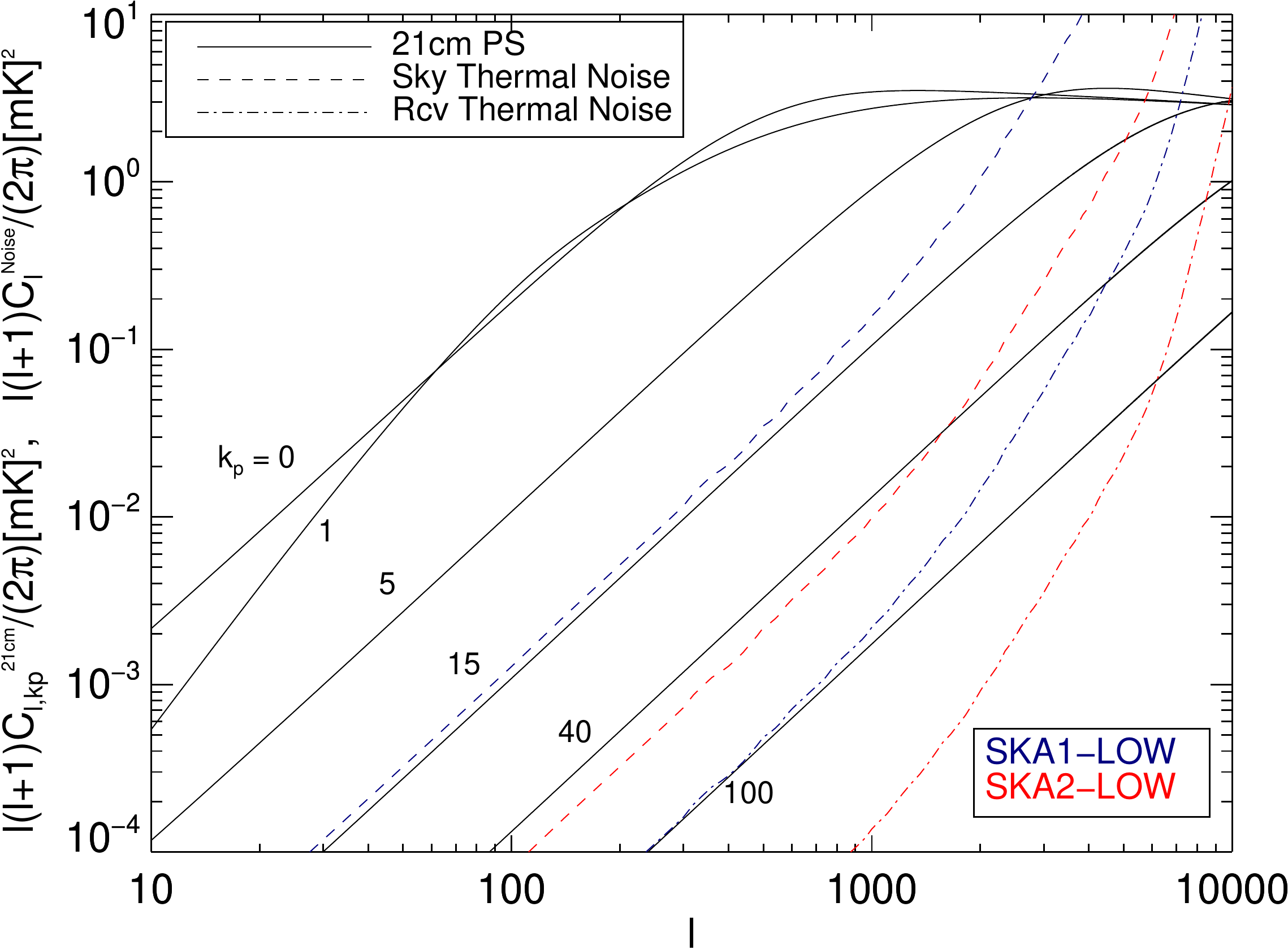}
\caption[Theoretical21cmPS]{The 21\,cm angular power spectra $C_{\ell,k_p}$ for several values of $k_p$ at $z_s=8$. The thick dashed lines are the sky noise power spectra, while the dashed-point lines are the receiver noise power spectra. The latter are produced assuming the SKA1-Low (blue) and SKA2-Low (red) R2 settings described in Section~\ref{sec:ThNoiseComp}.} 
\label{fig:21cmPS}
\end{figure}

We have included the redshift space distortion term in which $f = d\ln{D}/d\ln{a}\simeq\Omega_{\rm m}(z)^{0.55}$ with $D$ the linear growth rate. The cosine of the angle formed by the parallel component of the wavevector $\bm{k}$ and the wavevector itself is denoted $\mu_k = k_\parallel / k $. We will also assume that these fluctuations can be modelled with a Gaussian random field. In Figure~\ref{fig:21cmPS}, the power spectra (\ref{eqn:21cmPS}) are shown for different $k_p$, where we have assumed a bandwidth of $\Delta\nu = 5\mbox{ MHz}$ ($\Delta z = 0.286$) centered around a fiducial source redshift of $z_s=8$. Depending on the noise model (which will be specified in the next sections), modes beyond some $k_p^{\rm max}$ are dominated by noise and thus not useful for detecting lensing.

At the high redshifts and the resolution considered in this study, redshift space distortions can be modelled assuming $f\approx 1$. As discussed in \cite{MaoShapiro12}, redshift distortions have non negligible effects on the 21 cm power spectrum.  We also assume that no patchy reionization has occurred. The actual temperature distribution is likely to be considerably more complicated because of non-uniform ionization and segregation between baryons and dark matter. These are cases our code is designed to handle, but will be investigated in future work and combined with more detailed reionization simulations.

\subsection{Lensing Reconstruction in Fourier Space}
\label{sec:LensRec}
If the bandwidth is small and the redshift is relatively high, to a good approximation the matter within the band does not contribute significantly to the lensing of that band, \textit{i.e.} there is no self-lensing. In this case, the correlation between brightness temperature modes can be derived from equation~(\ref{eqn: LensField}), 
\begin{equation}
\bigg\langle \tilde{T}_{\bm{l},k_p}\tilde{T}^\star_{\bm{l}-\bm{L},k'_p} \bigg\rangle = \bm{L}\cdot\left[\bm{l} C_{l,k_p} + \left(\bm{L}-\bm{l}\right)C_{l-L,k_p}\right]\Phi(\bm{L}) \delta^K_{k_p k'_p}.
\end{equation}
(for $\bm{L} \neq \bm{0}$)  \citep{HuOkamoto02}. 
We can then define a discrete quadratic estimator for the gravitational potential of the form
\begin{equation}
\hat{\Phi}_{\bm{L}} = \sum_{\bm{l},k_p}{f(\bm{l},\bm{L},k_p)\tilde{T}_{\bm{l},k_p}\tilde{T}^\star_{\bm{l}-\bm{L},k_p}},
\end{equation}
in which the form of the filter $f(\bm{l},\bm{L},k_p)$ depends on the kind of source we are analysing and its statistical properties. Lensing induces correlations between different modes that would otherwise be uncorrelated. In the case of a Gaussian temperature field, an unbiased and optimal kernel can be derived by requiring $\langle\hat{\Phi}(\bm{L})\rangle = \Phi(\bm{L})$, and minimising its variance. The resulting estimator is 
\begin{equation}
\label{eqn:potest}
\hat{\Phi}_{\bm{L}} = \frac{N^{\hat{\Phi}}_{L}}{2\Omega_s}\sum_{\bm{l},k_p}{\left[\frac{\bm{L}\cdot\bm{l}\,C_{l,k_p} + \bm{L}\cdot\left(\bm{L}-\bm{l}\right)C_{l-L,k_p}}{C^{T}_{l,k_p}C^T_{l-L,k_p}}\right]\tilde{T}_{\bm{l},k_p}\tilde{T}^\star_{\bm{l}-\bm{L},k_p}}
\end{equation}\citep{ZahnZaldarriaga06}. The variance of this estimator is  
\begin{equation}
\label{eqn:SigPlusNoise}
\bigg\langle \hat{\Phi}_{\bm{L}} \hat{\Phi}_{\bm{L}}^\star \bigg\rangle = \Omega_s \left(N^{\hat{\Phi}}_{L} + C^{\Phi\Phi}_{L} \right),
\end{equation}
with $N^{\hat{\Phi}}_{L}$ being the lensing reconstruction noise. For the optimal estimator with a Gaussian source field this is
\begin{equation}
\label{eqn:EstNoise}
N^{\hat{\Phi}}_{L} = \left\{\frac{1}{2\Omega_s}\sum_{\bm{l},k_p}{\frac{\left[\bm{L}\cdot\bm{l}\,C_{l,k_p} + \bm{L}\cdot\left(\bm{L}-\bm{l}\right)C_{l-L,k_p}\right]^2}{C^{T}_{l,k_p}C^T_{l-L,k_p}}}\right\}^{-1},
\end{equation}
where $C^{T}_{l,k_p} = C_{l,k_p} + N_{l}^{\rm{Sky}} + N_l^{\rm{Rcv}}$ is the total observed power spectrum that includes the sky and receiver noises. In deriving this expression, and the optimal form of the kernel, the fourth order correlations of the field are required. These are easily found for a Gaussian field, but for a more complicated and realistic source field the lensing noise will need to be found numerically with simulations like the ones described in this paper. 

Expressions for the estimator and noise, for both the deflection and convergence fields, are trivially found using the Fourier space formulae $\hat{\bm{\alpha}}_{\bm{L}} = i \bm{L}\hat{\Phi}_{\bm{L}}$ and $\hat{\kappa}_{\bm{L}} = -(L^2/2) \hat{\Phi}_{\bm{L}} $. Moreover, $N_{L}^{\hat{\kappa}} = (L^4/4) N^{\hat{\Phi}}_{L} = (L^2/4) N^{\hat{\alpha}}_{L} $. 
These results can be linked to the continuous result by making the substitution $\Omega_s \rightarrow(2\pi)^2\delta(\bm{0}) $. 

Note that equation~(\ref{eqn:EstNoise}) is of the form $N^{\hat{\Phi}}_{L} = 1/\sum_{k_p}{\left[N^{\hat{\Phi}}_{L,k_p}\right]^{-1}}$, a result of the different $k_p$ modes being uncorrelated. Adding more $k_p$ modes reduces the total noise, but, as pointed out in \cite{ZahnZaldarriaga06}, only the first $\sim 20~k_p$ modes contribute to the lensing reconstruction. This is because of the monotonically decreasing behaviour of $C_{l,k_p}$ on all scales of interest as shown in Figure~\ref{fig:21cmPS}. For high values of $k_p$ the signal is well below the thermal noise level so these modes do not contribute to the estimator. Moreover, even if we could use a bigger number of $k_p$ modes than the ones allowed by Figure~\ref{fig:21cmPS}, the estimator noise would reach a maximum low level because of 21~cm field intrinsic fluctuations. Hence, the estimator noise saturates at $k^{\rm max}_p \sim 20$ for $z_s=8$ and $\Delta\nu = 5\mbox{ MHz}$ in this case. This effect will be clearly demonstrated in Section~\ref{sec:Results} for our particular model.

\subsubsection{Faster Lensing Estimator}
\label{sec:FastEst}
Estimator~(\ref{eqn:potest}) is computationally slow to calculate.
As shown in \cite{Anderes13}, \cite{LewisChallinor06} and \cite{CarvalhoMoodley10} for the analogous 2D CMB case, the estimator can be interpreted as a convolution in Fourier space which is equivalent to a real space product.  Calculating the product in real space allows one to take 
advantage of Fourier Transforms (FFTs) methods\footnote{This will be true in the full-sky representation too, since the azimuthal integrals can be treated similarly.}  such as FFTW\footnote{http://www.fftw.org/} to do the sums. Extending their derivation for $k_p$ modes, we have:
\begin{equation}
\label{eqn:FastEst}
\hat{\Phi}_{\bm{L}} = - \frac{N^{\hat{\Phi}}_{L}}{\Omega_s} \left(i\bm{L}\right) \cdot \sum_{k_p}{\left[\sum_{\bm{\theta}}{e^{-i\bm{L}\cdot\bm{\theta}}F_{\bm{\theta}}\bm{\nabla G}_{\bm{\theta}}}\right]_{k_p}} = - \frac{N^{\hat{\Phi}}_{\bm{L}}}{\Omega_s} \left(i\bm{L}\right) \cdot \sum_{k_p}{\bm{H}_{\bm{L},k_p}},
\end{equation}
where $\bm{H}$ is defined here and $F$ and $G$ are 2D angular space maps of the input 21\,cm intensity temperature field, 
defined by applying the following high-pass filters in Fourier space
\begin{equation}
\label{eqn:FiltFunc}
F_{\bm{l},k_p} = \frac{\tilde{T}_{\bm{l},k_p}}{C^{T}_{l,k_p}}, \qquad G_{\bm{l},k_p} = \frac{C_{l,k_p}\tilde{T}_{\bm{l},k_p}}{C^{T}_{l,k_p}}.
\end{equation}
In this way every $k_p$ contribution to the estimator is computed individually, by filtering the input fields and multiplying their inverse DFTs in real space.

As pointed out by \cite{LewisChallinor06}, seen from this point of view, the estimator measures the correlations in the product of  two Wiener filtered fields, the temperature gradient field, $\bm{\nabla G}(\bm{\theta})$, and the small-scale weighted field $F(\bm{\theta})$.

\subsection{Including the Beam}
\label{sec:BeamedEst}

In order to simulate more accurately the observational effects of a real telescope we include a beam. The beam smooths the signal coming from scales that are small with respect to the beam resolution scale (multipoles $L>L_{\rm cut}$) and the estimator can be modified to take this into account. 

The observed point will have a sky noise contribution $n^{\rm Sky}_{\bm{x}}$ and a receiver noise contribution $n^{\rm Rcv}_{\bm{x}}$, so that
\begin{equation}
\tilde{\mathcal{T}}_{\bm{x}} = \sum_{\bm{x}'}{W_{xx'} \left( \tilde{T}_{\bm{x}'} + n^{\rm Sky}_{\bm{x}'} \right) } + n^{\rm Rcv}_{\bm{x}},
\end{equation}
with Fourier transform 
\begin{equation}
\tilde{\mathcal{T}}_{\bm{l}, k_p} = W_{l}\left(\tilde{T}_{\bm{l}, k_p} + n^{\rm Sky}_{l}\right) + n^{\rm Rcv}_{l}.
\end{equation}
The beaming function $W_l$ could depend on frequency, but here we will assume it does not and that it generates no spurious correlations among $k_p$ modes. From this the discrete quadratic estimator can be found following the procedure outlined in Section~\ref{sec:LensRec}. We find
\begin{equation}
\label{eqn:BeamPotEst}
\hat{\phi}_{\bm{L}} = \frac{\mathcal{N}^{\hat{\phi}}_{L}}{2\Omega_s}\sum_{\bm{l},k_p}{\left\{\frac{W_lW^\star_{|l-L|}\left[\bm{L}\cdot\bm{l}\,C_{l,k_p} + \bm{L}\cdot\left(\bm{L}-\bm{l}\right)C_{|l-L|,k_p}\right]}{\mathcal{C}^{T}_{l,k_p}\mathcal{C}^T_{|l-L|,k_p}}\right\}\, \tilde{\mathcal{T}}_{\bm{l},k_p}\tilde{\mathcal{T}}^\star_{\bm{l}-\bm{L},k_p}}
\end{equation}
with $\mathcal{C}^T_{l,k_p} = |W_l|^2\left( C_{l,k_p} + N_{l,k_p}^{\rm Sky} \right) + N^{\rm Rcv}_{l,k_p}$. The estimator noise will consequently be modified into
\begin{equation}
\label{eqn:BeamEstNoise}
\mathcal{N}_{L}^{\hat{\phi}} = \left\{\frac{1}{2\Omega_s}\sum_{\bm{l},k_p}\frac{|W_l|^2|W_{|l-L|}|^2\left[\bm{L}\cdot\bm{l}\,C_{l,k_p} + \bm{L}\cdot\left(\bm{L}-\bm{l}\right)C_{|l-L|,k_p}\right]^2}{\mathcal{C}^T_{l,k_p}\mathcal{C}^T_{|l-L|,k_p}}\right\}^{-1}.
\end{equation}
This beamed discrete estimator noise is easily computable in a reasonable amount of time by parallelizing the innermost sums in the latter equation. So, if we re-define our filters Eqs.(\ref{eqn:FiltFunc}) as
\begin{equation}
\mathcal{F}_{\bm{l},k_p} = \frac{W_l\tilde{\mathcal{T}}_{\bm{l},k_p}}{\mathcal{C}^T_{l,k_p}}, \qquad \mathcal{G}_{\bm{l},k_p} = \frac{W_lC_{l,k_p}\tilde{\mathcal{T}}_{\bm{l},k_p}}{\mathcal{C}^T_{l,k_p}},
\end{equation}
we can find the beamed version of Eq. (\ref{eqn:FastEst}), namely
\begin{equation}
\label{eqn:BeamEst-2}
\hat{\phi}_{\bm{L}} = - \frac{\mathcal{N}^{\hat{\Phi}}_{L}}{\Omega_s} \left(i\bm{L}\right) \cdot \sum_{k_p}{\bm{\mathcal{H}}_{\bm{L},k_p}},
\end{equation}
where $\bm{\mathcal{H}}_{\bm{L},k_p}$ is again the Fourier transformed vectorial field formed by multiplying the inverse transformed Fourier $\mathcal{F}_{\bm{l},k_p}$ with the inverse Fourier transformed gradient of $\mathcal{G}_{\bm{l},k_p}$. A detailed derivation is presented in Appendix \ref{sec:AppA}.

For multipoles $L>L_{cut}$, the estimator noise diverges because of the dominance of thermal noise at those scales and the smoothing of structure by the beam. Explicitly incorporating the beam allows us to avoid aliasing and pixelization effects. The value of $L_{\rm cut}$ will be specified in Section~\ref{sec:BeamMaps}, and will depend on the observed redshift and telescope design. A low $L$ cutoff, $L_{\rm min}$ reflecting the finite field-of-view can also be incorporated into the beam.  We choose here to allow the boundaries of the simulated maps to implicitly impose this cutoff at $L_{\rm min} \simeq 2\pi/\theta^{\rm max}$.

\section{Simulation Details}
\label{sec:SimDet}
In this Section we will describe the method used to perform the lensing reconstruction on simulated 21~cm temperature maps, giving particular attention to the SKA instrumental configuration. Since the reconstruction signal-to-noise is dependent on the telescope's specifications, the telescope design will be crucial for these observations. We first describe how we generate the Gaussian random temperature fluctuation field, the lensing potential field and how we combine them to get the lensed temperature field. Then we explain how to model the thermal noise components due to sky contamination and instrumentation. Finally we apply an explicit model for the beam.

\begin{figure*}
\centering
\subfigure[Unlensed 21 cm simulated box]{\label{fig:Unl21box}\includegraphics[scale=0.35]{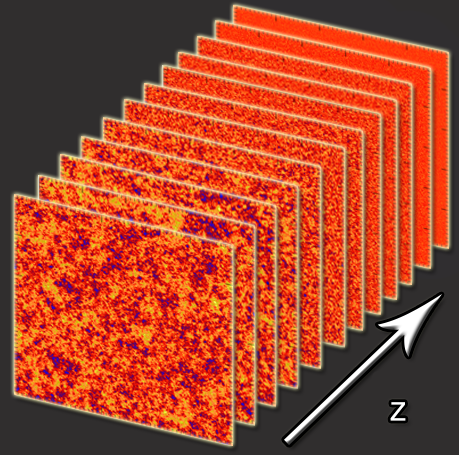}} \quad
\subfigure[Lensed 21 cm simulated box]{\label{fig:Len21box}\includegraphics[scale=0.35]{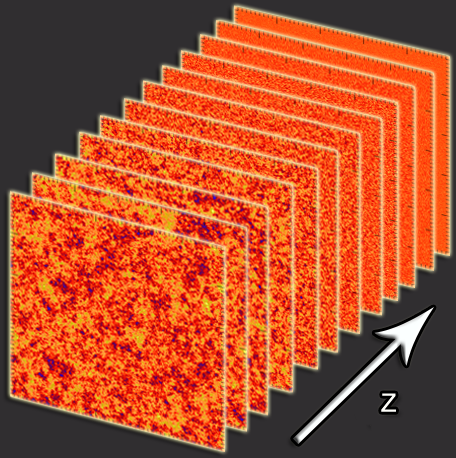}} \quad
\subfigure[Thermal noise simulated box]{\label{fig:ThNbox}\includegraphics[scale=0.35]{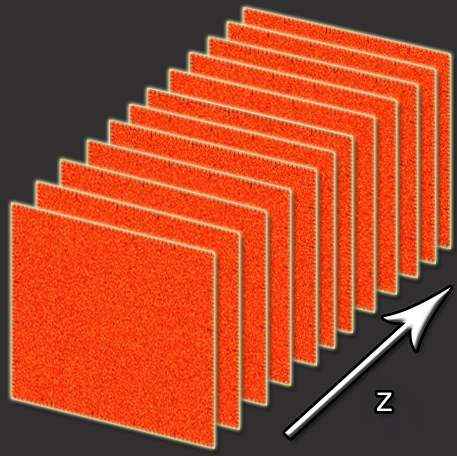}} \\
\subfigure[Unlensed 21 cm slice for $k_p=3$]{\label{fig:Unl21map}\includegraphics[scale=0.23]{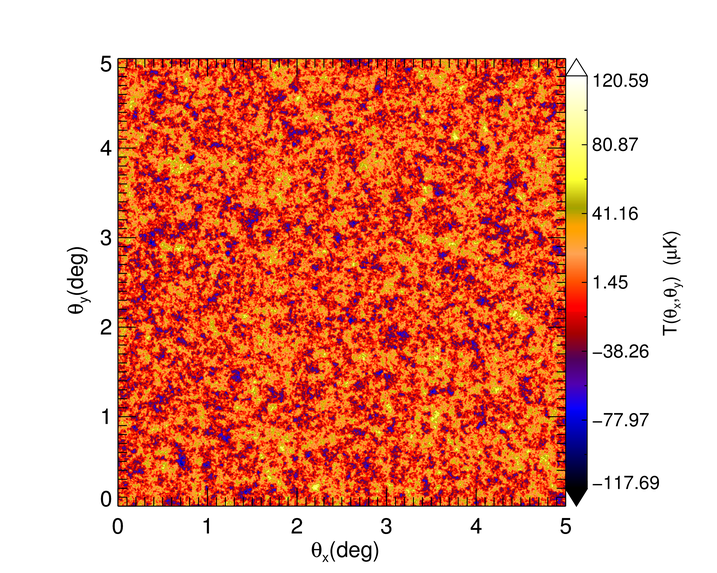}} 
\subfigure[Lensed 21 cm slice for $k_p=3$]{\label{fig:Len21map}\includegraphics[scale=0.23]{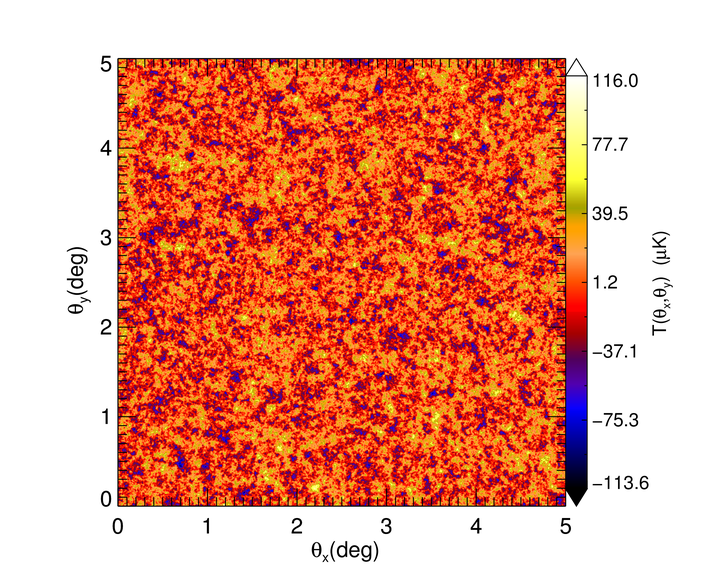}} 
\subfigure[Thermal noise slice for $k_p=3$]{\label{fig:ThNmap}\includegraphics[scale=0.23]{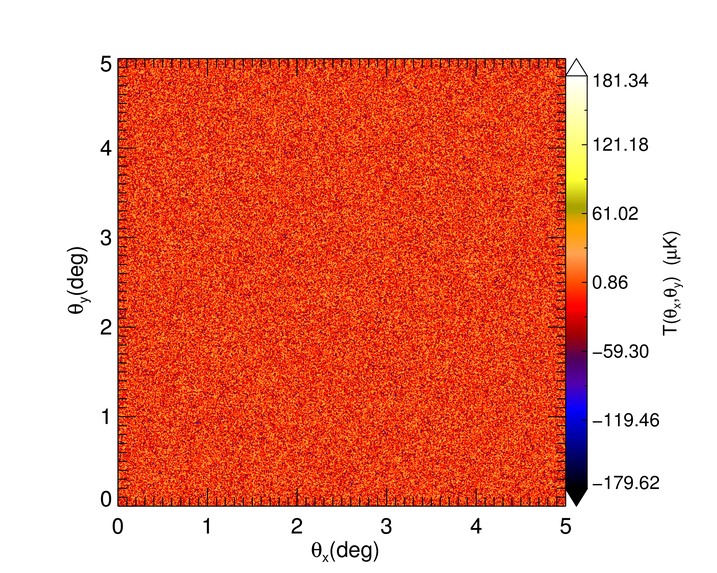}}
\caption{{\it Top}: Sample realisations of the simulated box centred around $z_s=8$ for every simulated component. The angular area is $\Omega_s = 5\degree\times 5\degree$ and $N_{\rm side}=650$. {\it Bottom}: The $k_p=3$ map extracted from the cubes are shown in the bottom panels.}
\label{fig:Cubes}
\end{figure*}

\subsection{Discrete Modeling of 21\,cm Field }
\label{sec:21cmMod}
For testing, the 21\,cm brightness temperature field is taken to be a Gaussian random field.  We generate, at each $k_p$, $l$  mode, a real and imaginary parts from
\begin{equation}
\label{eqn:GRF}
\mathcal{R}(T_{\bm{l}, k_p}) \propto G_1 \sqrt{\frac{C_{l,k_p}}{2}} \qquad \mathcal{I}(T_{\bm{l}, k_p}) \propto G_2 \sqrt{\frac{C_{l,k_p}}{2}},
\end{equation}
where $G_{1,2}$ are two random normally distributed numbers with null mean and unitary standard deviation. The full intensity temperature 3D field reality condition requires that $T_{-\bm{l}, - k_p} = T^\star_{\bm{l}, k_p}$, but because of FFTW storage convention only half a cube (the positive $k_p$ spectrum) is really needed to efficiently perform FFTs. By doing this we can take into account the correlations between maps simulated at different $z$.

As discussed in Section~\ref{sec:21cmfield}, we use Eq.~(\ref{eqn:21cmPS}) for the brightness temperature fluctuations power spectrum.
We approximate the non-linear matter power spectrum for structure formation using the \cite{PeacockDodds96} method although the lensing signal and noise are relatively insensitive to non-linear scales. The tests performed within this work justify this assumption.

In Figure~\ref{fig:Unl21box} we show a sample of our simulation boxes, where we can see the unlensed 21\,cm brightness temperature fluctuation field produced for several $k_p$. Modes with larger values of $k_p$ have less power and the signal quickly decays below the thermal noise level with increasing $k_p$, as shown in Figure~\ref{fig:21cmPS}. Because of this, we do not need to simulate a large number of $k_p$ maps, allowing the code to make maps more quickly.

\subsection{Lensing Maps}
\label{sec:LensMaps}
As we will show, we do not expect highly non-linear objects in the deflection potential to be detectable so we model the deflection field as a Gaussian random field in much the same way as we did the brightness temperature fluctuations. The potential field is generated analogously to what has been done in Section~\ref{sec:21cmMod}, but using the power spectrum (\ref{eqn:DefField}) and $C^{\Phi\Phi}_L = C^{\alpha\alpha}_L / L^2$ in Eq.~(\ref{eqn:GRF}). Then we produce the components of the deflection field in Fourier space, namely $\tilde{\bm\alpha}_{\bm L} = i\bm{L}\tilde{\Phi}_{\bm L}$. Lensed 21\,cm temperature brightness maps are produced applying, for each redshift map, a realisation of our randomly generated $x-$ and $y-$deflection field maps through bicubic interpolation of the values at the undeflected positions. When light rays are deflected outside the simulated source boundaries, periodic boundary conditions are applied by mirroring the source plane.
\begin{figure*}
\centering
\subfigure[Unlensed-Lensed Difference Map]{\label{fig:DiffMap}\includegraphics[scale=0.23]{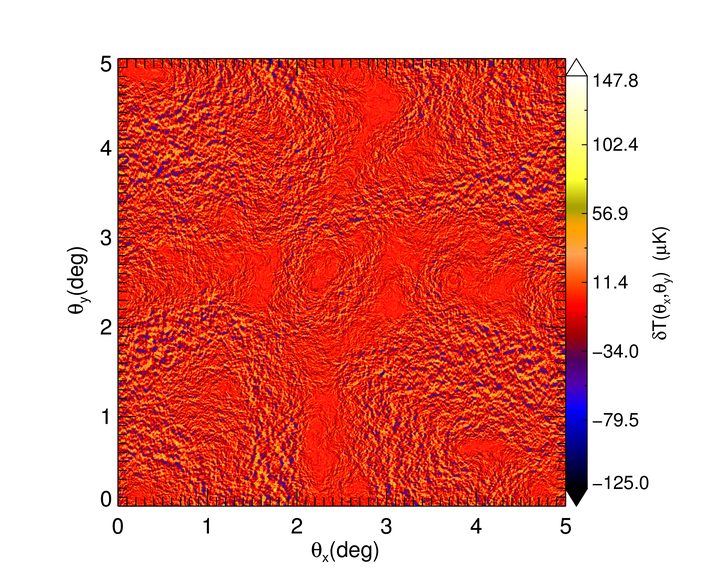}} \,
\subfigure[Potential Field]{\label{fig:LensPot}\includegraphics[scale=0.23]{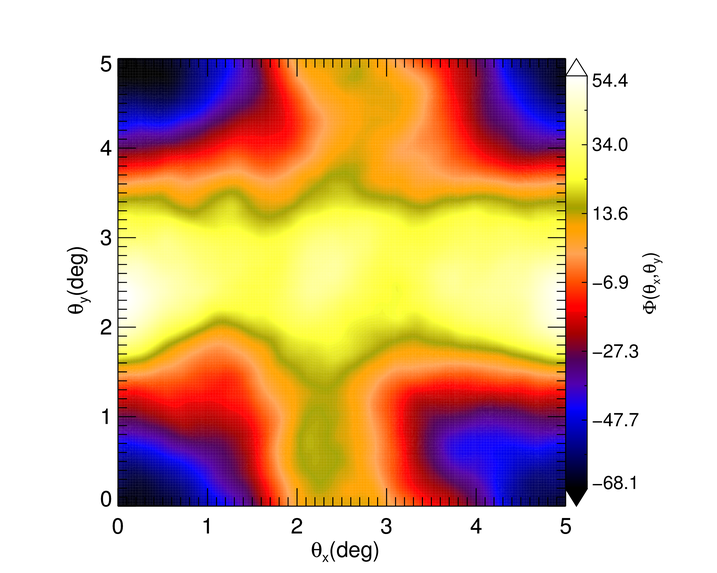}}  
\subfigure[Convergence Field]{\label{fig:LensKappa}\includegraphics[scale=0.23]{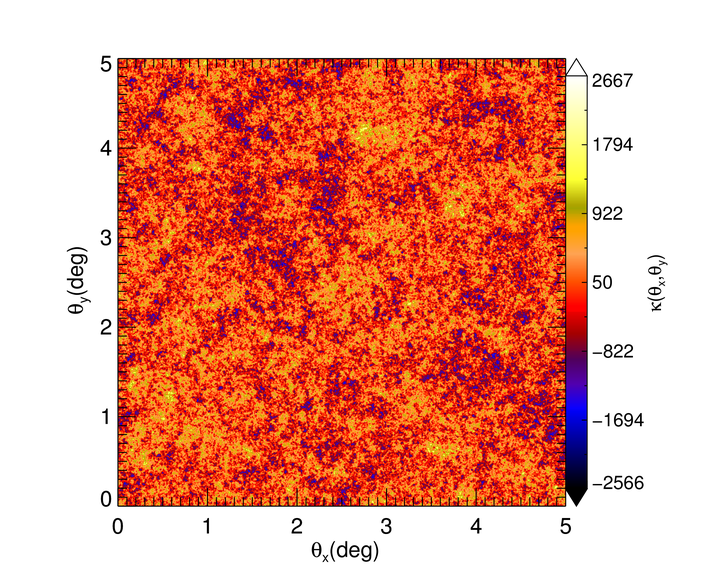}} 
\caption{{\it Left panel}: The map showing the difference between Figure~\ref{fig:Len21map} and Figure~\ref{fig:Unl21map}. {\it Central panel}: A sample realisation of the gaussian random potential field generating the deflection of 21~cm intensity points in Figure~\ref{fig:Len21map}. The potential values have been scaled by a factor $10^9$ in order to improve the readability of the colour bar. {\it Right panel}: The corresponding convergence field. The convergence values have been scaled by a factor $10^6$ in order to improve the readability of the colour bar. These maps have been computed using $N_{\rm side}=650$ and $\Omega_s = 5\degree\times 5\degree$.}
\label{fig:InputGRP}
\end{figure*}
Consider that the simulated rays will not intersect the source at the grid points on which the source itself was simulated.  As a result some interpolation is required. Bicubic interpolation of the source also removes visible pixelization artifacts in the lensed map. Another issue that has to be considered is the importance of structures in the source that are below the simulated resolution. In fact, lensed interpolation introduces scales below the ones allowed by map's dimensions. As pointed out by \citet{Lewis05} for CMB lensing, the importance of such scales can be addressed by downsampling a temperature field produced at a higher resolution (usually $3$ or $4$ times the desired resolution).  We find that after beam smoothing, section~\ref{sec:BeamModel}, the bicubic interpolation on a map of the same resolution as the image reconstruction is sufficient to produce accurate results. This was tested using downsampled intensity temperature maps. 
 
A sample box image (lensed) is shown in Figure~\ref{fig:Len21box}. It is hard, but not impossible, to see differences by eye between the unlensed and the lensed image, as one can notice by comparing Figures~\ref{fig:Unl21map} and~\ref{fig:Len21map}. The difference between the unlensed and the lensed image can be seen in detail in Figure~\ref{fig:DiffMap}. The potential and the corresponding convergence field realisation used to deflect the temperature points are shown in Figure~\ref{fig:InputGRP} for $N_{\rm side}=650$ and $\Omega_s = 5\degree\times 5\degree$. 

\subsection{Thermal Noise Component }
\label{sec:ThNoiseComp}
The lensing estimator and noise, given respectively by Equations~(\ref{eqn:potest}) and (\ref{eqn:EstNoise}), include a total power spectrum contribution $C^T_{\ell,k_p}$ which depends on the thermal noise power spectrum as well as the intrinsic 21~cm fluctuations.  The noise in this estimator is sensitive to the particular telescope model that is used. Here we will take into account the current model for the SKA-Low thermal noise which includes a realistic description of the array density distribution in visibility space. 

\subsubsection{The Thermal Noise Angular Power Spectrum}
A pair of elements in an interferometer separated by a baseline of length $D$ will measure a visibility $V(\bm{U},\nu)$, where $\bm{U}$ is the vector in visibility space and $U = \lvert\bm{U}\rvert = D/\lambda$. The resolution in visibility space defines the Field of View (FoV) of the telescope, namely $U_{\rm min} = \Diff2 U = 1/\Omega_s \sim D_{\rm min}^2/\lambda^2$, with $D_{\rm min}$ the interferometer element diameter which in the case of SKA-Low is a station containing a certain number $N_{\rm ant}$ of antennae. The  visibility space is interpretable as a Fourier dual space, and its relation to the multipole space is $ \bm{U} = \bm{l}/(2\pi)$, so that $(2\pi)^2\Diff2 U =\Diff2 l = L^2_{\rm min} $. The maximum observable visibility is hence set by the baseline maximum length $U_{\rm max} = L_{\rm max}/(2\pi) = D_{\rm max}/\lambda(z)$.

We can define the noise power spectrum in visibility space for an interferometer in the Rayleigh-Jeans limit by averaging all visibilities falling in one visibility space resolution, for a bandwidth $\Delta\nu$ centered around the redshift $z+1 = \nu_{21}/\nu$. That is
\begin{equation}
\label{eqn:28}
C_U^{\rm N} = \left[\frac{\lambda^2(z) T_{\rm sys}}{A_{\rm eff}}\right]^2\frac{\Diff2 U}{N_{\rm pol}\Delta\nu \,t_{\bm{U}}},
\end{equation}
\citep{ZaldarriagaFurlanetto04}, where $\lambda(z) = \lambda_{21}(1+z)$ and $A_{\rm eff}$ is the effective area of one station. $A_{\rm eff}$ is usually defined as $A_{\rm eff} = \varepsilon \pi D_{\rm min}^2/4$, with $\varepsilon$ the antenna efficiency, usually a number $0.7\lesssim \varepsilon\lesssim 1$. $N_{\rm pol}$ is the number of polarisation channels which can be added incoherently \citep{Morales05}.  $t_{\bm{U}}$ is the observation time per visibility pixel,
\begin{equation}
\label{eqn:29}
t_{\bm{U}} = \Diff2 U \, n(\bm{U},\nu) \, t_{\rm p}  = \Diff2 U \, n(\bm{U},\nu) \frac{t_o}{N_{\rm p}} = \Diff2 U \, n(U,\nu) \frac{t_o N_{\rm b} \Omega_s}{S_{\rm area}}.
\end{equation}
Here we have included the possibility to observe several sky patches, using different pointings $N_{\rm p} = S_{\rm area}/\Omega_s$ to scan a given sky area $S_{\rm area}$, and using a certain number $N_{\rm b}$ of beams per station with FoV $\approx\Omega_s\approx \lambda/D_{\rm station}$ observed within a time $t_{\rm p}$ per pointing.  This allows for increasing the number of independent measurements  in a given total observational time $t_o = N_{\rm p} t_{\rm p}$, since the number of observed modes is increased by a factor $S_{\rm area}/(N_{\rm b}\Omega_s)$. Note that $S_{\rm area} > N_{\rm b}\Omega_s$. For EoR observations we will consider $N_{\rm p} = 1$, and so the observing time per pointing will coincide with the total observation time.

The averaged baseline number density (over a 24 hrs period) is denoted as $n(\bm{U}, \nu)$, and it is usually a function of $(U,\nu)$ due to rotational invariance in visibility space given by a circularly symmetric baseline distribution \citep{VillaescusaViel14}. Its normalization will be frequency dependent, since $\int n(U,\nu) \,\Diff2 U =N_{\rm stat}(N_{\rm stat}-1)/2 $, with $N_{\rm stat}$ the number of stations forming the considered baseline within a diameter $D_{\rm max}$.  Hence, considering that the integral is constant and the maximum and minimum visibilities are frequency dependent, the number density needs to be scaled from a fiducial curve if it has to be computed at different frequencies. More details about this will be given in Section~\ref{sec:MultiBand}.


For an aperture array, the effective area of the station will be constant below a critical frequency $\nu_{\rm c}$, \textit{i.e.} when the array is dense, while above $\nu_{\rm c}$ it scales with frequency as
\begin{equation}
\label{eqn:31}
A_{\rm eff}(\nu) = A_{\rm eff}(\nu_{\rm c})\left\{ 
\begin{array}{rl}
\left(\nu_c/\nu\right)^2 & \mbox{ for} \:\nu\geq \nu_c \\
1 & \mbox{ for}\: \nu<\nu_c .
\end{array}
\right.
\end{equation}
Moreover the FoV scales at any frequency as $\Omega_s (\nu) = \Omega_s(\nu_{\rm c})\left(\nu_{\rm c}/\nu\right)^2$.
Hence, using  definitions ~(\ref{eqn:31}) and ~(\ref{eqn:29}) we can write Eq.~(\ref{eqn:28}) as
\begin{equation}
\label{eqn:32}
C_{l,\Delta\nu}^{\rm N} = \left[\frac{\lambda^2(z)}{A_{\rm eff}(\nu_{\rm c})}\left(\frac{\nu}{\nu_{\rm c}}\right)^2\right]^2\frac{T^2_{\rm sys}(\nu)}{N_{\rm pol}\Delta\nu \,t_o \, n\left[U = l/(2\pi), \nu\right]}.
\end{equation}

In the case of a uniform antenna density distribution, $n(U,\nu) \approx N^2_{\rm stat} \lambda^2(z)/\left(2\pi D^2_{\rm max}\right)$ and neglecting the frequency dependence of the effective area and the FoV, (\ref{eqn:32}) reduces to the widely used flat angular power spectrum
\begin{equation}
\label{eqn:ThNoisePS}
C_{l,\Delta\nu}^{\rm N} = \frac{(2\pi)^3 T^2_{\rm sys}}{t_o\Delta\nu  f^2_{cov} L_{\rm max}^2}
\end{equation}
\citep{FurlanettoOhBriggs06}. Within this approximation the channel polarisation contribution and the frequency scaling of the station area have been neglected. $f_{\rm cov}$ is the total collecting area of the telescope divided by $\pi (D_{\rm max}/2)^2$, the aperture covering fraction, while the highest multipole that the array is able to probe at the observed wavelength $\lambda_{\rm obs}(z)$ is $L_{\rm max} (z) = 2\pi D_{\rm max}/\lambda_{\rm obs}(z) = 2\pi U_{\rm max}$. 

Finally, the noise consists of one component coming from the sky and another one coming from the instrumentation, whose combined impact is taken into account by $T_{\rm sys}$. At such low frequencies the most important source of astrophysical noise is galactic synchrotron emission which produces a representative sky temperature of 
\begin{equation}
T_{\rm Sky} \simeq 1.1\times 60 \left(\frac{\nu_{obs}}{300\mbox{ MHz}}\right)^{-2.55}\mbox{ K}
\end{equation}
\citep{Dewdney13}. The receiver noise power spectrum is computed analogously by following Eq.~(\ref{eqn:ThNoisePS}) and setting $T_{\rm Rcv}$, the receiver temperature, in the place of $T_{\rm Sky}$. This contribution is added after the inclusion of the beam; however, it only becomes important at low redshifts. It is assumed that this contribution is uncorrelated with the signal and sky noise terms.  In Figure~\ref{fig:ThNbox} we show a sky noise component cube, where a different noise realisation is produced for each channel. 

\subsubsection{SKA1 and SKA2-Low Specifications}
\label{sec:3.3.2}
The design of SKA-Low is not yet finalized. The most complete official design document is \citet{Dewdney13}, but the recent rebaselining modified these specifications \citep{McPherson15}. De-scoping halved the number of receiving stations, but the frequency sensitivity is relatively unaffected with respect to the original design plan, because of the dense core array. We will take this into account by reducing the baseline density function by a factor of four, $n_{\rm descop}(U,\nu) = n(U,\nu)/4$, and so increasing the thermal noise power spectrum level by the same factor.

In this work we consider a SKA1-Low design with $D_{\rm min} = 35$ m diameter stations, and $N_{\rm stat} = 433$ within a maximum baseline of $D_{\rm max} = 4$ km. Generally, baselines larger than $4-5$ km do not contribute much to the total sensitivity for our purposes  although large baselines are used for calibration and foreground source removal. The critical frequency is $110$ MHz, and the values for the effective area and field-of-view (FoV) at $\nu_{\rm c}$ are $A_{\rm eff}(\nu_{\rm c}) = 925 \mbox{ m}^2$ and $\Omega_s(\nu_c) = 27\mbox{ deg}^2$ respectively. The receiver temperature is set to be $T_{\rm rcv} = 40$ K. SKA-Low is also assumed to have two polarization channels, and we will consider a single-pointing observation ($N_{\rm p} = 1$) performed at EoR redshifts. The fiducial baseline density function has been provided at $z=8$ (corresponding to a central frequency of $157.82$ MHz) (J. Pritchard, private communication).

Observational times and bandwidths are given in Table~\ref{tab:SKAconfigs}, where we define the hypothetical R0, R1, and R2 survey strategies The total frequency range explored by SKA-Low is $50-350\mbox{ MHz}$, but these bandwidths are the one used in the lensing estimator so that there can be multiple lensing maps centered on different source redshifts.

\begin{table}
\centering
\begin{tabular}{|c|c|c|}
\hline
   & $\Delta\nu\mbox{ [MHz]}$    & $t_{o}\mbox{ [hrs]}$ \\ \hline
R0 & $8$   & 1000         \\ \hline
R1 & $5$ & 1000        \\ \hline
R2 & $5$ & 2000        \\ \hline
\end{tabular}
\caption{The considered SKA simulation settings for this study at $z=8$. For every case we considered a telescope baseline diameter of $D=4\mbox{ km}$, while the single station element has a diameter of $35$ m. For every case we assume a total frequency range of $50-350\mbox{ MHz}$ and the critical frequency is $\nu_{\rm c} = 110$ MHz.}
\label{tab:SKAconfigs}
\end{table}

For the R1 and R2 strategies we have used $\Delta\nu = 5\mbox{ MHz}$. This bandwidth $\Delta\nu$ is sufficiently thin to have good resolution over a certain redshift range, which is good for exploring EoR epoch, but thick enough so that correlations between bands can be ignored. As discussed in Section~\ref{sec:21cmfield}, we will assume that the lensing, angular size distance and the statistical properties of the source do not change within $\Delta\nu$. The single channel resolution for SKA-Low is $\delta\nu = 100$ kHz. There is a maximum number of detectable $k_p$ modes which will depend on the ratio between the bandwidth and the frequency resolution in a single channel \citep{ParsonsPoberMcQuinn12}. Note that \cite{PritchardIchiki15} have assumed $B=8\mbox{ MHz}$ and $t_{o}=1000$ hours, which corresponds to our R0 survey strategy. The R2 survey strategy has been introduced to keep a comparable thermal noise level to R0 and have the possibility to stack more frequency bands within a given redshift range. 

SKA-Low Phase 2 has still to be formally defined, but we will assume a total collecting area that is four times the one expected for SKA1-Low. This will cause the thermal noise level to be a factor 16 lower. We do not include multiple beams, although it could have as many as $N_{\rm b} \sim 10$ beams simultaneously. Following \citet{Santos:2015gra}, we increase the sensitivity of this instrument also by decreasing the receiver noise to $15$ K, although this does not cause a big change in the total system noise due to sky noise domination at these frequencies.  

Sky and receiver noise power spectra for the R2 configuration are shown in Figure~\ref{fig:21cmPS}, where they are compared with the brightness temperature angular power spectrum for different $k_p$ modes. 

\subsection{Modeling The Beam}
\label{sec:BeamModel}
\begin{figure}
\centering
\subfigure{\label{fig:BL21box}\includegraphics[scale=0.45]{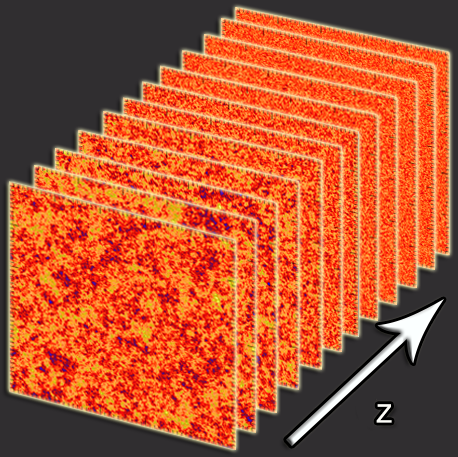}}
\subfigure{\label{fig:BL21map}\includegraphics[scale=0.33]{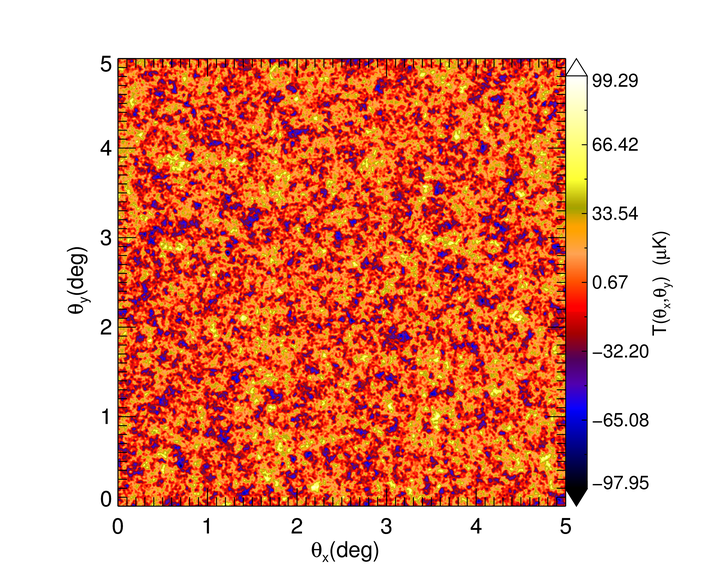}} 
\caption{{\it Top}: A sample realisation of the simulated beamed box centred around $z_s=8$,  for different redshift channels within one frequency bandwidth. The angular area is $\Omega_s = 5\degree\times 5\degree$ and $N_{\rm side}=650$. {\it Bottom}: One map extracted from the above cube. The beam smoothing is $2.5$ times the basic map resolution, namely $\Delta b = 1.15\mbox{ arcmin}$.}
\label{fig:CubesB}
\end{figure}

 In Section~\ref{sec:BeamedEst} we described the effect of the beaming function, which is to smooth the Fourier frequencies near the characteristic beam frequency $L_{\rm cut}$ corresponding to a beam resolution $\sigma$.  We will make the approximation that this is constant within $\Delta \nu$ although this can be easily relaxed. We use a simple Gaussian beam
\begin{equation}
W_l = e^{-l(l+1)\sigma^2/2},
\end{equation}
with $\sigma = b\Delta\theta / \sqrt{8\ln {2}}$; the $b$ parameter quantifies the beam resolution with respect to the pixelization of the simulated map $\Delta\theta$, namely $b = L_{\rm Nyq}/L_{\rm cut}$. 
The top panel of Figure~\ref{fig:CubesB} is a realisation of our beamed simulation box, whose $k_p=3$ slice is shown in the bottom panel. The suppression of the smallest scale modes with respect to Figure~\ref{fig:Len21map} is hard to notice.
\label{sec:BeamMaps}
Assuming the SKA-Low instrument modelled in Section~\ref{sec:3.3.2}, the typical beam resolution is $\Delta b = b\Delta\theta = 1.15\mbox{ arcmin}$. This value is set by $ \Delta b = \sqrt{2}\pi/L_{\rm cut}$ for a square grid\footnote{The $\sqrt{2}$ factor comes from considering a square grid, in which the Nyquist mode is $L_{\rm Nyq} = \sqrt{L_x^2 + L_y^2} = \sqrt{2}\Delta L N_\perp/2$, with $\Delta L = 2\pi/\theta_{\rm max}$ the resolution in Fourier space. The extension for a rectangular grid is obvious.}, where $L_{\rm cut} = L_{\rm max}$, the maximum multipole which is observed by the interferometer baseline.

\subsubsection{Aliasing and the Beam}
\label{sec:AliaBeam}
When using the fast lensing estimator, (\ref{eqn:BeamEst-2}), we find that when the beam cut off, $\sigma$, is close to the resolution of the image or when no beam is taken into account, spurious aliasing effects occur that cause the lensing signal and noise to disagree with the input signal and the analytically calculated noise. The quadratic estimator is a convolution of filtered fields and there will be a visible aliasing effect if the beam resolution is too close to the Nyquist frequency, resulting in a contamination due to already existing frequencies mirrored around the Nyquist frequency. Incorporating a beam solves this problem for our estimator because it acts as a low-pass filter that reduces the aliased contamination coming from high frequency modes\footnote{Another way to solve this problem is by padding the temperature field in Fourier space with a sufficient number of null arrays. This may be computationally expensive, especially for multi-dimensional arrays. Alternative FFTW efficient methods that do not involve padding in Fourier space have been developed. Interested readers can consult \citet{BowmanRoberts10}.}. This problem is much less prominent for CMB lensing because in that case there is relatively little power in the high frequency modes. To reduce memory usage and computational time it is advantageous to keep $L_{\rm Nyq}$  as small as possible while avoiding this aliasing problem. We found that the beam resolution has to be bigger than $2.5\Delta\theta$: this means that $L_{\rm Nyq} \geq 2.5 L_{\rm cut}$ to eliminate this effect. Tests of this limit are discussed in Section~\ref{sec:AliaTest}.

\section{Results}
\label{sec:Results}
In this Section we investigate the possibility of reconstructing high-quality images of the weak lensing potential field and present our results. We then discuss how the telescope and survey parameters influence these results as well as various tests of the performance of the estimator derived in Section~\ref{sec:BeamedEst}.

\subsection{Single-Band Reconstruction}
\label{sec:Imaging}
We will first consider a single-band measurement made with a frequency bandwidth $\Delta\nu$, whose values are specified in Table~\ref{tab:SKAconfigs}, centered around the observational redshift $z=8$, corresponding to an observed central frequency of $157.82$ MHz. The FoV at this frequency is determined by the size of the telescope's smallest element and is $\Omega_s = 13\mbox{ deg}^2$, hence the angular size per map side is $\theta_{\rm side} = 3.6\degree$. This survey area corresponds to an observed fraction of the sky $f_{\rm sky} = 3.15\times 10^{-4}$. The maximum probed multipole at such a redshift is $L_{\rm max}  = L_{\rm cut}\simeq 13230$. Because of the aliasing effect explained in Section~\ref{sec:AliaBeam} we need to generate the 21~cm intensity source field with a $L_{\rm Nyq}$ that is at least 2.5 times the beam cut off,  $L_{\rm cut}$. This is $L_{\rm Nyq} = 33092.6$, which corresponds to an angular resolution of $27.7$ arcsec. The number of pixels per map side is $N_{\rm side} = \sqrt{2} L_{\rm Nyq}/\Delta l = 468$, with $\Delta l = L_{\rm min}= 2\pi/\theta_{\rm side} = 100$ the resolution in Fourier space.

We generate our 21~cm temperature brightness maps and lens using a single realisation of the lensing potential field, following the simulation method described in Sections~\ref{sec:21cmMod} and ~\ref{sec:LensMaps}. Then we add the thermal noise component modelled in Section~\ref{sec:ThNoiseComp} to the lensed temperature map and smooth the field with the beam function introduced in Section~\ref{sec:BeamModel}. In the final step we add a different realisation of the receiver noise contribution for each $k_p$ mode which is valid under the assumption that the noise level is constant within the band. Using the estimator described in Section~\ref{sec:BeamedEst} with the sum over a given number of $k_p$ modes we obtain maps of the estimated lensing potential. The maximum number of combined frequency channels per frequency band is taken to be $k_p^{\rm max} = 20$ throughout the paper, unless otherwise stated.

\begin{figure}
\centering
\includegraphics[scale=0.37]{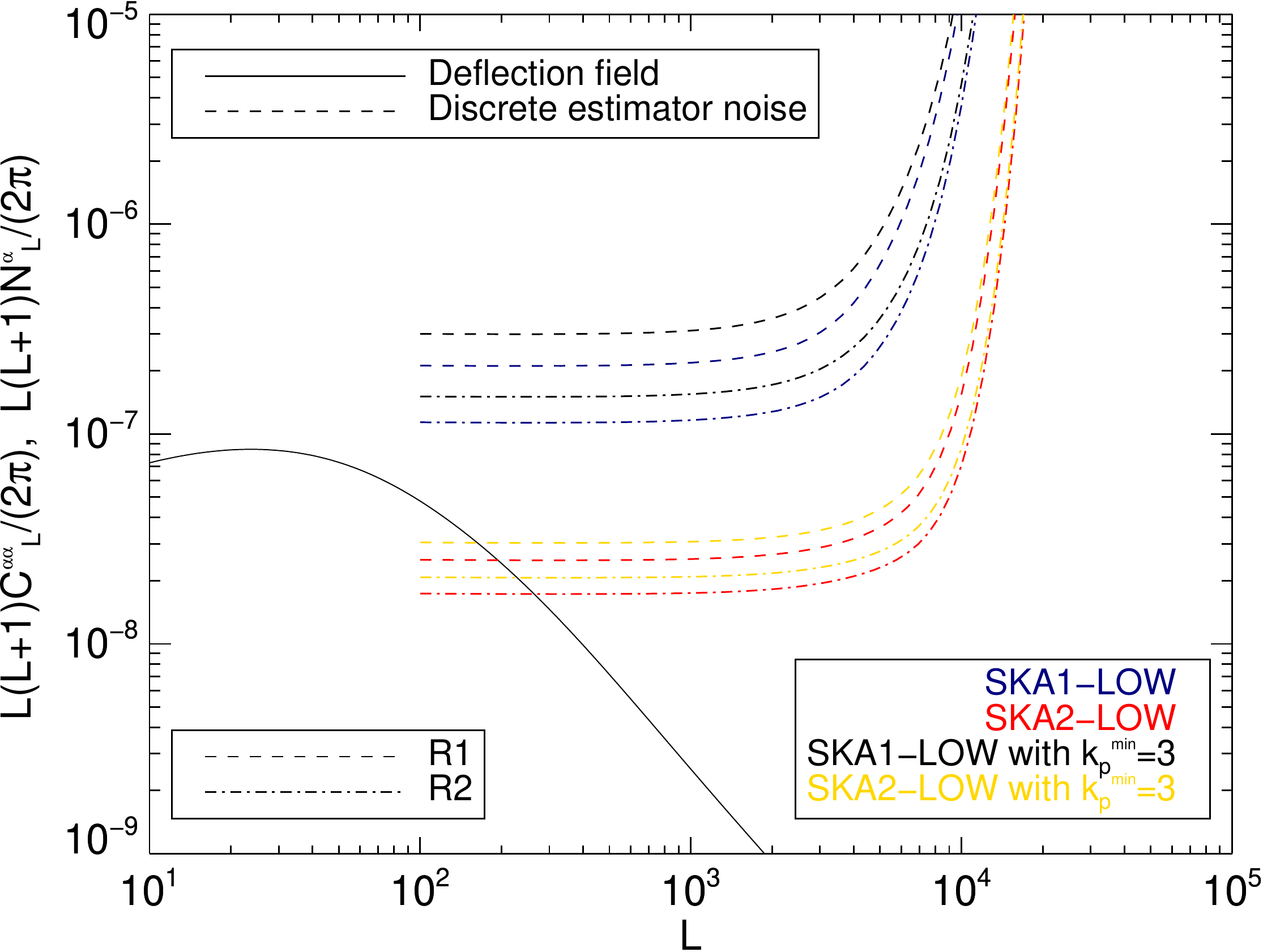}
\caption
{The discrete estimator noise for SKA1-Low (blue), SKA2-Low (red), SKA1-Low with $k_p^{\rm min}=3$ (black), and SKA2-Low with $k_p^{\rm min}=3$ (gold), with choices for observation time and bandwidth listed in Table~\ref{tab:SKAconfigs} and for the non-flat thermal noise power spectrum introduced in Section~\ref{sec:ThNoiseComp}. The simulated sky area is $\Omega_s = 3.6\degree\times 3.6\degree$ and the beam has a resolution of $1.15\mbox{ arcmin}$ at $z=8$. The R0 survey strategy results are not plotted because they produce an estimator reconstruction noise level close to the R2 one. The R1 configuration is denoted by dashed lines while the R2 by dashed-dot lines.}
\label{fig:EstNvarSKA-New}
\end{figure}

The signal-to-noise of the lensing reconstruction depends on several parameters of the telescope configuration and survey strategy.
For a fixed source redshift like $z_s = 8$, the behavior is mainly driven by the thermal noise, shown in Figure~\ref{fig:21cmPS}, which can be considerably different if the collecting area, observational time, or observation bandwidth is changed\footnote{Varying $L_{\rm max}$ also has an influence on the estimator noise level, but this effect is important only if we vary the source redshift (as we will see in the next section) or the telescope's maximum baseline length (which is considered fixed in this work).}. To have a preliminary idea of the reconstruction quality, we compute the discrete estimator reconstruction noise for the survey strategies introduced in Table~\ref{tab:SKAconfigs}. The results are shown in Figure~\ref{fig:EstNvarSKA-New}, where we have omitted the R0 results as they are similar to the R2 results. 

\begin{figure*}
\centering
\subfigure[Input Gaussian random potential]{\label{fig:GRP468}\includegraphics[scale=0.23]{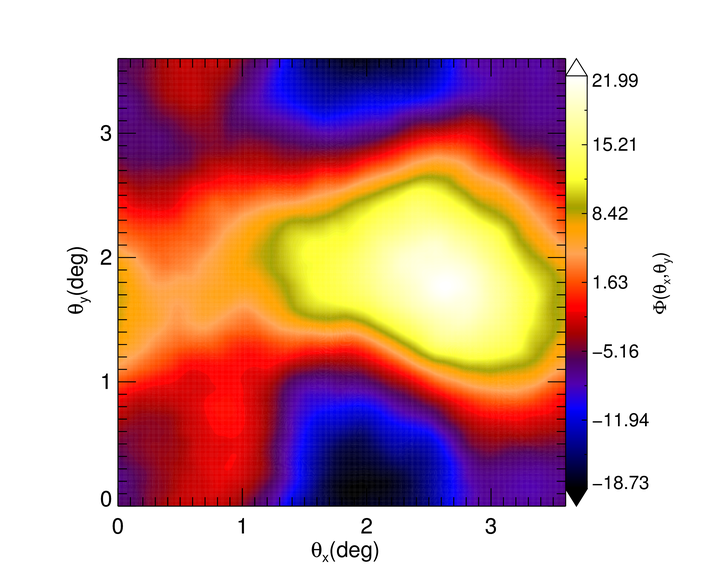}} 
\subfigure[Recovered estimator noise image]{\label{fig:RecEstN468}\includegraphics[scale=0.23]{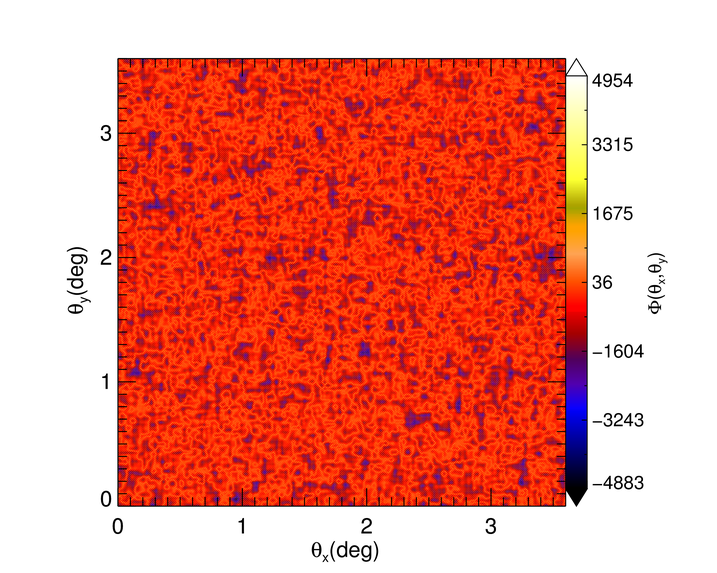}} 
\subfigure[Recovered potential image]{\label{fig:RecGRP468}\includegraphics[scale=0.23]{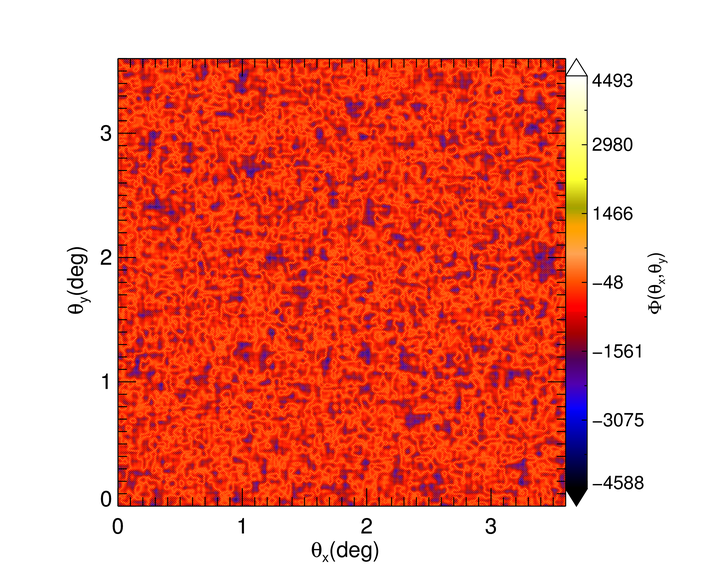}} 
\caption{\textit{Left panel}: the input potential field used to deflect the input 21~cm maps. \textit{Central panel}: the recovered estimator noise image obtained by computing the estimator with the unlensed temperature map. \textit{Right panel}: the recovered estimator image obtained by computing the estimator using the lensed 21~cm temperature map. The images are computed for $N_{\rm side} = 468$, $\Omega_s = 3.6\degree\times 3.6\degree$, $z=8$, $\Delta b = 1.15\mbox{ arcmin}$, $k_p^{\rm min} = 3$, and the SKA2-Low R2 configuration. The potential values in these maps have been scaled by a factor $10^9$ in order to improve the readability of the colour bars. }
\label{fig:NonUnifNoiseReconstr}
\end{figure*}

In figure~\ref{fig:EstNvarSKA-New}, we also show the results obtained considering $k_p^{\rm min} = 3$, in order to take into account the possible removal of the first $k_p$ modes caused by foreground cleaning techniques.  The exact number of $k_p$ modes to be removed will depend on the specific foreground removal technique \citep{McQuinnZahnZaldarriaga06,LiuTegmark12}. The noise is increased by omitting these modes, but not drastically. In general, the more serious the foreground contamination is, the higher is the number of $k_p$-modes to remove. This is an issue that needs to be investigated more deeply in the future and that our simulation pipeline is designed to handle.

High fidelity maps are possible from a single frequency band when the noise in figure~\ref{fig:EstNvarSKA-New} is below the expected signal power spectrum, so when typical fluctuations are detected with high signal-to-noise. For every SKA1-Low case the noise is well above the signal. This means that for a single frequency band detection made with the SKA1-Low instrument it is not possible to get high-fidelity images of the reconstructed lensing mass distribution. The situation will be different if multiple frequency bands are stacked as will be considered in Section~\ref{sec:MultiBand}. On the other hand, the results for SKA2-Low configurations are considerably more optimistic. A high fidelity imaging of the underlying mass distribution should be possible for the SKA2-Low experiment even in our worst case R1 with $k_p^{\rm min}=3$, whose noise level crosses the signal power spectrum at $L\sim 200$. Comparing the R1 and R2 survey strategies, it can be noted how using more observational time does not substantially improve the overall signal-to-noise.
For such a single-band measurement, we expect to recover images which are heavily contaminated by small-scale noise due to the small number of large lensing modes that are below the noise level. This is confirmed by looking at Figure~\ref{fig:NonUnifNoiseReconstr}, where we show the input potential, the estimator noise image, and the recovered image for the SKA2-Low R2 model with $k_p^{\rm min}=3$. 

\begin{figure}
\centering
\includegraphics[scale=0.37]{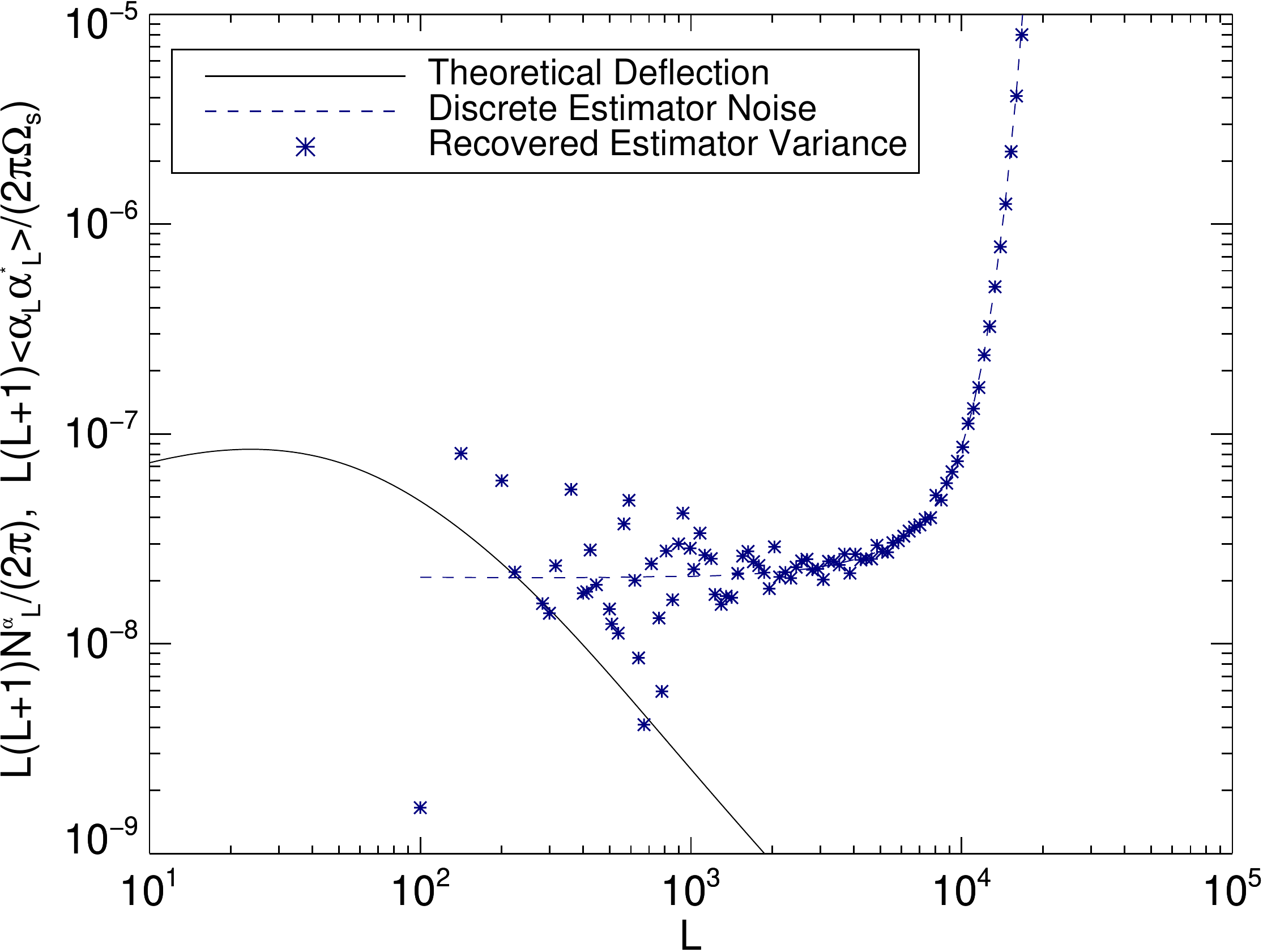}
\caption
{The recovered estimator square amplitude in Fourier space from the reconstructed potential image Figure~\ref{fig:RecGRP468}.}
\label{fig:RecEstPS468}
\end{figure}

In  Figure~\ref{fig:NonUnifNoiseReconstr}, the small-scale noise overwhelms the signal, making it nearly indistinguishable from the noise image. The reconstruction noise image is obtained by using the non-lensed 21~cm temperature map in the estimator. Figure~\ref{fig:RecEstPS468} is shown to underline that the recovered square amplitude of the modes in Fourier space follows the theoretical profile, but the high small-scale noise appearing in the recovered image is due to the estimator reconstruction noise feature at high multipoles.  

\subsubsection{De-noising the Reconstructed Image}
\label{sec:Wiener}
The findings of the previous Section suggest that the image of the recovered potential can be visualised if a proper de-noising procedure is applied to remove the small-scale noise contaminated map. For this purpose a Wiener filter has been used to unveil the reconstructed potential image. The Wiener filtering is an optimal method, \textit{i.e.} it minimizes the estimated image variance, and is often used for deconvolutions or images degraded by additive noise and blurring caused by a Point Spread Function (PSF). 
This approach requires the second-order stationarity and the statistical independence of signal and noise processes which are satisfied in our test case.  Moreover there must not be correlations between signal and noise, \textit{i.e.} the noise has to be additive. This means that the input image is $S(i,j) = s(i,j) + N(i,j)$, where $s(i,j)$ is the uncontaminated image and $N(i,j)$ is the additive noise. 

\begin{figure*}
\centering
\subfigure[De-noised estimated potential]{\label{fig:GRPWien}\includegraphics[scale=0.23]{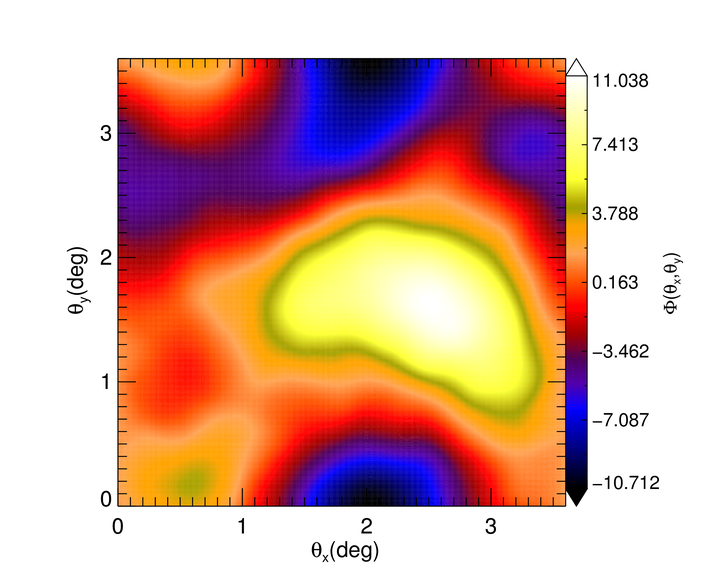}} \,
\subfigure[De-noised Fourier square amplitude]{\label{fig:GRPWienPS}\includegraphics[scale=0.24]{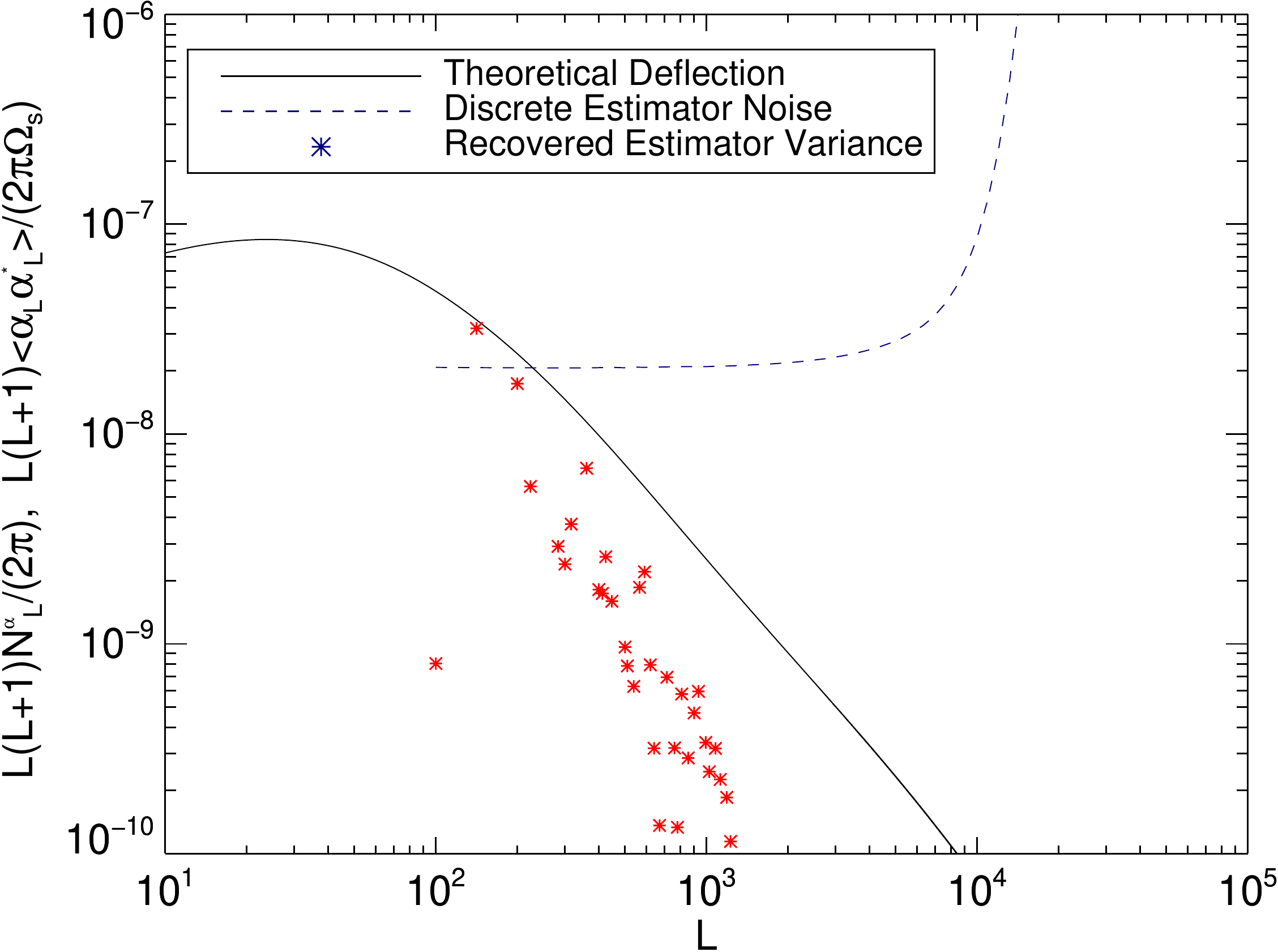}} \,\,
\subfigure[Fidelities for estimated images]{\label{fig:FidelityWien}\includegraphics[scale=0.24]{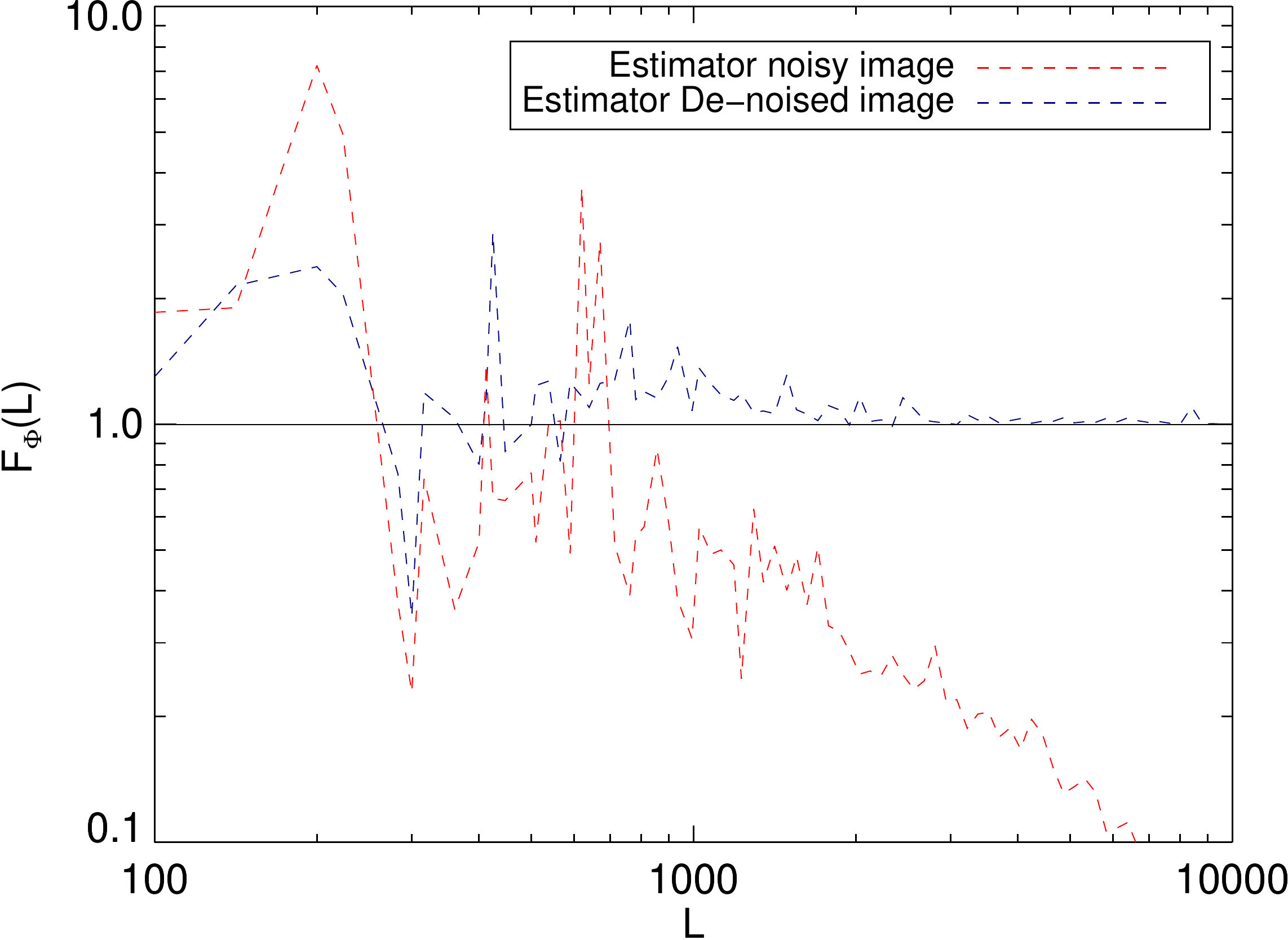}} 
\caption{\textit{Left panel}: the de-noised potential field estimator obtained by applying a Wiener filter to image~\ref{fig:RecGRP468}. The recovered potential values have been scaled by a factor $10^9$ in order to improve the readability of the colour bar. \textit{Central panel}: the recovered de-noised estimator squared amplitude in Fourier space (red star points) compared to the input deflection field power spectrum (black solid line) and to the discrete estimator noise (blue dashed line). \textit{Right panel}: the fidelity of the Wiener de-noised estimated potential (blue dashed line) compared to the noisy one (red dashed line). The straight line with $F_\phi (L) = 1$ helps to distinguish modes with good fidelity from the ones with bad fidelity.}
\label{fig:DeNoisedImages}
\end{figure*}

Thus, if we have the Fourier transform of a pixeled image, namely $S(l,m)$, the estimated image is $\hat{S}(l,m) = W(l,m) S(l,m)$, in which the Wiener filter is defined as
\begin{equation}
\label{eq:weiner_filter}
W_L = \frac{1}{1+ \frac{\mathcal{N}^{\hat{\phi}}_L}{C^{\phi\phi}_L}},
\end{equation}
where we can substitute Eqs.~(\ref{eqn:DefField}) and ~(\ref{eqn:BeamEstNoise}) for signal and noise respectively. We can see that when SNR $ = C^{\phi\phi}_L/\mathcal{N}^{\hat{\phi}}_L  \gg 1$, the filter is one, while for SNR $\ll 1$ we have $W_L \rightarrow 0$. If we apply this filter to our reconstructed potential image Figure~\ref{fig:RecGRP468}, we obtain Figure~\ref{fig:GRPWien}, which looks like a high-fidelity smoothed version of the input image \ref{fig:GRP468}. This is more quantified in Figure~\ref{fig:GRPWienPS}, which illustrates the recovered Fourier square amplitude of the image. We can see that much of the noise have been filtered out and that the image has been smoothed on small scales. 

Note that it is possible to use fewer $L$-modes to reconstruct the lensing potential instead of using a de-noising filter. The Weiner filter just provides a systematic way of down weighting modes that are dominated by noise. At $z_s=8$ it can be seen that the estimator noise crosses the deflection field power spectrum at $L\leq 300$, so the image effective resolution will be $\Delta\theta\geq 51\mbox{ arcmin}$. 

\subsubsection{Fidelity of the Reconstruction}
In order to quantify the accuracy of the reconstructed images we define the ``Fidelity" $F_\phi (L)$ as
\begin{equation}
F_\phi (L) = \left[\frac{\big\vert\hat{\phi}_{L}-\phi_{L}\big\vert}{\big\vert\phi_{L}\big\vert}\right]^{-1},
\end{equation}
the local fractional difference between the estimated $\hat{\phi}_{L}$ and the true potential $\phi_{L}$ at every mode $\bm{L}$. Bigger fidelities correspond to better reconstruction of the lensing potential. The resulting fidelities for the estimated potential image Figure~\ref{fig:RecGRP468} and its de-noised version Figure~\ref{fig:GRPWien} are shown in Figure~\ref{fig:FidelityWien}, in red and blue dashed lines respectively. The straight line with $F_\phi (L) = 1$ helps to distinguish modes with good fidelity from the ones with bad fidelity. Note how the Wiener filter tends to reach the limit of $S/N \rightarrow 1$ at $L>3000$ or so. This happens because for modes that are dominated by noise, (\ref{eq:weiner_filter}) the signal power spectrum is too small, leading to $F_\phi (L) = 1$.

The image fidelity improves in the intermediate range $300\lesssim L\lesssim 2000$ because of the absence of reconstruction noise, while in the noisy image the fidelity gets worse and worse as $L$ increases. Thus, the overall fidelity of the de-noised reconstructed potential is bigger than $1$ for almost all the modes involved in the reconstruction.

It is worth to remember that the forecasts presented here depend on our rather simple model for reionization and the distribution of HI at high redshifts, Section~\ref{sec:21cmfield}. If the true reionization history varies a great deal from what we have assumed, for example reionization is extended over a large redshift range\footnote{Observations suggest that the EoR ended at least at redshifts $z>6$ \citep{Zaroubi13}.} or ends well before z=8, then these forecasts will not be valid, since the estimator will not be optimal. We will extend this work to more complicated reionization scenarios in the future, and, for simplicity, we keep assuming that EoR has been a uniform process for redshifts around $z_s = 8$.

\begin{figure*}
\centering
\subfigure[Input potential field case (a)]{\label{fig:GRP650}\includegraphics[scale=0.23]{GRP_650_5.png}} 
\subfigure[Estimator Noise map case (a)]{\label{fig:EstN650}\includegraphics[scale=0.23]{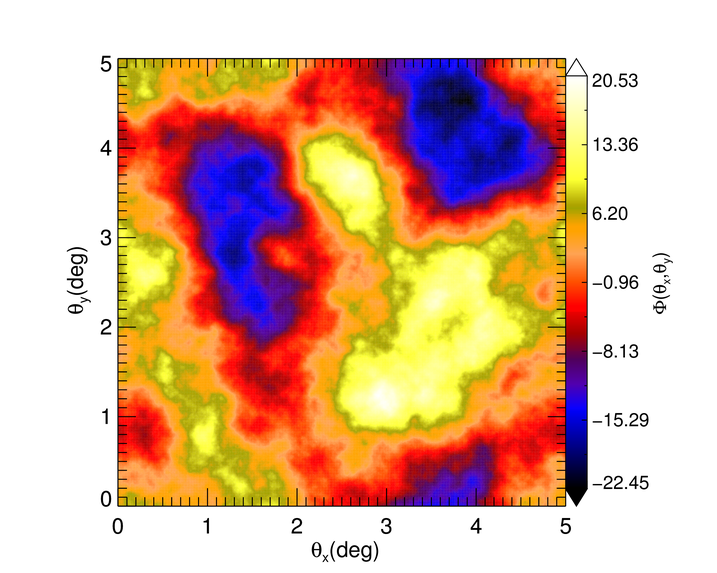}} 
\subfigure[Estimated potential map case (a)]{\label{fig:Rec650}\includegraphics[scale=0.23]{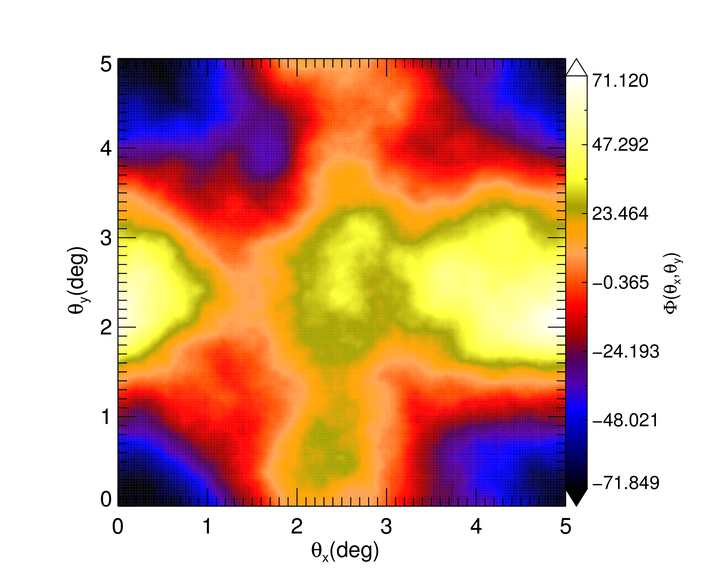}}
\subfigure[Input potential field case (b)]{\label{fig:GRP1300}\includegraphics[scale=0.23]{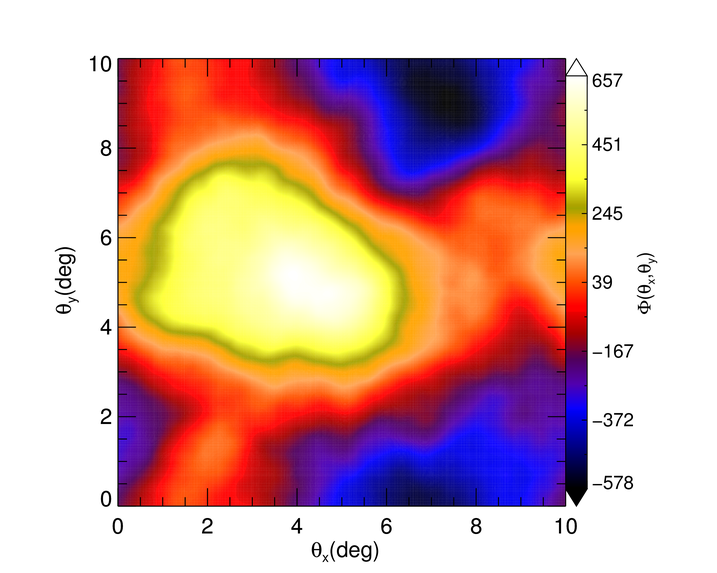}} 
\subfigure[Estimator Noise map case (b)]{\label{fig:EstN1300}\includegraphics[scale=0.23]{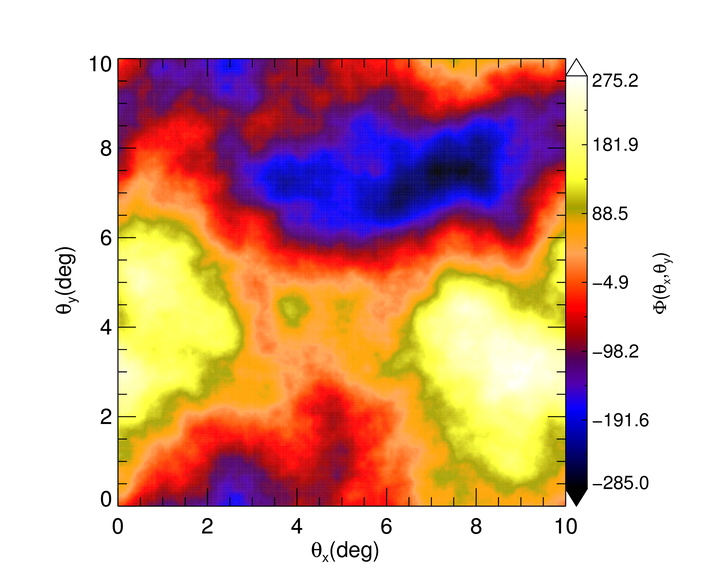}} 
\subfigure[Estimated potential map case (b)]{\label{fig:Rec1300}\includegraphics[scale=0.23]{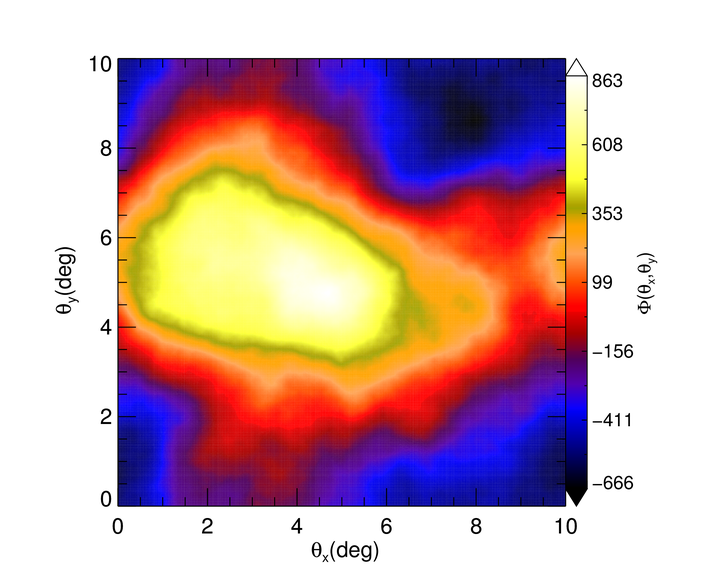}}
\subfigure[Input potential field case (c)]{\label{fig:GRP2600}\includegraphics[scale=0.23]{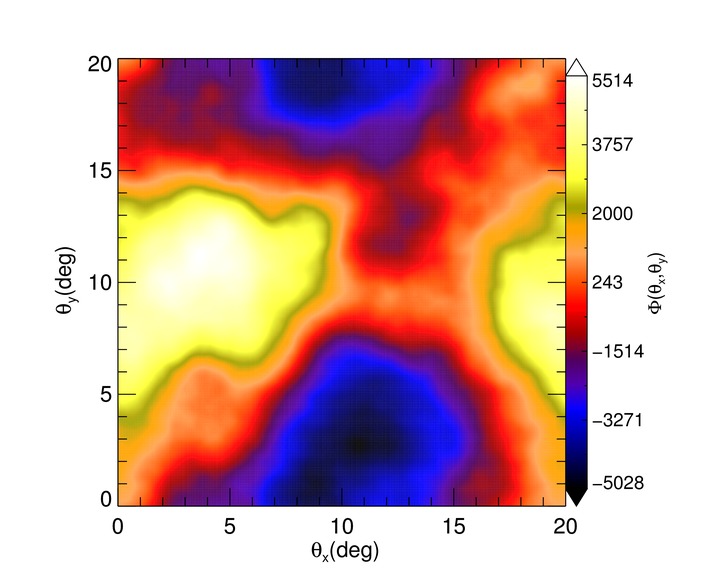}} 
\subfigure[Estimator Noise map case (c)]{\label{fig:EstN2600}\includegraphics[scale=0.23]{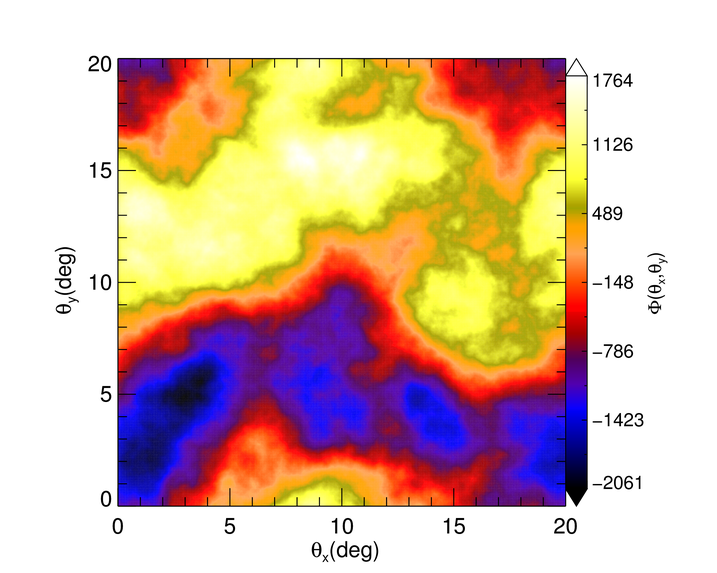}} 
\subfigure[Estimated potential map case (c)]{\label{fig:Rec2600}\includegraphics[scale=0.23]{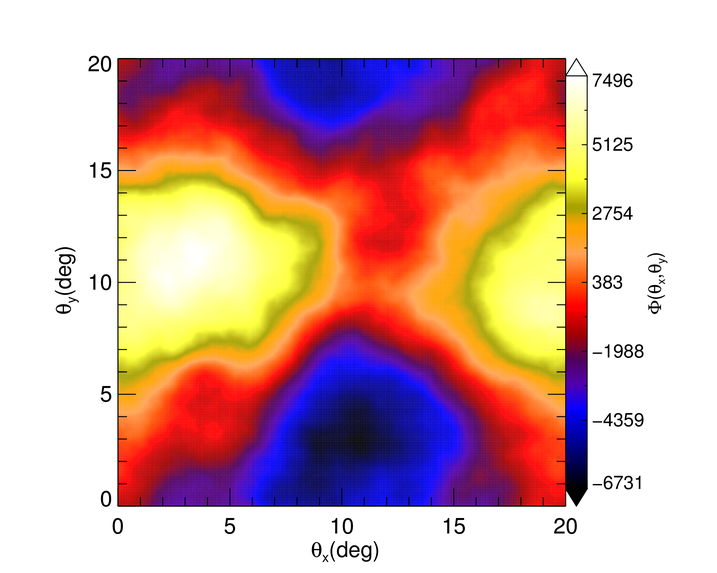}}
\caption{Reconstructed potential images from a realisation of the input 21 cm source box with beam cut off $L_{\rm cut} = 13237$ at $z_s=8$. Each one of the three rows corresponds to the cases listed in Table~\ref{tab:cases}. On the left panel we see the input potential field we wish to reconstruct. On the middle panel we see the estimator noise image produced by our estimator without any input lensing signal. On the right panel we show the reconstructed potential using the contribution of $20$ $k_p$ modes. For every case we used a SKA2-Low configuration with choices for observational time and bandwidth denoted as R2 in Table~\ref{tab:SKAconfigs}. The considered thermal noise power spectrum models a uniform density array distribution. The potential values in these maps have been scaled by a factor $10^9$ in order to improve the readability of the colour bars.}
\label{fig:SingleSim}
\end{figure*}

\subsection{Testing the Estimator}
All the tests performed on the estimator will be described in this section. For simplicity, we use the thermal noise power spectrum with uniform density array distribution Eq.~(\ref{eqn:ThNoisePS}) in these tests and use the SKA2-Low R2 survey strategy, with covering fraction $f_{\rm cov} = 0.095$. Using the more realistic thermal noise spectrum has very little effect on the lensing reconstruction noise in this case so it is sufficient for testing the estimator and gauging its dependencies on important parameters.

\label{sec:4.2.1}
\begin{figure*}
\centering
\subfigure[Recovered estimator variance for case (a).]{\label{fig:RecEstPS1_650}\includegraphics[scale=0.24]{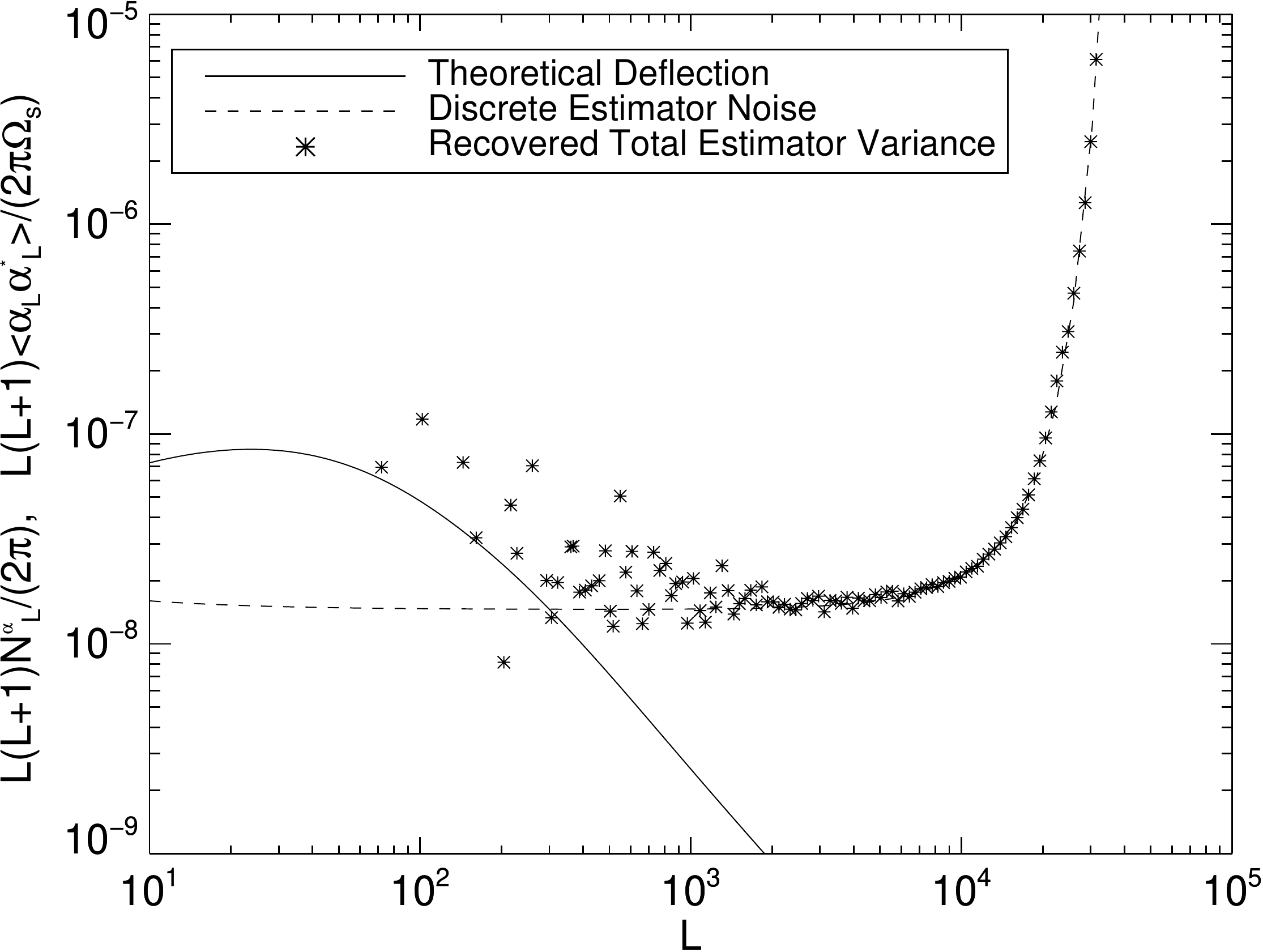}} \qquad
\subfigure[Recovered estimator variance for case (b).]{\label{fig:RecEstPS1_1300}\includegraphics[scale=0.24]{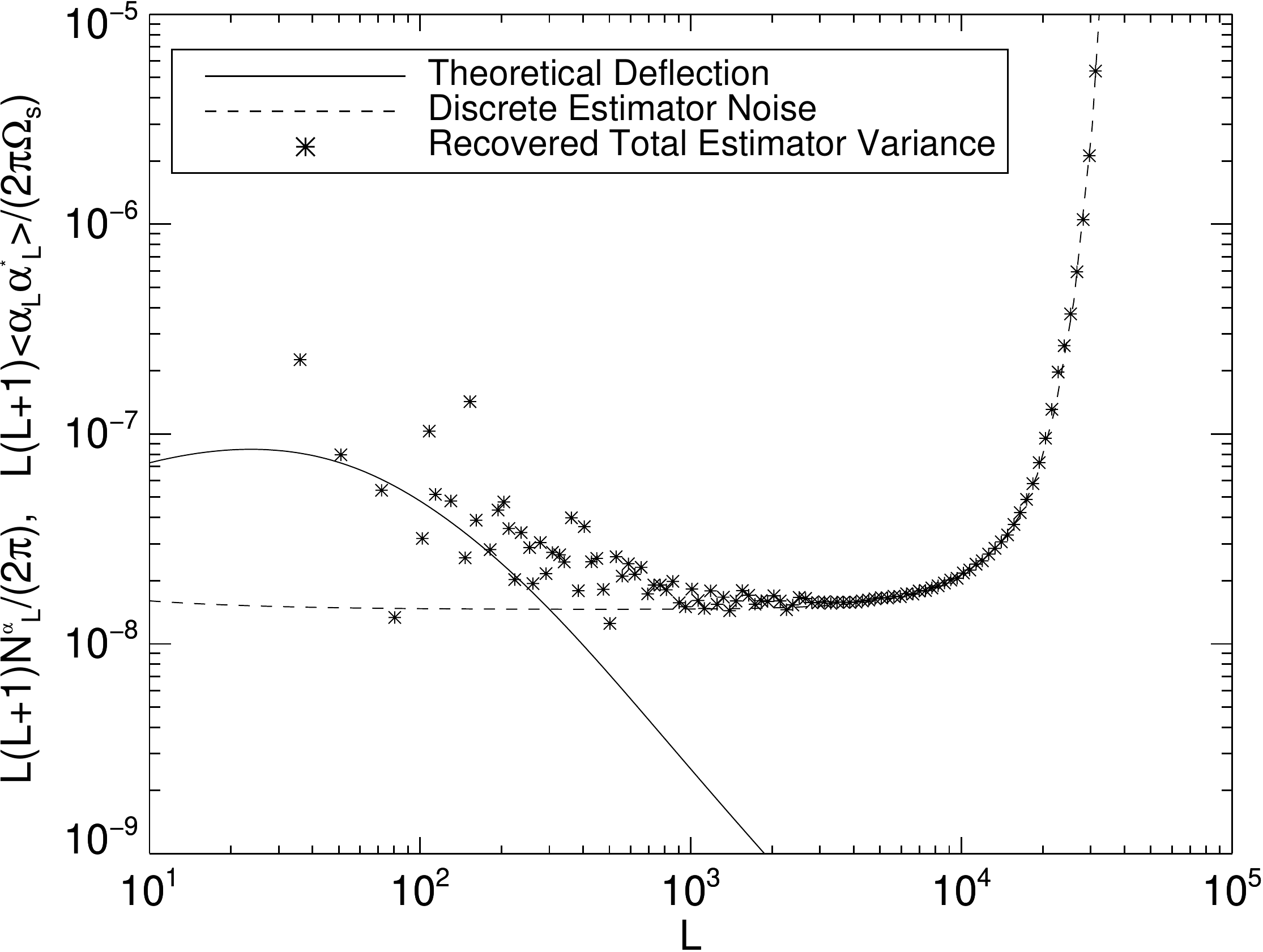}} \qquad
\subfigure[Recovered estimator variance for case (c).]{\label{fig:RecEstPS1_2600}\includegraphics[scale=0.24]{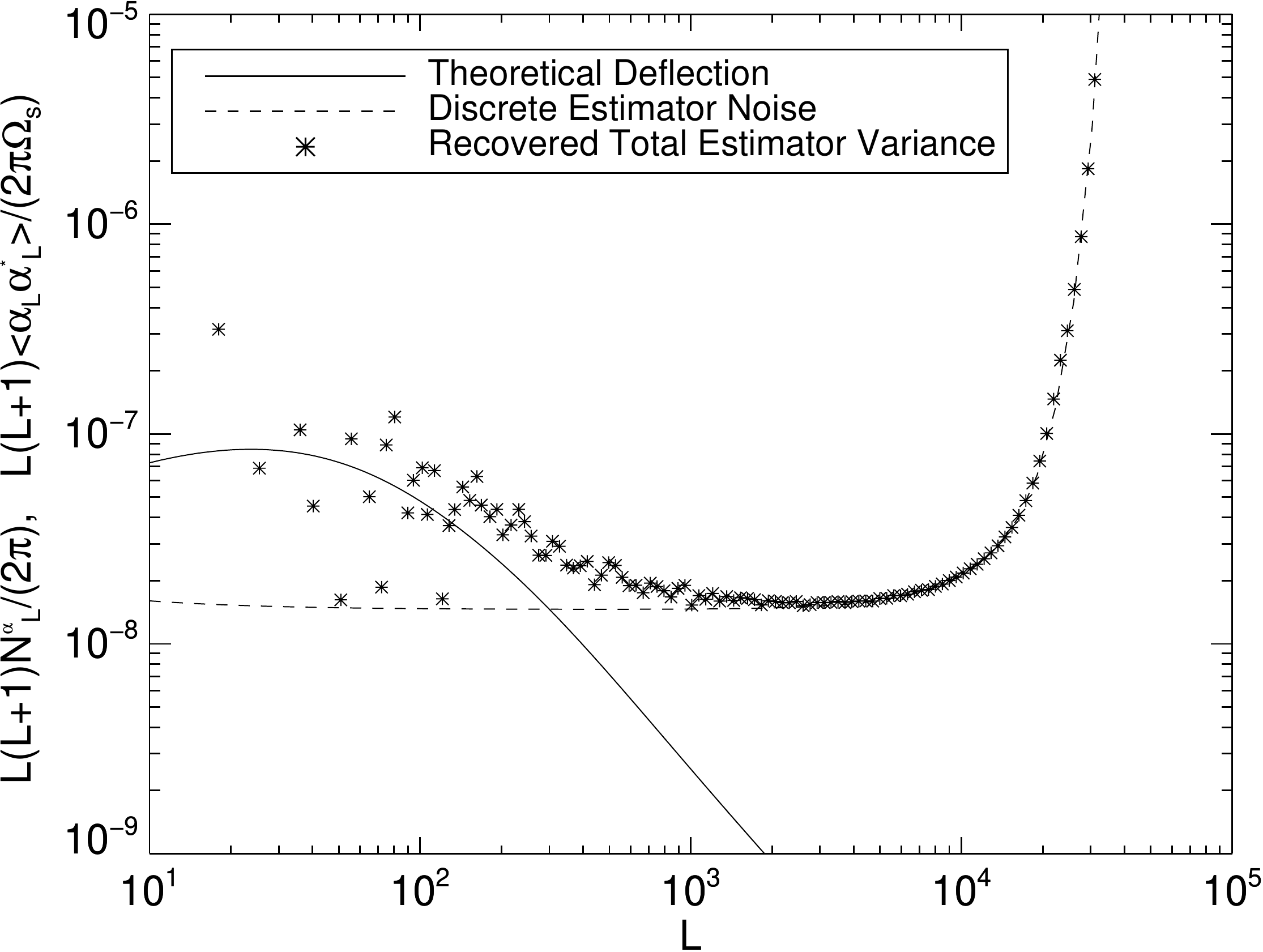}}
\caption{Recovered estimator Fourier space squared amplitudes for the considered settings listed in Table~\ref{tab:cases}, corresponding to images Figures~\ref{fig:Rec650}, \ref{fig:Rec1300}, and \ref{fig:Rec2600} respectively. The potential values in these maps have been scaled by a factor $10^9$ in order to improve the readability of the colour bars.}
\label{fig:RecPS1}
\end{figure*}

\subsubsection{Testing Dependency on the Surveyed FoV}

In this section we study the dependency of the reconstruction's performance on the observed sky area. The FoV will affect the Fourier space resolution $\Delta l$,  see Table~\ref{tab:cases}. $N_{\rm side}$ changes in order to keep the same $L_{\rm cut}$ and $L_{\rm Nyq}$ set in Section~\ref{sec:3.3.2}. Thus, the temperature and potential images always have the same angular resolution as at the beginning of Section~\ref{sec:Imaging}. 

The results for the three cases listed in Table~\ref{tab:cases} are presented in Figure~\ref{fig:SingleSim}, where for each row, the input potential field, the recovered pure estimator noise image and the recovered potential field are shown.
For case (a) we are able to recover a noisy version of the input map shape; in this case a large number of available modes are under the noise level. As we increase the map dimensions in cases (b) and (c), more large scale modes become available and a better image is recovered. 

We measure the Fourier space modes square amplitude from these recovered potential maps, as seen in Figure~\ref{fig:RecPS1}. The low-$L$ modes are the ones mostly involved in signal reconstruction.  Despite the high signal-to-noise, there are few of them so the sample variance is high in the power spectrum. We will see in Section~\ref{sec:MultiRealiz} that the correct average is recovered over a large number of realizations. We see that the total reconstruction signal-to-noise does not change too much from case (a) to case (c) and thus is not strongly dependent on the FoV.

\begin{table}
\centering
\begin{tabular}{|c|c|c|c|c|}
\hline
   & $\Omega_s$    & $N_{\rm side}$ & $\Delta l$ & $f_{\rm sky}$ \\ \hline
(a) & $5\degree\times 5\degree$   & 650  & 72  &   $6.06\times 10^{-4}$    \\ \hline
(b) & $10\degree\times 10\degree$ & 1300 & 36 &  $2.42\times 10^{-3}$      \\ \hline
(c) & $20\degree\times 20\degree$ & 2600 & 18    &  $9.7\times 10^{-3}$   \\ \hline
\end{tabular}
\caption{The three considered simulation settings for this study on FoV dependency. For every case we have $L_{\rm cut} = 13237$, $L_{\rm Nyq} = 2.5 \times L_{\rm cut}$, $z_s=8$ and a SKA2-Low R2 survey strategy.}
\label{tab:cases}
\end{table}

\begin{figure}
\centering
\includegraphics[scale=0.36]{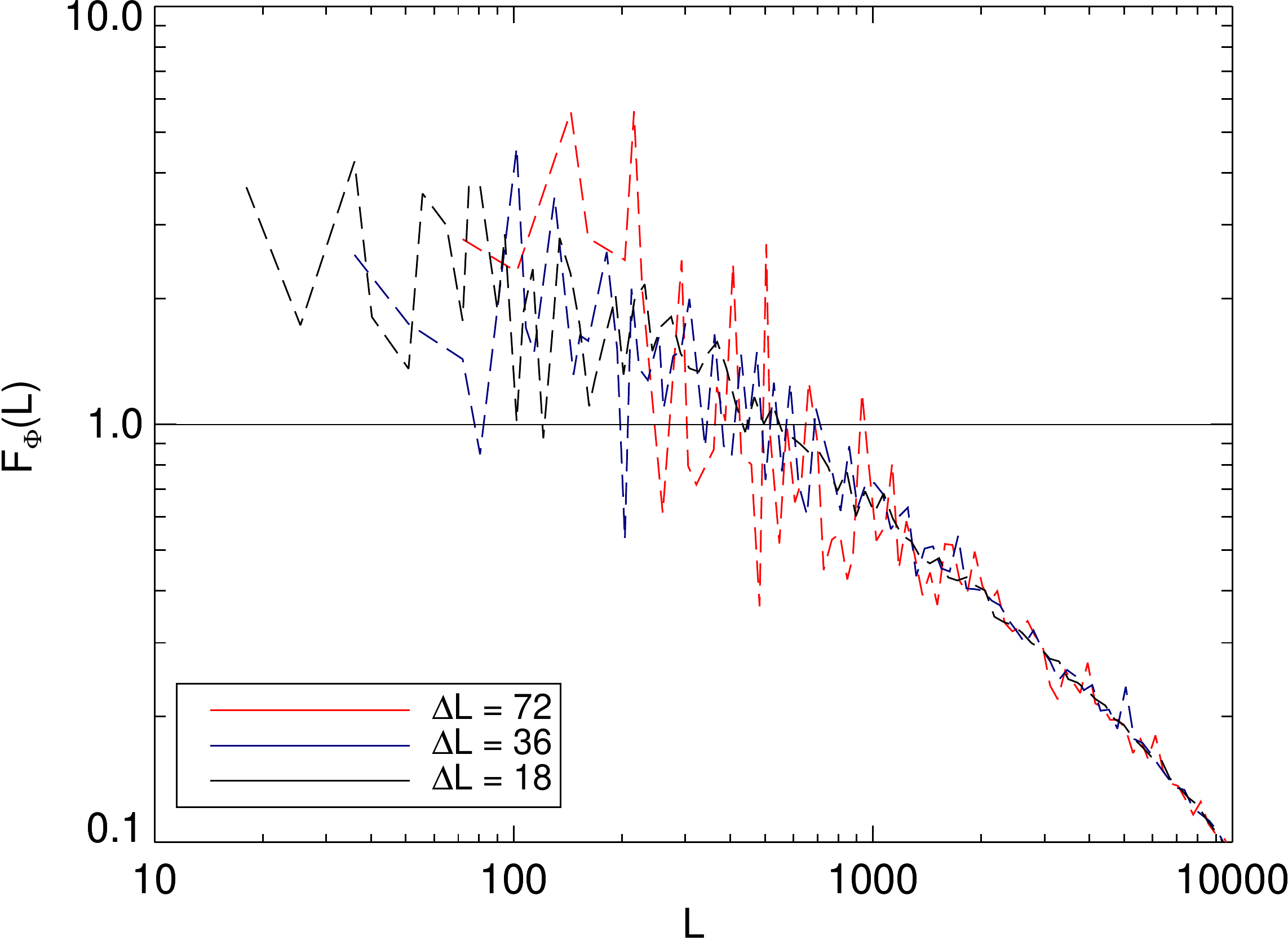}
\caption[Fidel]
{The fidelity of the reconstructed lensing potential images as a function of the multipole mode $L$. Red, blue, and black dashed lines are for Table~\ref{tab:cases} cases (a), (b), and (c) cases, respectively. The telescope model is SKA2-Low R2 and a noise power spectrum having a uniform array distribution is assumed. The straight line with $F_\phi (L) = 1$ helps to distinguish modes with good fidelity from the ones with bad fidelity.} 
\label{fig:Fidel}
\end{figure}

The last statement can be further tested considering the fidelities of the three cases listed in Table~\ref{tab:cases} which are shown in Figure~\ref{fig:Fidel}. As we increase the map's dimensions, going from case (a) (red dashed line) to case (c) (black dashed line), the fidelity of the images does not considerably change. We get better images only because bigger large scale modes are more available from case (a) to case (c). To give a more quantitative idea about this, we report the number of modes $n_i$ that have Fidelity bigger than one in the first three columns of Table~\ref{tab:Nmodes}. These are computed for the ranges $L\leq 200$, $200<L\leq 500$, and $500<L\leq 1000$, and considering for each row the cases listed in Table~\ref{tab:cases}. Then we report the total number $n_{\rm tot}$ of simulated mode and the fractional number of modes $f_i = n_i/n_{\rm tot}$ for the $L$-ranges considered before. It is clearly seen how the number of well reconstructed modes increases as larger scale modes become available from case (a) to case (c) for the single $L$-range. On the other hand, the fraction of modes in the considered ranges is always more or less constant, because the number of modes in a range $\delta L$ is weighted by the total number of modes, which increases from case (a) to case (c), leading to almost constant fidelities in all three cases.

If a sky mosaicking is performed to increase the number of available modes over the ones allowed by the FoV the reconstructed images will look more similar to the input potential field, but the overall reconstruction quality is not really improved. We can get a better image only because we are using bigger large-scale modes which are over the noise level. This technique can especially be used at lower redshifts, using SKA-Mid. We plan to explore this case in future work.

\begin{table*}
\centering
\begin{tabular}{|c|c|c|c|c|c|c|c|}
\hline
   & $n_{L\leq 200}$    & $n_{200<L\leq 500}$ & $n_{500<L\leq 1000}$  & $n_{\rm tot}$ & $f_{L\leq 200}$ & $f_{200<L\leq 500}$ & $f_{500<L\leq 1000}$  \\ \hline
(a) & 10  & 31  & 53 & 211900   & $4.72\times 10^{-5}$ & $1.46\times 10^{-4}$ & $2.5\times 10^{-4}$     \\ \hline
(b) & 37 & 115 & 194 & 846300 & $4.37\times 10^{-5}$ & $1.36\times 10^{-4}$ & $2.3\times 10^{-4}$         \\ \hline
(c) & 133 & 464 & 731  & 3382600&   $3.93\times 10^{-5}$ & $1.37\times 10^{-4}$ & $2.16\times 10^{-4}$      \\ \hline
\end{tabular}
\caption{The columns are the number of modes that have Fidelity bigger than 1 for $L\leq 200$, $200<L\leq 500$, and $500<L\leq 1000$, the total number of available modes, and the equivalent fractional number of modes with Fidelity bigger than one for each of the considered $L$-ranges. Each row corresponds to the three cases listed in Table~\ref{tab:cases}.}
\label{tab:Nmodes}
\end{table*}

\subsubsection{Testing Aliasing Contamination}
\label{sec:AliaTest}
\begin{figure}
\centering
\includegraphics[scale=0.36]{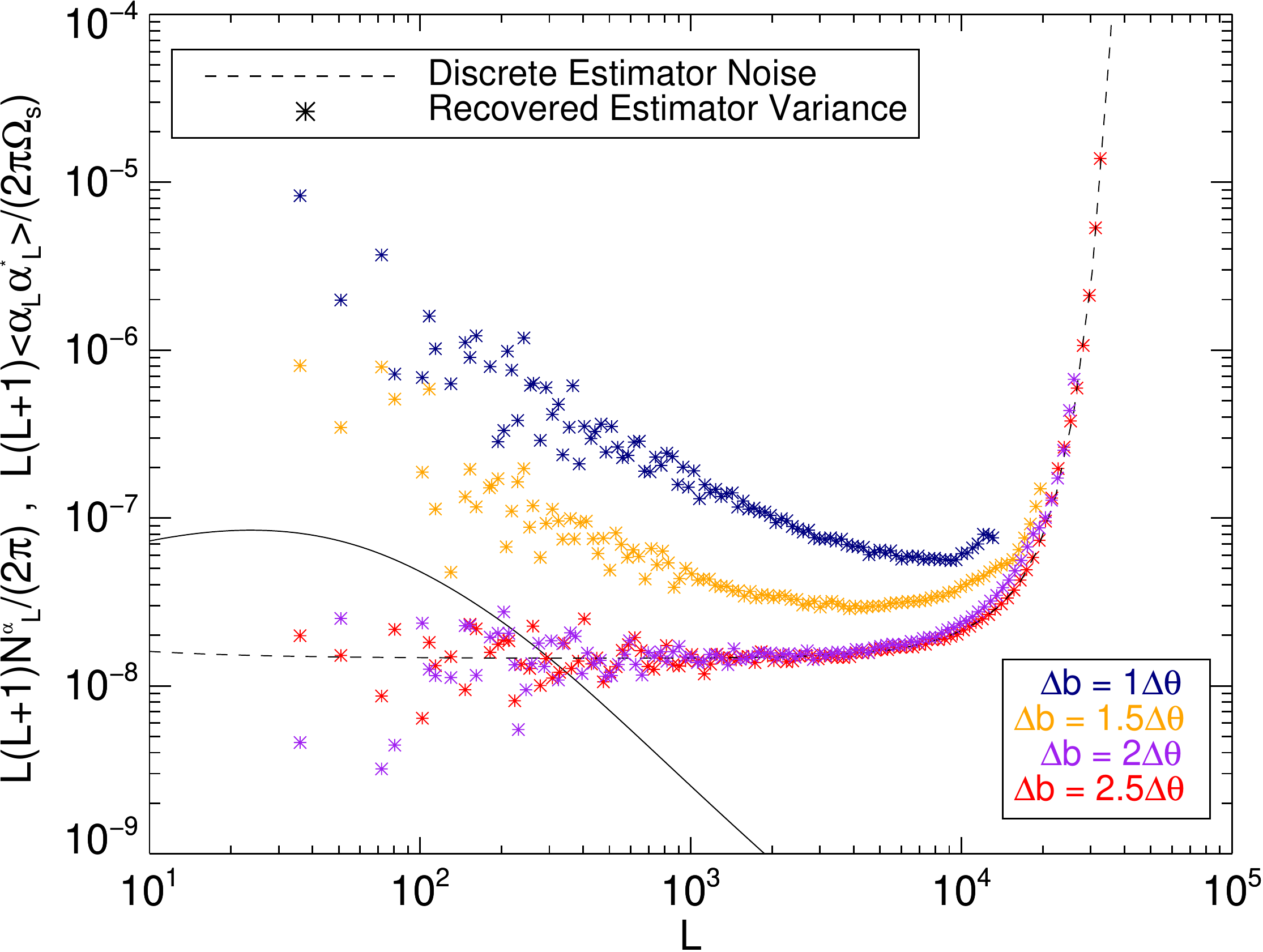}
\caption[RecoveredVarEst]
{The recovered estimator variance (star points) in absence of lensing signal compared to the discrete estimator noise (solid lines), since $\langle\alpha_L\alpha^\star_L\rangle / \Omega_s = N_L^\alpha$. Produced maps have $\Omega_s =10\degree\times 10\degree$, $L_{\rm cut}=13237$ and the estimator is recovered with $k_p^{\rm max}=20$ for $z_s=8$. We vary the distance between our fixed $L_{\rm cut}$ and the Nyquist frequency $L_{\rm Nyq}$ by changing the ratio $b$ between the beam resolution and the resolution of the simulation. We used a SKA2-Low R2 configuration and we assumed a noise power spectrum having a uniform array distribution.}
\label{fig:RecVarEst}
\end{figure}

As discussed in Section~\ref{sec:AliaBeam}, an important concern is the aliasing effect coming from the convolution performed in the real space estimator (\ref{eqn:BeamEst-2}). The aliasing of the slow estimator (\ref{eqn:BeamPotEst}) is negligible so a comparison between the two estimators is a good tool for determining how strong the aliasing effect is. We can visualise the aliasing using the variance of the estimator. In absence of lensing, we know that, because of Eq.~(\ref{eqn:SigPlusNoise}), the relation $\langle \hat{\phi}_{\bm{L}} \hat{\phi}_{\bm{L}}^\star \rangle = \Omega_s \mathcal{N}^{\hat{\phi}}_{L}$ has to be satisfied. We can see from Figure~\ref{fig:RecVarEst} that below a certain ratio $b$ between cut and Nyquist frequency, aliasing causes spurious power to be distributed over the simulated frequency range.  In order to avoid aliasing in our reconstruction simulation, we need to have a Nyquist frequency that is at least $2.5 \times L_{\rm cut}$. This is the reason for $\Delta b \geq 2.5 \Delta\theta$. It can be seen how, for smaller values of this ratio, the aliasing effect becomes more important.

The validity of the rule $L_{\rm Nyq}\geq 2.5 L_{\rm cut}$ can be investigated for different redshifts. Setting $z_s=7$, even though the Universe is unlikely to have been completely neutral at this time, the estimator noise level is lower and we have $L_{\rm cut} \simeq 14885$. The signal-to-noise is slightly higher than the $z_s=8$ case. The beam's resolution is higher, namely $\Delta\theta \sim 1.03\mbox{ arcmin}$. This means that we might need larger grids to avoid aliasing, with $L_{\rm Nyq} \sim 37165.56$  and with $N_{\rm side} = 730, \, 1460, \mbox{ and }2920\mbox{ pixels}$ for field-of-views of the cases (a), (b), and (c), respectively. If higher redshifts are considered, $L_{\rm cut}$ will be lower: for example at $z_s=10$, $L_{\rm cut} \simeq 10855$. More details on behavior of the estimator's noise at various redshifts will be given in Section~\ref{sec:MultiBand}. 

For the moment we limit this discussion to test aliasing for simulated high redshifts. Thus, instead of keeping the same grid dimension and change $\Delta b$, we have considered to fix the value for the ratio $b$ to $b =2.5$ and used a smaller square grid with $N_{\rm side} = 450$, considering $z=12$ and $L_{\rm cut} = 9160$. We have seen that the resulting estimator power spectrum in absence of lensing signal is weakly aliased with respect to the discrete estimator noise level. So a value of $L_{\rm Nyq} = 2.5 L_{\rm cut}$ is not enough to ensure the estimator to be aliasing-free, but $b$ needs to be slightly higher. This is due to the approaching of the characteristic beam scale $L_{\rm cut}$ to the one in which the power spectrum begins to bend (see Figure~\ref{fig:21cmPS} for clarity), causing the presence of more power at scales closer to the Nyquist frequency. Consequently for a smaller redshift like $z\approx 6$, a smaller $b$ is enough to avoid aliasing in reconstructed images.

\subsubsection{Multi-Realisation Reconstruction}
\label{sec:MultiRealiz}
\begin{figure*}
\centering
\subfigure[Averaged estimator for case (a).]{\label{fig:1000RecGRP650}\includegraphics[scale=0.228]{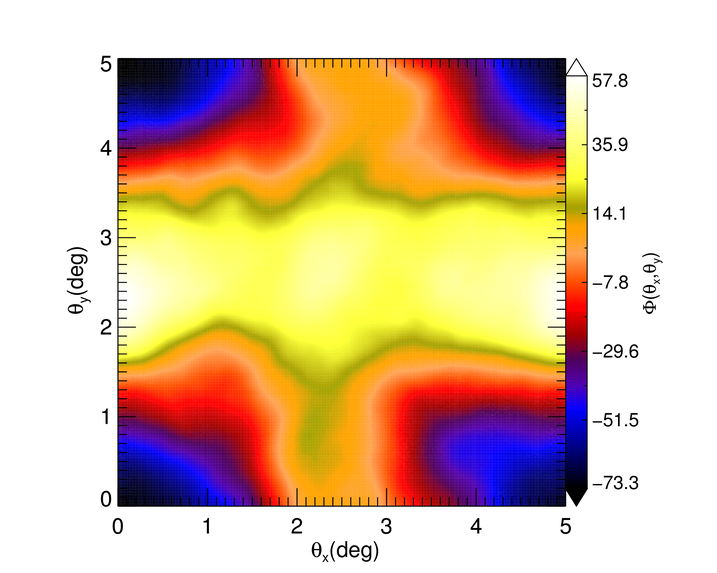}}\,\,
\subfigure[Averaged estimator for case (b).]{\label{fig:1000RecGRP1300}\includegraphics[scale=0.228]{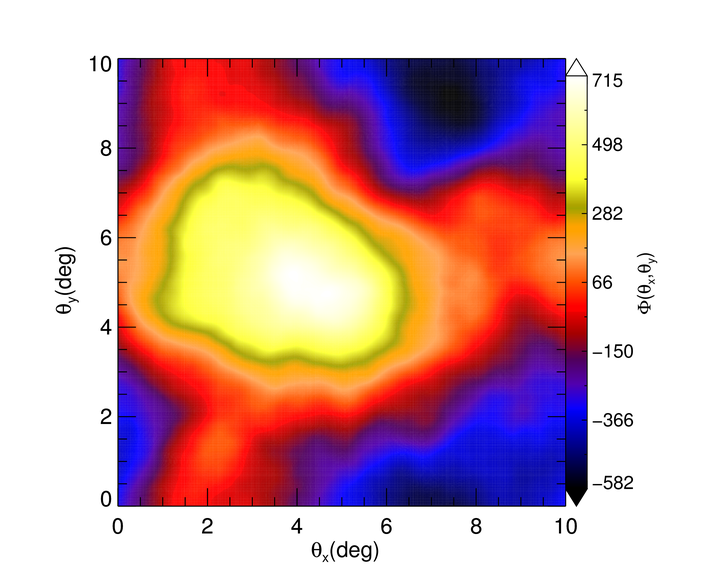}} \,\,
\subfigure[Averaged estimator for case (c).]{\label{fig:1000RecGRP2600}\includegraphics[scale=0.228]{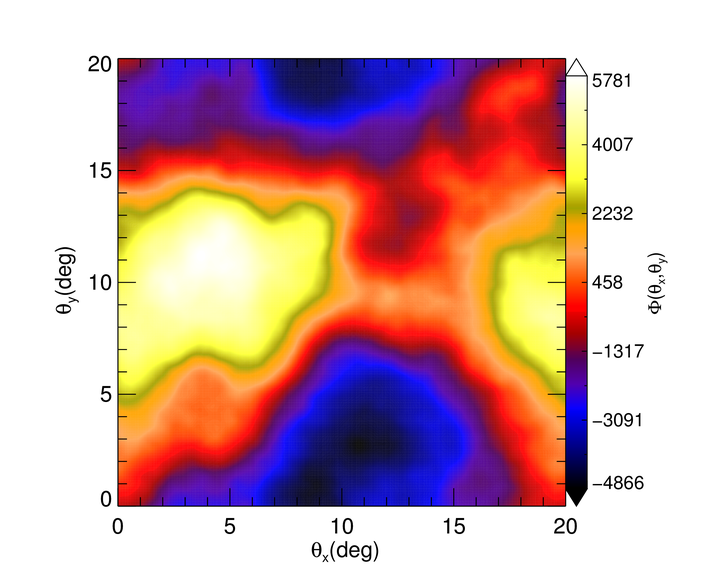}}
\subfigure[Recovered power spectra for case (a)]{\label{fig:1000RecPS650}\includegraphics[scale=0.238]{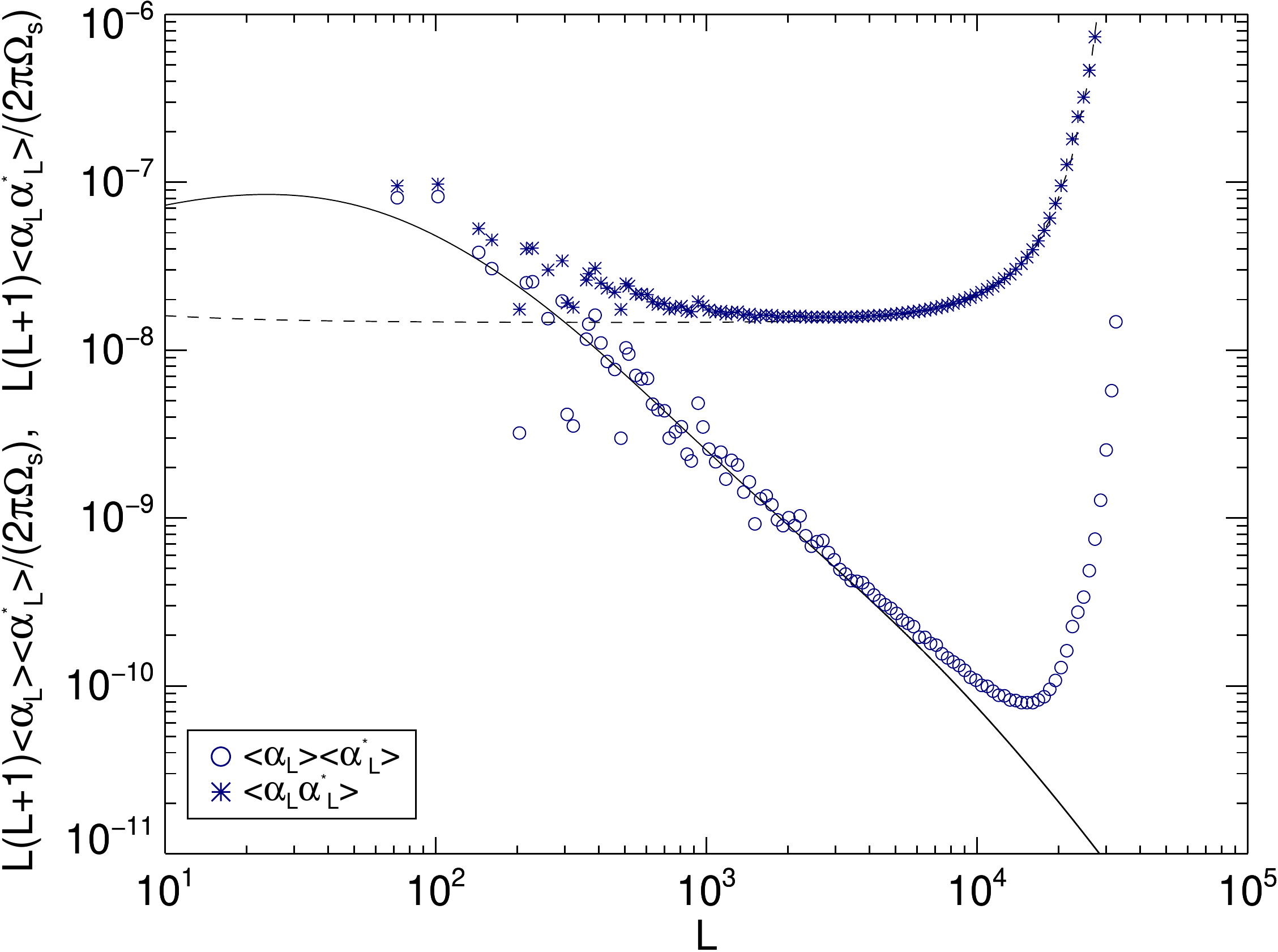}} \qquad
\subfigure[Recovered power spectra for case (b)]{\label{fig:1000RecPS1300}\includegraphics[scale=0.238]{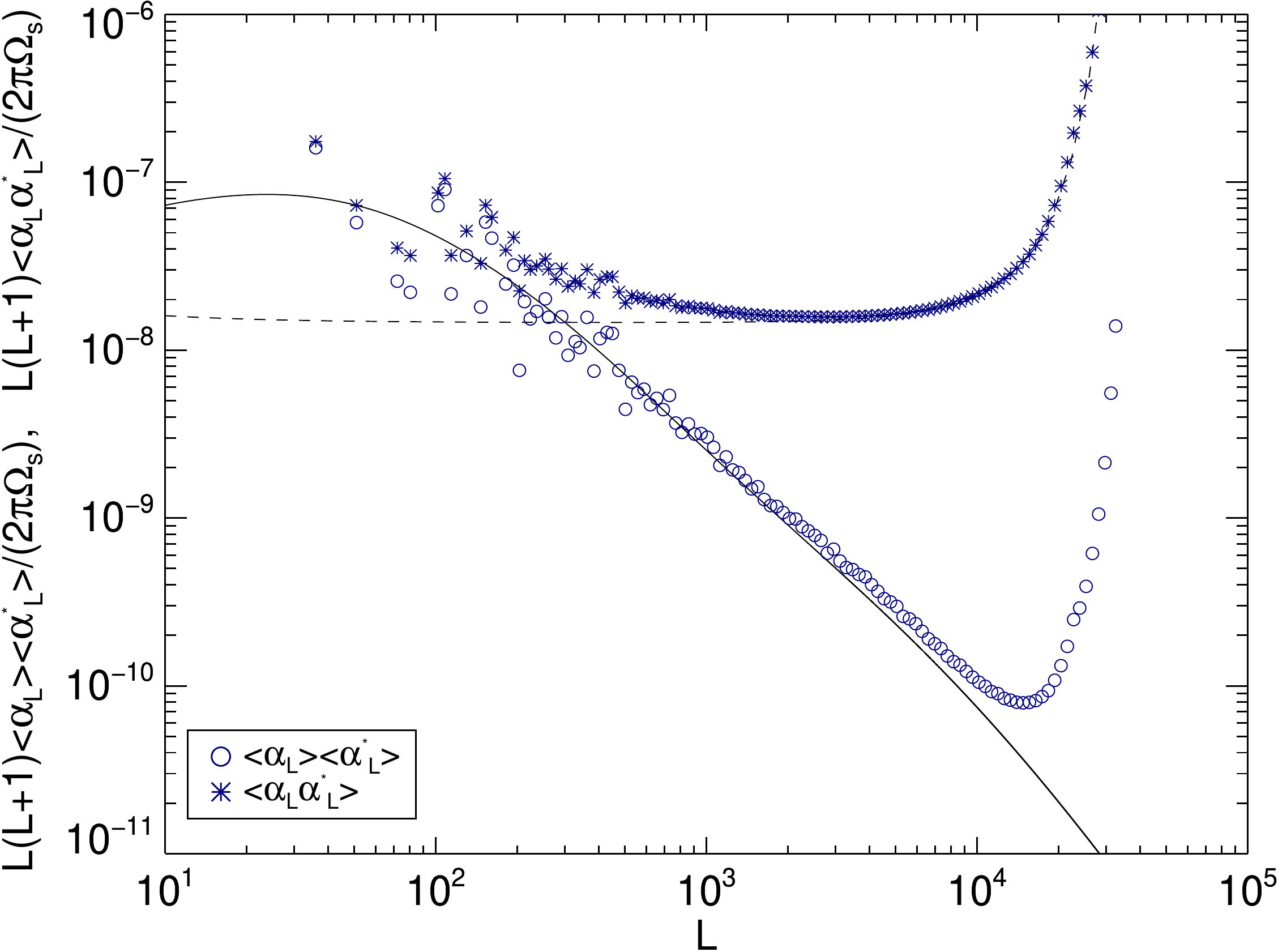}} \qquad
\subfigure[Recovered power spectra for case (c)]{\label{fig:1000RecPS2600}\includegraphics[scale=0.238]{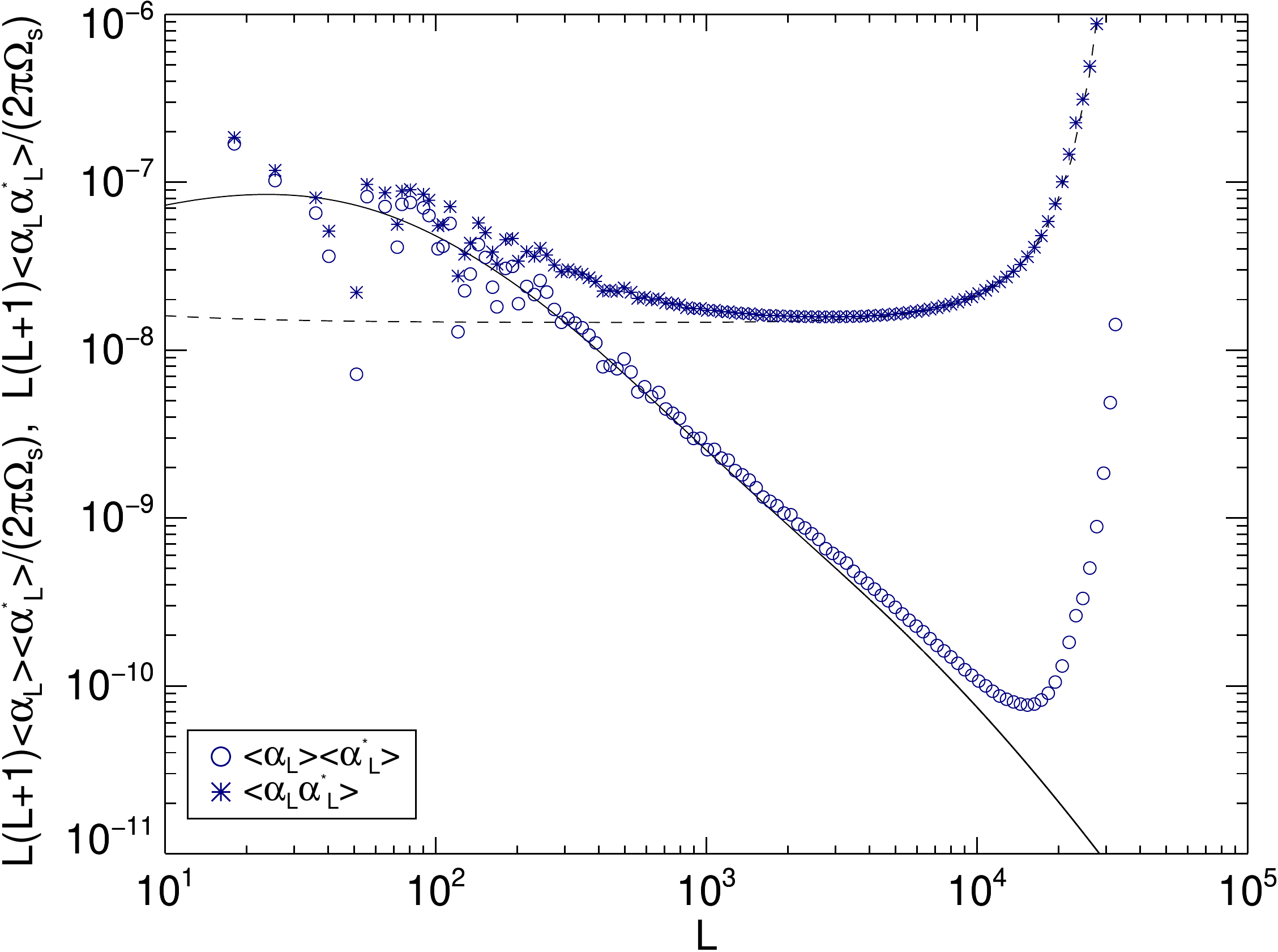}} 
\caption{\textit{Top:} reconstructed images from $N_{\rm Sim}=1000$ realisations of the input 21 cm source box for the three cases listed in Table~\ref{tab:cases}. The potential values in these maps have been scaled by a factor $10^9$ in order to improve the readability of the colour bars. \textit{Bottom:} the power spectrum of the overall estimator image is displayed for every case, together with the averaged power spectrum over $N_{\rm Sim}$. We have used a SKA2-Low R2 configuration with a uniform array distribution for the thermal noise power spectrum.}
\label{fig:MultiSim}
\end{figure*}

The estimator Eq.~(\ref{eqn:BeamEst-2}) must be unbiased after a large number of realisations, \textit{i.e.} the estimated potential field is equal to the true one ($\phi_{\bm L} = \langle\hat\phi_{\bm L} \rangle$) while keeping the same realisation of the input lensing field. Thus we tested the validity of this property by generating $N_{\rm sim}$ realisations of the input 21~cm source field. The estimator has then been produced for each realisation as described in Section~\ref{sec:Imaging}, being careful of generating always different random realisations for the thermal noises within every single source realisation. We are always using a power spectrum for the SKA2-Low R2 model assuming a uniform array distribution. Then we summed these estimators to produce the total one $\sum_{N_{\rm sim}} \hat\phi_{\bm{L}, \rm sim}$ shown in Figure~\ref{fig:MultiSim}, for each of the three cases listed in Table~\ref{tab:cases} and for $N_{sim} = 1000$. The recovered potential converges quite quickly towards the input one for all of the considered FoVs. The reader can compare these images with the input ones presented in Figures~\ref{fig:GRP650}, \ref{fig:GRP1300}, and \ref{fig:GRP2600}. Case (c) provides a better result using less realisations than case (a) because of the higher number of available large-scale modes.

The total reconstruction noise decreases as $N_{\rm sim}$ increases. Hence this simple test could also give us a preliminary idea about the potential of what in this work we called \textit{multi-band detection}. By this we mean an estimated lensing potential measurement made with a certain number of bandwidths each centered around several source redshifts. These data collected in different redshift channels, can be stacked together in order to produce a low-noise reconstructed image. To support this idea, let us explain what has been displayed in the bottom row of Figure~\ref{fig:MultiSim}, where we show the recovered Fourier space modes square amplitude of the total estimator after $N_{\rm sim}$ realisations $\langle \hat{\alpha}_{L}\rangle\langle \hat{\alpha}^\star_{L}\rangle/\Omega_s$ (purple circles) and the one resulting from the sum of every individual recovered variance $\langle \hat{\alpha}_{L}\hat{\alpha}^\star_{L}\rangle/\Omega_s$ (purple stars). The former converges to the input power spectrum Eq.~(\ref{eqn:DefField}), because the estimator noise is averaged out when several realisations are added and it is hence decreased by a factor $N_{\rm sim}$. The second quantity is instead converging to the sum of the lensing signal plus the estimator noise, as expected from Eq.~(\ref{eqn:SigPlusNoise}): in this case only the sample variance error within every considered $L$-bin decreases with respect to the one displayed in Figure~\ref{fig:RecPS1}. 

\begin{figure}
\centering
\includegraphics[scale=0.36]{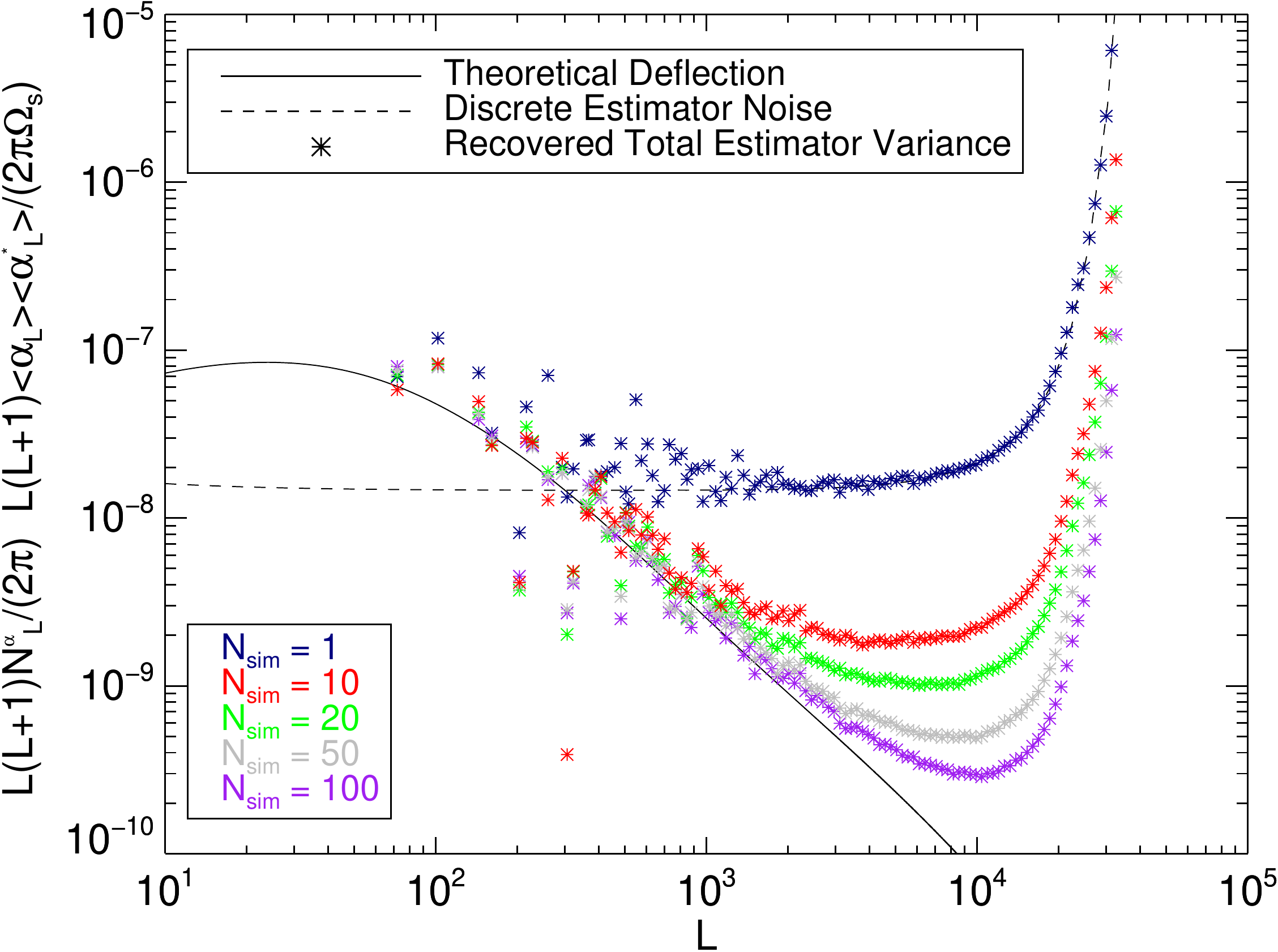}
\caption[RecPSMulti_650]
{The recovered Estimator power spectrum as the number of source and noise realisations increases up to $N_{\rm sim} = 100$. This plot is produced for the small map cases, with $\Omega_s = 5\degree\times 5\degree$ and using a SKA2-Low R2 configuration with a uniform array distribution for the thermal noise power spectrum.}
\label{fig:RecPSMulti_650}
\end{figure}

In order to better appreciate the behavior pictured by the purple circles shown in the bottom row of Figure~\ref{fig:MultiSim}, we plot the recovered power spectrum of the total estimator for several realisations up to $N_{\rm sim} = 100$ in Figure~\ref{fig:RecPSMulti_650}. Here the noise statistically decreases by a factor $N_{\rm sim}$. Regarding multi-band detection, this means that with only $10$ bands we should be able to have a larger number of modes above the noise signal in the intermediate range $100\leq L \leq 1000$. Therefore the same behavior could be expected when multiple $5\mbox{ MHz}$ bands are stacked together to fit a given redshift range, as anticipated by the analytic estimates made in \citep{PourtsidouMetcalf15} for their SKA-Low flat thermal noise model. We note that the multi-realisation approach explored in this section relies on a few crude approximations: the estimator noise will be slightly different from band to band, being higher for high redshifts, because noises, sources and $L_{\rm cut}$ depend on $z$. Moreover, the single estimators need to be renormalised by the estimator reconstruction noise in that band and weighted by the total-band estimator. This topic will be discussed in Section~\ref{sec:MultiBand}, when a proper treatment for this case will be performed.

\subsubsection{Tests with Strong Lensing Toy Models}
We have performed further tests in order to verify the validity of our code; for example by computing the recovered 21~cm and noise power spectra or the effective Gaussianity of the sources. In particular, we have checked our lensing simulation routines by applying various lensing models to our source field (in particular lensing toy models like the Singular Isothermal Sphere (SIS) potentials or Point Mass Potential) obtaining artifacts-free images and plausible strong lensing effects in the lensed temperature map. Then, we used these lensed maps to recover the input lensing potential. The estimator in this strong regime gives interesting results, allowing us to have further insights about the estimator itself and the weak lensing assumption Eq.~(\ref{eqn: LensField}). In fact, the slope of the potential is recovered correctly up to a certain scale, until the approximation of Eq.~(\ref{eqn: LensField}) breaks down, and contributions from higher order terms are needed to recover the input potential. At these scales the lensing gradient is not small anymore, and such higher-order terms become more and more important, irrespectively of the magnitude of the temperature gradient. 

\subsection{Multi-Band Reconstruction}
\label{sec:MultiBand} 
Encouraged by the results obtained in Section~\ref{sec:MultiRealiz} and considering the multi-band estimator described in \citep{MetcalfWhite09}, we performed a simulation involving several estimators computed at different frequency (redshift) bands stacked togheter using the channels available from the next generation 21~cm radio telescopes. As shown in the last Section, the combination of multiple frequency bands can aid the reconstruction of the underlying lensing potential, by statistically lowering the estimator reconstruction noise level. Here we will adopt the same SKA1 and SKA2-Low configurations modelled with the non-uniform array distribution noise power spectrum Eq.~(\ref{eqn:32}) described in Section~\ref{sec:3.3.2} and used to get the single-band results obtained in Section~\ref{sec:Imaging}. The investigated range of observed redshifts is $ z_c = 7- 11.6$, corresponding to a frequency range of $\nu_{c} = 177.55 - 112.55$ MHz.
 
\subsubsection{Noise Weighted Total-Band Estimator}
In this section we will introduce a combined multi-band noise weighted estimator that can be applied to our simulated maps. Each bandwidth $\Delta\nu$ is centered around a given source redshift $z_c$ within the range $\Delta z$. \cite{PourtsidouMetcalf15} calculated the lensing reconstruction noise stacking $10$ frequency bins of $8 \, {\rm MHz}$ bandwidth spanning the redshift range $z_c \simeq 6.5-11$ using SKA1-Low and SKA2-Low parameters. This noise is lower than the one obtained using a single band and shown in Figure~\ref{fig:EstNvarSKA-New}, because the total estimator noise is
\begin{equation}
\label{eqn:totBandEstN}
\mathcal{N}_L^{\rm tot} = \left[\sum_\nu \frac{1}{\mathcal{N}_{L,\nu}^{\hat\phi}}\right]^{-1}.
\end{equation}
This behaviour can be understood from the multi-realisation study performed in Section~\ref{sec:MultiRealiz}: as we stack frequency bands, the 21~cm source signal will be averaged out together with the thermal noises, and the estimator noise will go down by a factor $N_{\nu}$, the number of stacked frequency bands. This is shown in Figure~\ref{fig:RecPSMulti_650}. The combined discrete quadratic estimator is hence noise-weighted, namely
\begin{equation}
\label{eqn:totBandEst}
\hat\phi_{\bm L}^{\rm tot} = \mathcal{N}_L^{\rm tot}\sum_\nu {\frac{\hat\phi_{\bm{L},\nu}}{\mathcal{N}^{\hat\phi}_{L,\nu}}},
\end{equation}
where each single-band estimator Eq.~(\ref{eqn:BeamEst-2}) contributes for every frequency band and multipole to the total-band map.

\subsubsection{Frequency Dependencies}
\label{sec:zDep}
\begin{figure}
\centering
\includegraphics[scale=0.36]{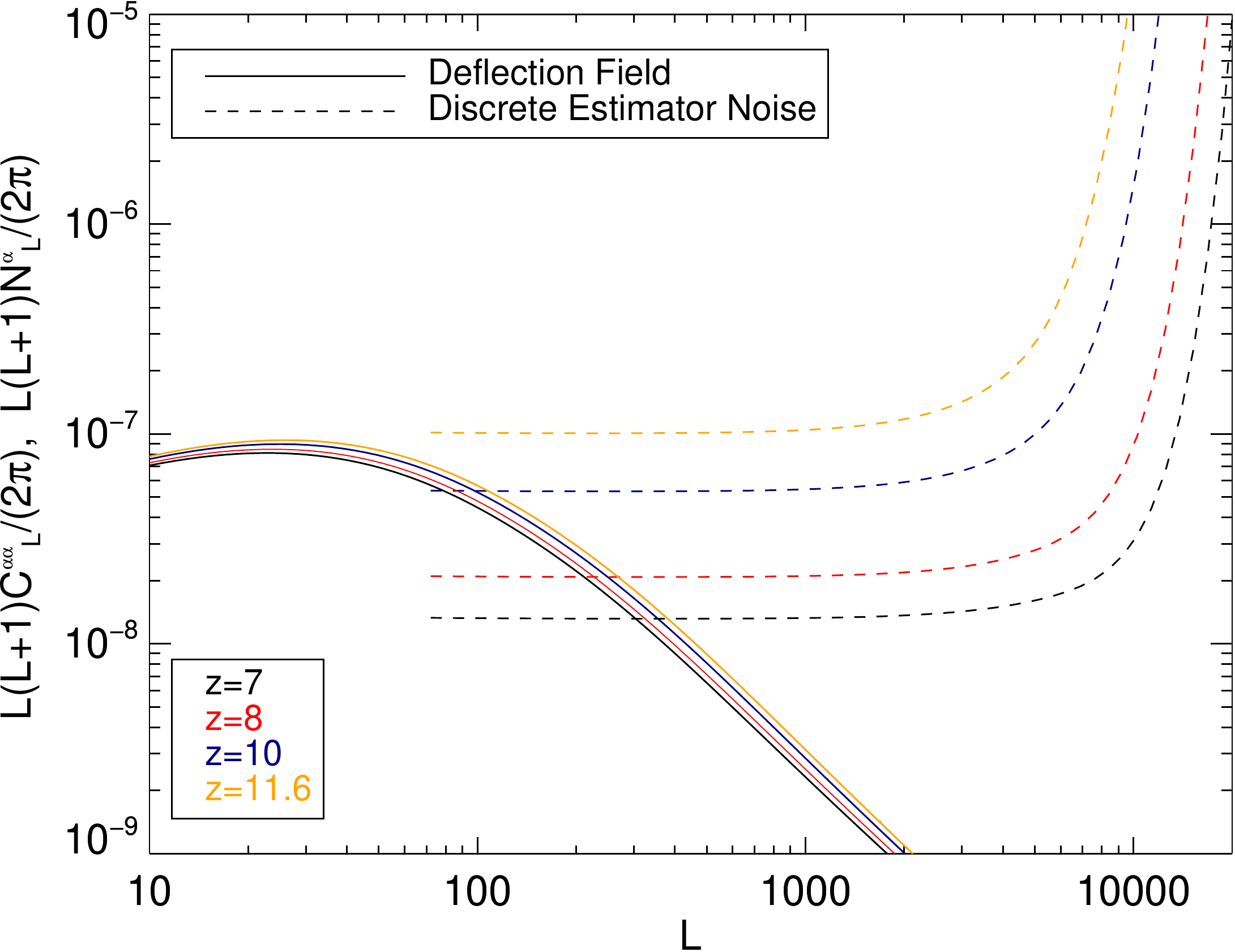}
\caption[EstNZ]
{The deflection field and the discrete estimator noise (dashed lines) are shown for several redshifts, from $z_c=7$ to $z_c=11.6$, for a $5\degree\times 5\degree$ survey with $l_{\rm min} = 72$. The adopted telescope model is a SKA2-Low R2, as listed in Table~\ref{tab:cases}, with $k_p^{\rm min} =3$ and $k_p^{\rm max} = 20$.}
\label{fig:EstNZ}
\end{figure}

The redshift (frequency) dependence of $L_{\rm cut}$ and thermal noise needs to be taken into account, when multiple frequency/redshift bands are stacked. The higher the redshift, the higher the thermal noise and the lower $L_{\rm cut}$ will be. This is shown in Figure~\ref{fig:EstNZ}, where the discrete estimator noise is substantially varying in the redshift range from $z_c=7$ to $z_c=11.6$, for a $5\degree\times 5\degree$ survey with $l_{\rm min} = 72$. The adopted telescope model in this case is SKA2-Low R2 with thermal noise power spectrum Eq.~(\ref{eqn:32}), $k_p^{\rm min} =3$ and $k_p^{\rm max} = 20$.

When the thermal noise is computed at different central frequencies, apart from the explicit frequency-dependent terms, we need to scale the baseline density $n(U,\nu)$ for different frequencies, as explained in Section~\ref{sec:ThNoiseComp}, because its integral over visibilities has to be constant, while minimum and maximum visibilities change. Given a fiducial baseline density $n_f(U,\nu_f)$ at a fiducial frequency $\nu_f$, we can write, up to first order, the scaling relation to any other frequency as
\begin{equation}
n(U,\nu) = \frac{\nu_f}{\nu} n_f\left(U\frac{\nu_f}{\nu}, \nu_f\right),
\end{equation}
where $U = D/\lambda = D\nu/c$.

It is assumed in this work that the lensing signal is not substantially varying between the first source redshift and the last one. Considering the deflection field power spectrum computed at different redshifts in Figure~\ref{fig:EstNZ}, we see that this approximation is valid across a substantial redshift range. In this range it is also assumed that Eq.~(\ref{eqn:21cmPS}) is valid and that our optimal estimator can be derived for a Gaussian field, considering the entire hydrogen budget to be fully un-ionised.

For our results here we will assume a redshift range $z_c= 7- 11.6$, corresponding to a frequency range of $\nu_{c} = 177.55 - 112.55$ MHz. We have increased $z_{\rm min}$ with respect to \citet{PourtsidouMetcalf15} in order to be more conservative about the EoR ending period (we will discuss this further in Section~\ref{sec:minZdisc}). In order to avoid aliasing, the Nyquist frequency is set to $2.5 L_{\rm cut}(z_c=7) \approx 37267$ corresponding to a resolution of $\approx 24.5\mbox{ arcsec}$ for a $25\mbox{ deg}^2$ map. For higher redshifts we keep this maximum frequency and we will vary $\Delta b = b\Delta\theta$, so that the simulation resolution for all the bands is set by the lowest redshift considered in the range. The range $z_c=7-11.6$ then corresponds to beams with resolutions going from $1.02\mbox{ arcmin}$ to $1.62\mbox{ arcmin}$.

Note that because of Equation~(\ref{eqn:totBandEstN}), the upper limit of this redshift range will not influence the total estimator noise level, since the estimator noise for $z_c\gtrsim 11$ turns to be considerably high. The discussion over the adopted lower limit will be postponed to Section~\ref{sec:minZdisc}.

As seen in Section~\ref{sec:ThNoiseComp}, the FoV is frequency dependent. A more complete description of the beam would indeed include a cutoff at large scales induced by the PSF of the telescope as already mentioned in Section~\ref{sec:BeamedEst}. A real telescope PSF needs to be handled numerically, since it can be a very complicated function for radio telescopes like the SKA \citep{SantosAlonso15}. This means that the estimator in Fourier space will have a different grid dimension at each band\footnote{In reality this would be true also within each frequency band, for each $k_p$ mode.}. While the lowest frequency band sets the resolution of the reconstructed image, the highest frequency band sets the total-band resolution in Fourier space through the FoV. The range $z_c=7-11.6$ corresponds to different FoVs going from $10.24$ to $25\mbox{ deg}^2$, which means $\Delta l = 112.5-72$, respectively.

On the other hand the estimator noise level does not greatly depend on the FoV (which sets the resolution in Fourier space), as shown from comparing the estimator noise levels in Figures~\ref{fig:EstNvarSKAMulti-New} and ~\ref{fig:EstNZ}. As seen in Section~\ref{sec:4.2.1}, the quality of the lensing reconstruction is almost unaffected by adding or leaving large scale-modes in the considered redshift range, since having different FoV would only add or subtract a few number of modes over the signal-to-noise level (see Figure~\ref{fig:Fidel}). However, a more complete beam expression would cut any contribution coming from a FoV bigger than the one set at $z_c^{\rm max}$. Thus, the general properties of the reconstruction do not change too much if we use a fixed instead of a varying FoV through the considered redshift range. 

To get the multi-band results we therefore consider that the FoV is set to be $\Omega_s = 5\degree\times 5\degree$ across each band, and we will keep assuming that the properties of the lensing and the telescope do not change within a single band. 

\subsubsection{Stacking Bands}
\label{sec:StackBands}
The number of stacked bands depends on the adopted bandwidth which, for a given central redshift $z_c$, corresponds to a redshift interval $\Delta z = (1+z_c)^2 \Delta\nu/\nu_{21}$. Starting from the first band centered at $z_c^{\rm min}$, the lower central redshift limit, the following band is found by decreasing the central frequency according to $\nu'_c (z_c) =\nu_c (z_c') = \nu_c (z_c) - \Delta\nu$, where $\nu(z) = \nu_{21}/(1+z)$. The new central frequency is thus extended between the limits $\nu_{\rm max, min} = \nu_c' \pm \left(\Delta\nu/2\right)$. These correspond to a new redshift interval $\Delta z = z_{\rm max} - z_{\rm min}$ with $z_{\rm max,min} = \left(\nu_{21}/\nu_{\rm min,max}\right)-1$. Hence the new central redshift is $z_c' =\left( z_{\rm min}+z_{\rm max}\right)/2$. For example, considering $\Delta\nu= 5\mbox{ MHz}$ and $7\leq z_c\leq 11.6$, we can stack $14$ bands in the frequency range $\nu_c = 177.55 - 112.55\mbox{ MHz}$. 

Consider that for a R0 survey strategy $\Delta\nu = 8\mbox{ MHz}$ $9$ bands can be stacked within the range $z=7-11.6$. If we use a thinner bandwidth, like $\Delta\nu = 3\mbox{ MHz}$, this number increases to $22$, reaching $66$ stacked bands for $\Delta\nu = 1\mbox{ MHz}$. The frequency band can be chosen as thin as possible until effects due to correlations between different Fourier modes show up. In \citet{MetcalfWhite09} it was found that the correlation between estimators at different frequency bands is not significant if $\Delta\nu \sim 1 \mbox{ MHz}$ or higher. This means that the statistical properties of the 21 cm radiation field and noise can be assumed as constant within a band. On the other hand, a very thin band increases the thermal and estimator noises. Moreover, the band choice poses limits on the maximum number of $k_p$ modes we can detect within a given band as well, since $k_p^{\rm max} = \Delta\nu/\delta\nu$, where $\delta\nu$ is the frequency resolution of one channel. The values of the adopted bandwidths in the considered survey strategies for this work are listed in Table~\ref{tab:SKAconfigs}.

\subsubsection{SKA1-Low and SKA2-Low Results}
\label{sec:4.3.4}
Resuming our considerations made in Sections~\ref{sec:zDep} and~\ref{sec:StackBands}, we will consider a $\Omega_s = 5\degree\times 5\degree$ survey area, which implies $\Delta l = 72$. Given the explored redshift range $z=7-11.6$, the smallest observable redshift fixes the Nyquist mode to $L\sim 37267$, since $L_{\rm Nyq} \gtrsim 2.5 L_{\rm cut}$, with $L_{\rm cut} = 14884.7$, corresponding to $\Delta b = b\Delta\theta =  1.02$ arcmin. Considering that at $z_{\rm max} = 11.6$, $L_{\rm cut} = 9430.89$, we will vary $b$ from band to band, reaching the final beam resolution at $z_{\rm max}$ of $\Delta b = 1.62$ arcmin. For each band we used $k_p^{\rm max}=20$ modes and $N_{\rm side} = 732$. 

\begin{figure}
\centering
\includegraphics[scale=0.37]{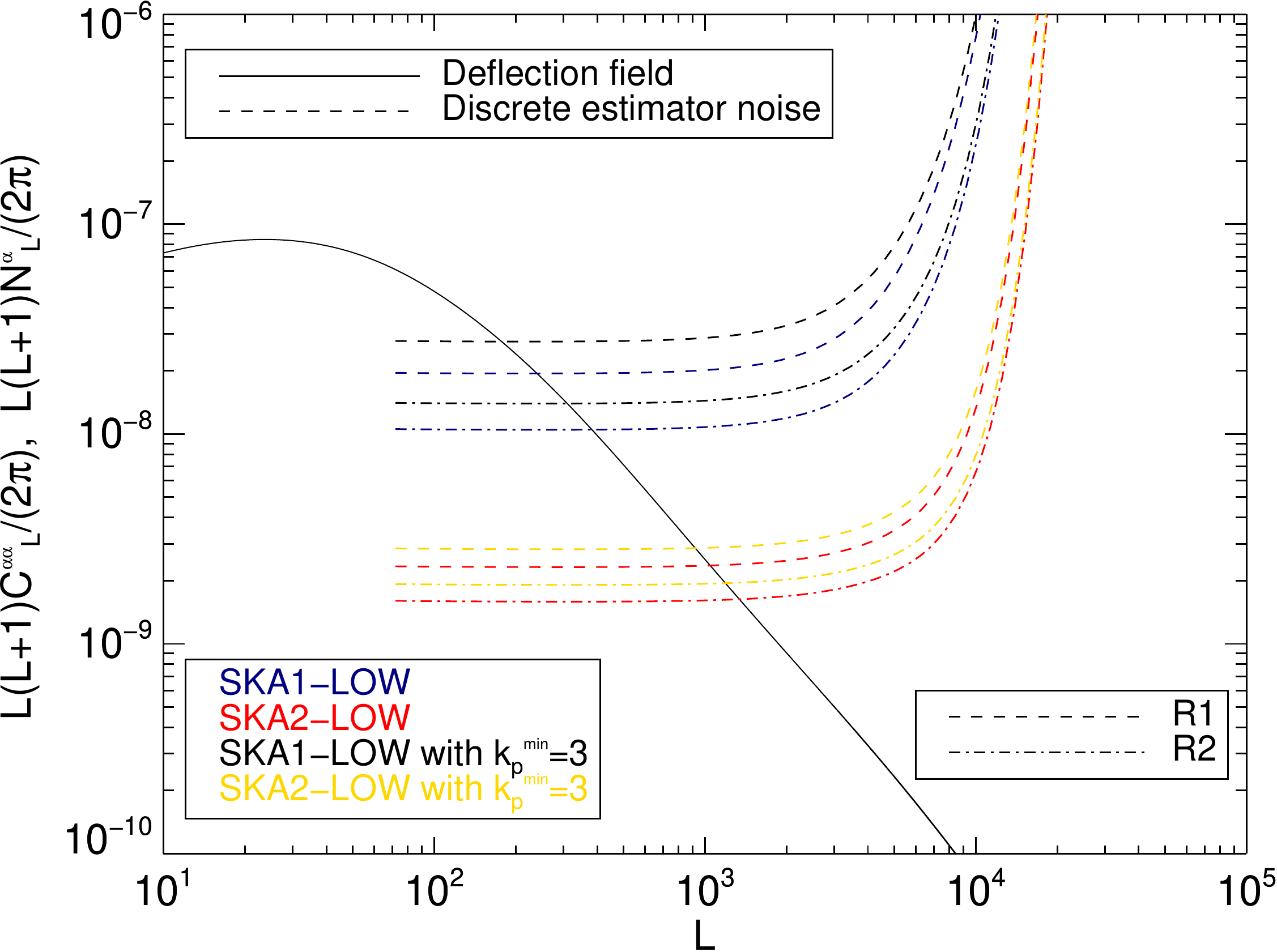}
\caption
{The multi-band discrete estimator noise for SKA1-Low (blue), SKA2-Low (red), SKA1-Low with $k_p^{\rm min}=3$ (black), and SKA2-Low with $k_p^{\rm min}=3$ (gold), with choices for observation time and bandwidth listed in Table~\ref{tab:SKAconfigs} and using the thermal noise power spectrum with non-uniform array distribution. The explored redshift range goes from $z=7$ to $z=11.6$. The simulated sky area is $\Omega_s = 5\degree\times 5\degree$ and $k_p^{\rm max}=20$. The R0 survey strategy results are not plotted because they produce an estimator reconstruction noise level close to R2 one. The R1 configuration is on dashed lines while the R2 is on dashed-dot lines.}
\label{fig:EstNvarSKAMulti-New}
\end{figure}

The computed multi-band estimator noise Eq.~(\ref{eqn:totBandEstN}) is presented in Figure~\ref{fig:EstNvarSKAMulti-New}, for SKA1 and SKA2-Low R1 and R2 telescope models including the power spectrum with non-uniform array distribution for both detector and sky noise. The R0 models is again not displayed to enhance the cleanliness of the plot, since it produces a result similar to the smaller bandwidth models. As in Section~\ref{sec:Imaging}, the effect of foreground contamination has been included considering the estimator noise computed with $k_p^{\rm min} = 3$. SKA2-Low configurations already give good results in the single-band case, as seen in Section~\ref{sec:Imaging}, and for the multi-band the forecasts are improved by more than an order of magnitude (a factor similar to the number of stacked bands). But the most interesting result comes from SKA1-Low detections, whose reconstruction noise level allows for high-quality imaging of the reconstructed lensing potential, with fidelity comparable to the one obtained for a SKA2-Low single-band experiment. Increasing the observational time from model R1 to model R2 causes the noise level to be decreased by nearly a factor 2. 

\begin{figure*}
\centering
\subfigure[Input potential field]{\label{fig:GRP732}\includegraphics[scale=0.23]{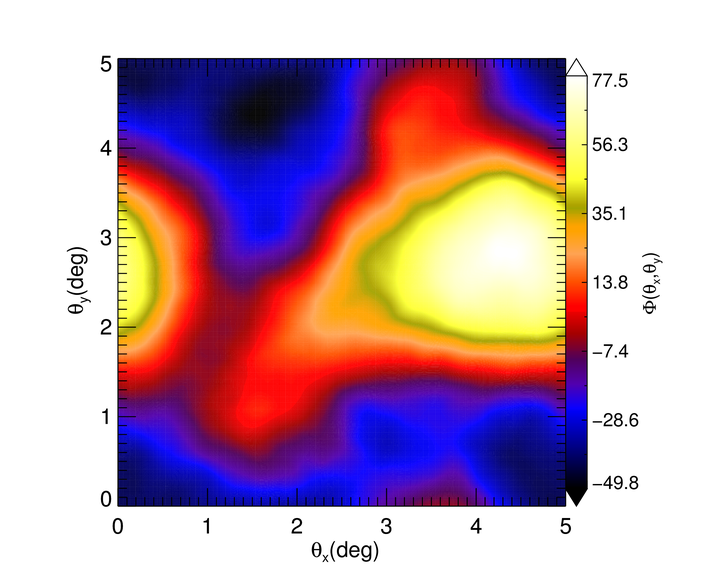}} 
\subfigure[Denoised estimator SKA1-Low R1]{\label{fig:RecGPRMultiSKA1R1-Wiener}\includegraphics[scale=0.23]{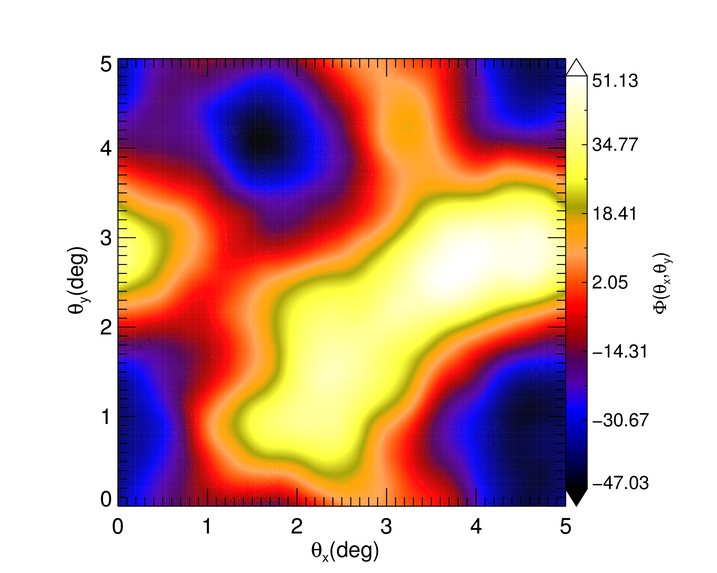}} 
\subfigure[Denoised estimator SKA1-Low R2]{\label{fig:RecGPRMultiSKA1R2-Wiener}\includegraphics[scale=0.23]{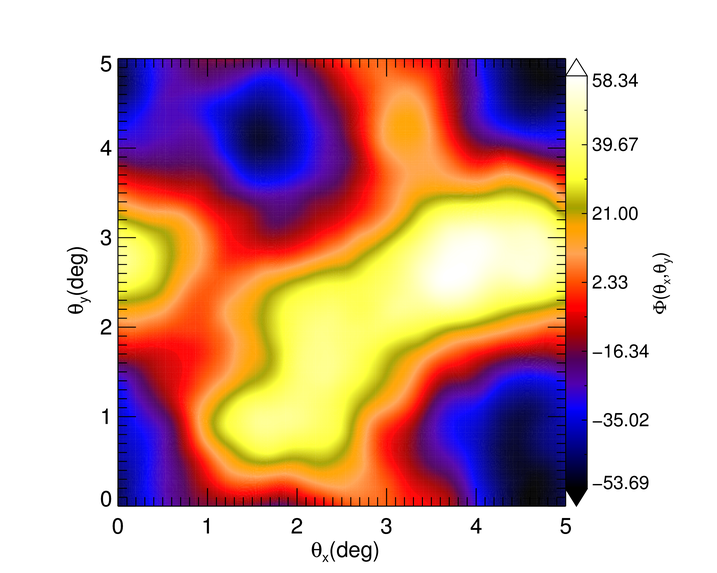}} \\
\subfigure[Denoised estimator SKA2-Low R1]{\label{fig:RecGPRMultiSKA2R1-Wiener}\includegraphics[scale=0.23]{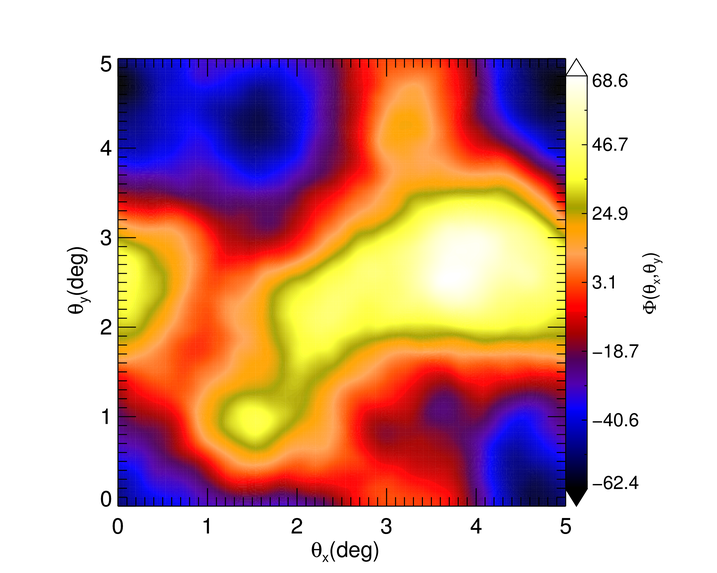}} \quad
\subfigure[Estimator variance for SKA1-Low R1]{\label{fig:RecEstPSMultiNewNoise}\includegraphics[scale=0.24]{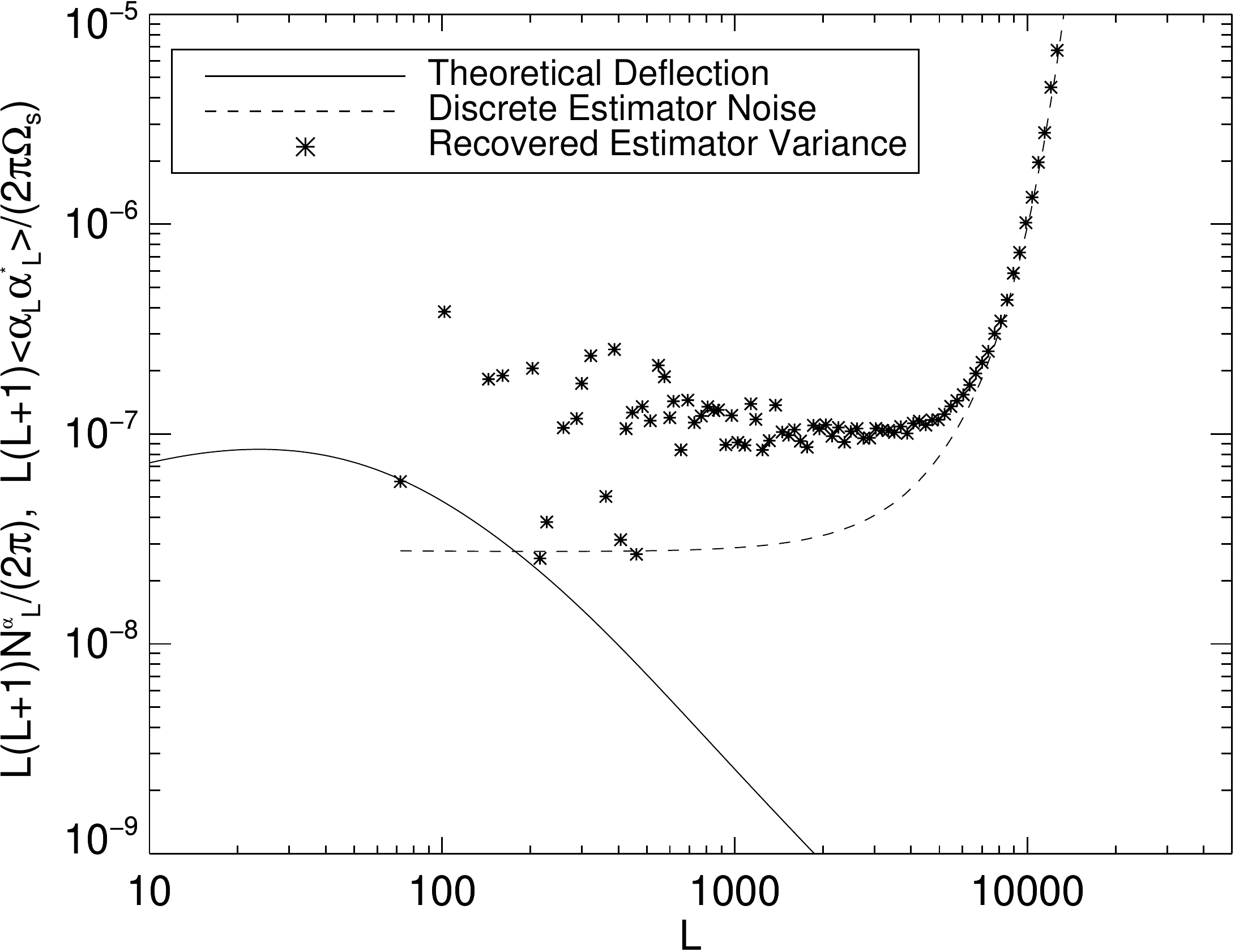}} \qquad
\subfigure[Fidelities for considered models]{\label{fig:FidelMultiNewNoise}\includegraphics[scale=0.24]{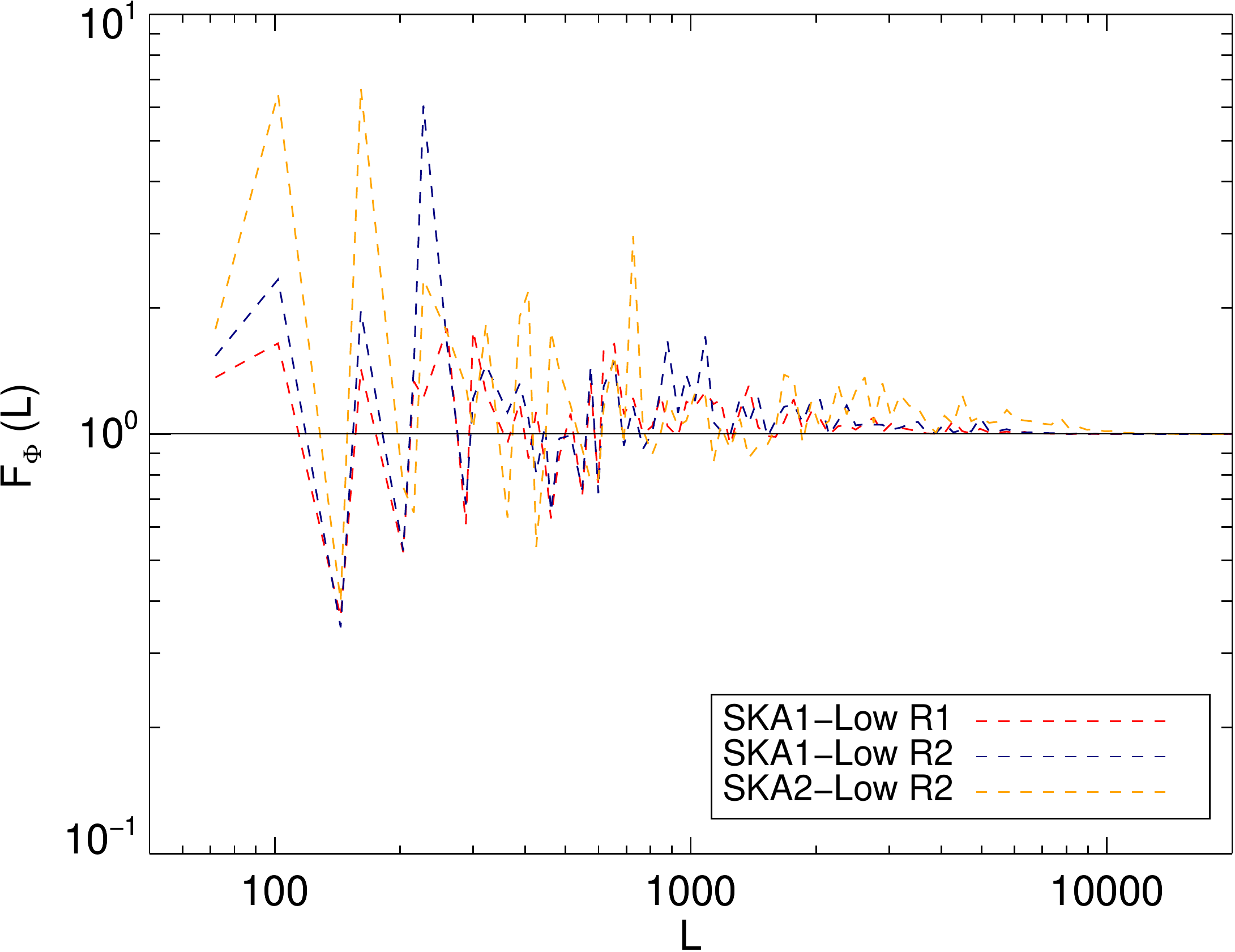}} 
\caption{\textit{From top to bottom and left to right}: the input potential field, the de-noised estimators for SKA1-Low R1, SKA1-Low R2, and SKA2-Low R1 models, the recovered Fourier space square amplitudes for non-de-noised SKA1-Low R1 image, and the fidelities for the Wiener filtered images above. The recovered estimators are computed for $N_{\rm side} = 732$, $\Omega_s = 5\degree\times 5\degree$, $k_p^{\rm min} = 3$, thermal noise model with non-uniform array distribution, and by combining 14 maps from the redshift range $z = 7-11.5$. Denoising is performed using the Wiener filter described in Section~\ref{sec:Wiener}. The input potential values and the estimated potential values have been scaled, in the images, by a factor $10^9$ in order to improve the readability of the colour bars.}
\label{fig:MultiEstimatorsNewNoise}
\end{figure*}

Following the procedure described in Section~\ref{sec:Imaging} we proceeded to compute the estimated potential images excluding the first 3 $k_p$ modes from the total-band estimator for each survey strategy and frequency band, being careful to keep the realisation for the 21 cm source fixed. So, with the discrete estimator in one band $\Phi_{\bm{L},\nu}$ given by Eq.~(\ref{eqn:BeamEst-2}), we applied Eq.~(\ref{eqn:totBandEst}) to get the total multiband discrete estimator for the potential field.  As for the single-band results, the resulting images are maps dominated by small-scale noise. Figure~{\ref{fig:MultiEstimatorsNewNoise}} shows the denoised maps computed following the Wiener filtering procedure described in Section~\ref{sec:Wiener}, which can be compared to the input potential map in Figure~\ref{fig:GRP732}. Here we have considered the SKA1-Low R1/R2 and SKA2-Low R1 models for our discussion. The differences between the SKA1-Low images can be barely noticed, mainly due to Wiener filter smoothing. On the other hand, the SKA2-Low model seems to reproduce the input structures with more accuracy than the two SKA1-Low models. The final resolution of the recovered images is set by the beam of the highest redshift band, because the modes $L > L_{\rm cut} \left(z_c^{\rm max}\right)$ belonging to other bands are smoothed and not used for the total-band reconstruction.

The recovered square amplitude of Fourier modes is recovered in Figure~\ref{fig:RecEstPSMultiNewNoise}, where the star points follow the multi-band analogous of Eq. (\ref{eqn:SigPlusNoise}) including the beam, namely
\begin{equation}
\label{eqn:SigPlusNoiseMulti}
\bigg\langle \hat\phi_{\bm L}^{\rm tot}\left(\hat\phi_{\bm L}^{\rm tot}\right)^\star\bigg\rangle = \Omega_s \left(C_L^{\phi\phi} + \mathcal{N}_L^{\rm tot}\right)
\end{equation}
confirming the behavior studied in Section~\ref{sec:MultiRealiz}. 

The fidelities for the examined SKA-Low models are displayed in Figure~\ref{fig:FidelMultiNewNoise}. Because of the Wiener filtering procedure, we see that the fidelity is generally above 1, and obviously the SKA2-Low model map has a better quality, explaining the accuracy with which the structures are reproduced in Figure~\ref{fig:RecGPRMultiSKA2R1-Wiener}. The two SKA1-Low reconstructed potentials seem to have similar fidelities, but the R2 model produces a slightly better image than R1 one and reconstructs with more fidelity all the modes $L\lesssim 1000$, as expected from Figure~\ref{fig:EstNvarSKAMulti-New}.

\subsubsection{Limits on Lower Central Redshift}
\label{sec:minZdisc}
As stated in Section~\ref{sec:zDep}, the upper bound on the considered redshift range does not considerably affect the signal-to-noise because the single reconstruction noises for $z\gtrsim 11$ are very high, leading to negligible contributions in the $\nu$-sum performed in Eq.~(\ref{eqn:totBandEstN}). The estimator reconstruction noise level of the multi-band approach mainly depends on the first central redshift that is chosen to define our total band. If EoR ended at earlier redshifts than our lower limit or if EoR is so patchy to make hydrogen not uniformly ionized at that lower redshift, a higher low-redshift range has to be considered (unless we use a non-optimal estimator for reconstruction in this case). If $z_c^{\rm min} = 8$ we should exclude $4$ frequency bands from the ones considered in Section~\ref{sec:StackBands} for R1 and R2 strategies, while we should exclude $3$ bands for R0 case. If we consider the SKA1-Low R1 case, this would lead to a $2.5$ factor increase of the reconstruction noise. These values would compromise the high-fidelity reconstruction for SKA1-Low if foreground contamination is serious and too many $k_p$ modes need to be discarded because of spurious frequency correlations left from a given foreground cleaning technique. This point clearly proves the crucial importance of investigating more realistic models of foreground removal and EoR physics for the observations simulated in this work.

\begin{figure*}
\centering
\subfigure[Single Band Errors]{\label{fig:DC_SB}\includegraphics[scale=0.245]{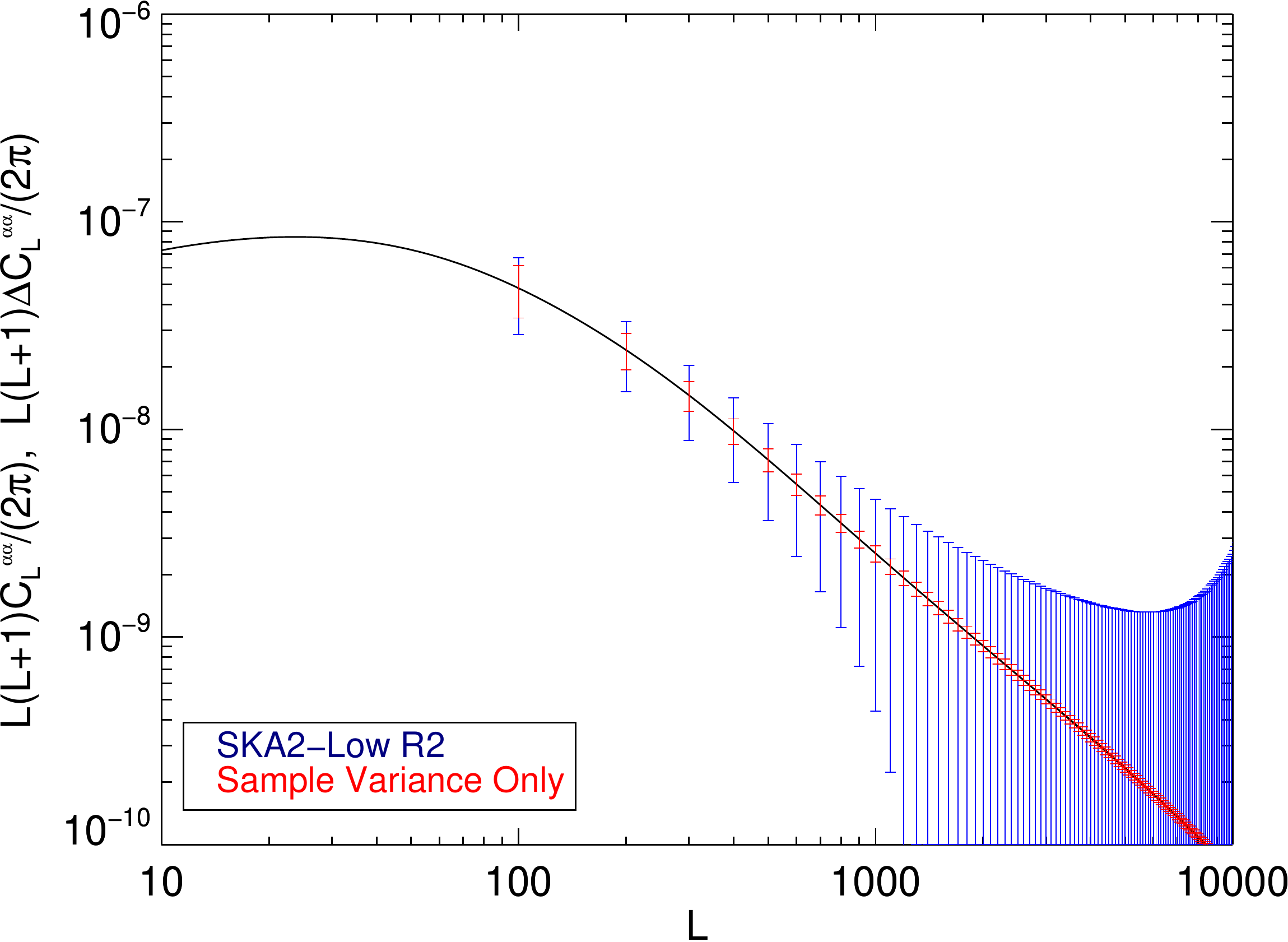}} \,
\subfigure[Single Band Fractional Errors]{\label{fig:DCR_SB}\includegraphics[scale=0.245]{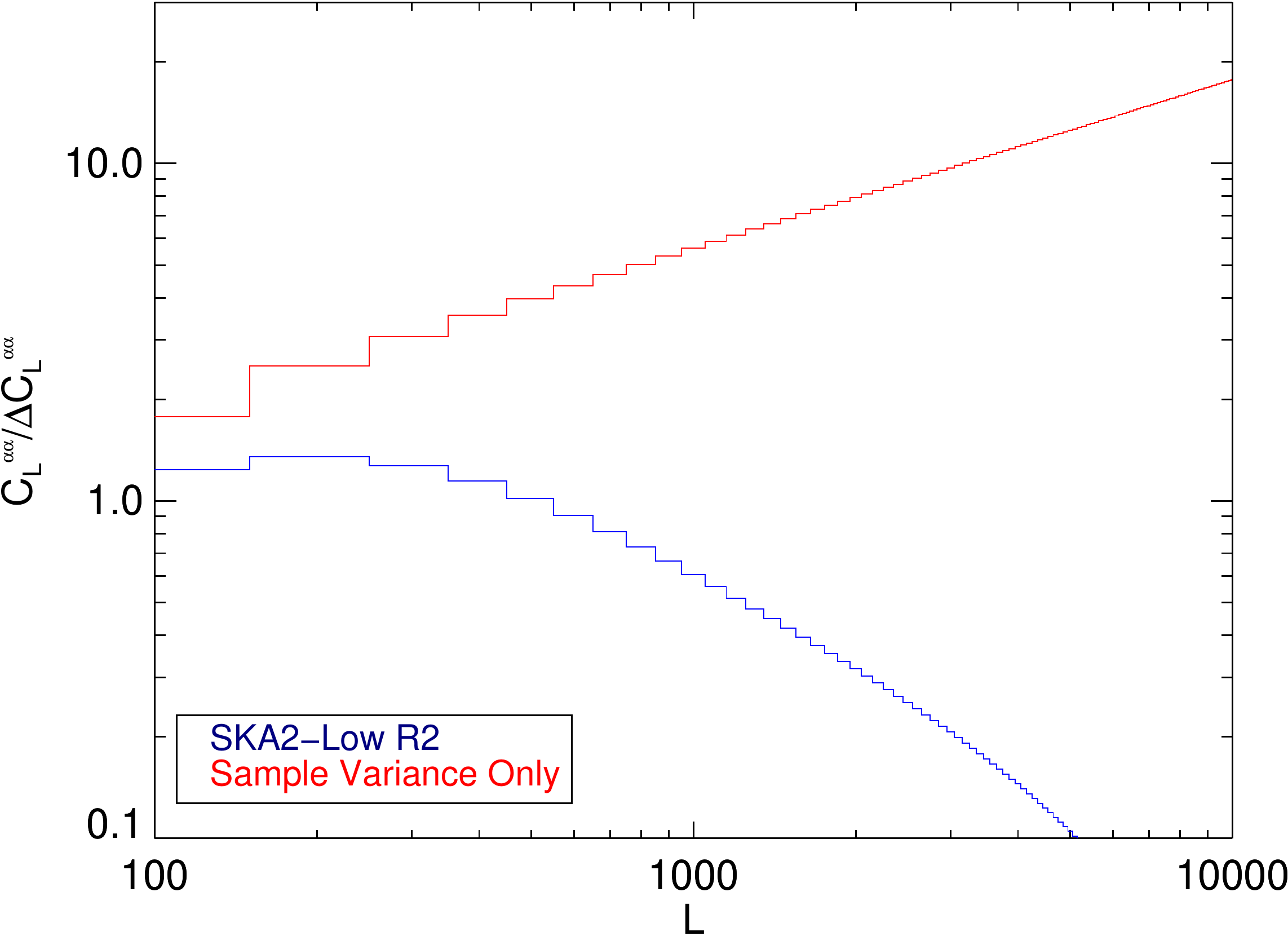}} \,
\subfigure[Multi-band Errors SKA-Low R1]{\label{fig:DC_MB_SKAR1}\includegraphics[scale=0.245]{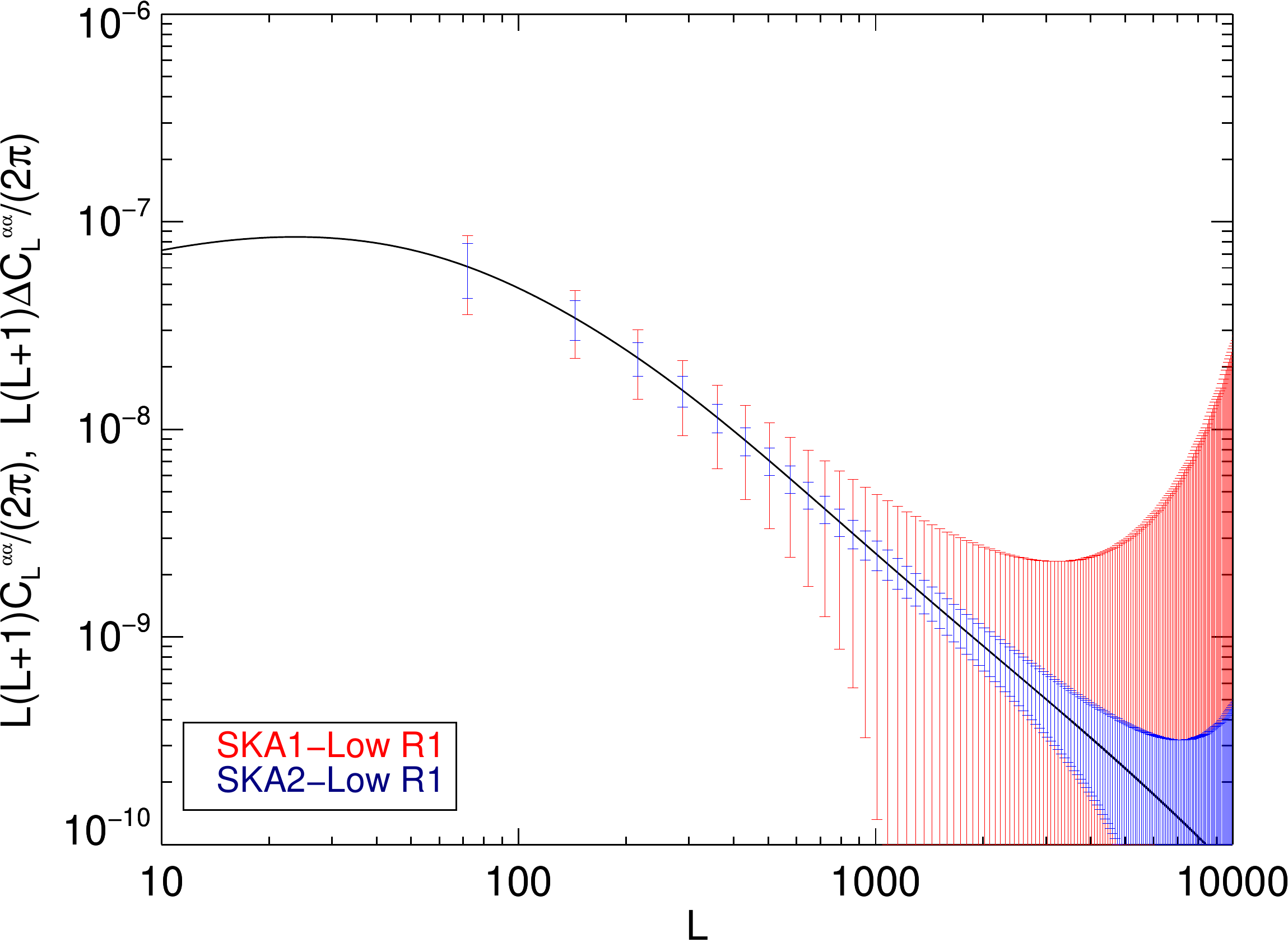}} \\
\subfigure[Multi-band Fractional Errors SKA R1]{\label{fig:DCR_MB_SKAR1}\includegraphics[scale=0.245]{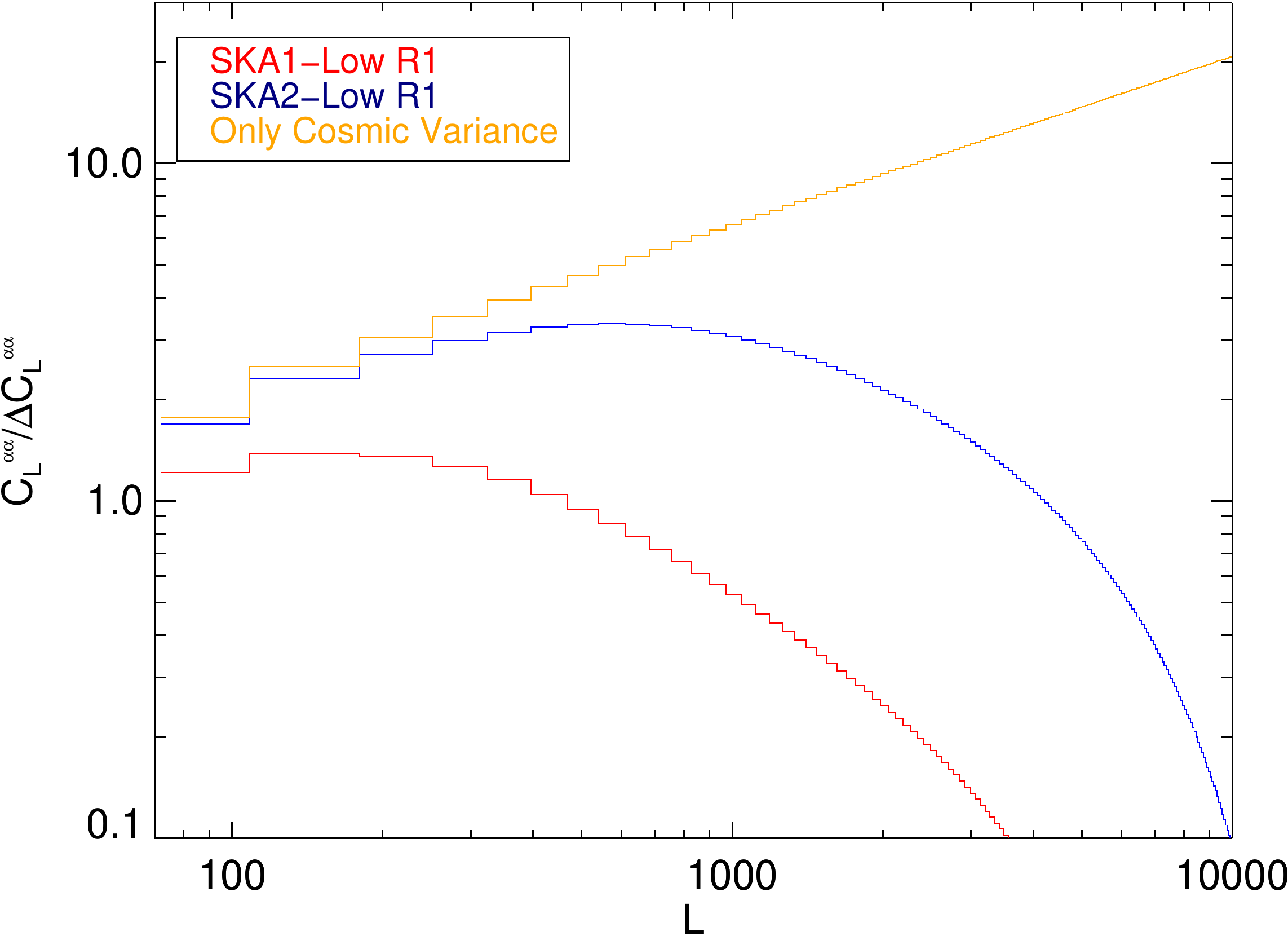}} \,
\subfigure[Multi-band Errors SKA1-Low R1/R2]{\label{fig:DC_MB_SKA1R1R2}\includegraphics[scale=0.245]{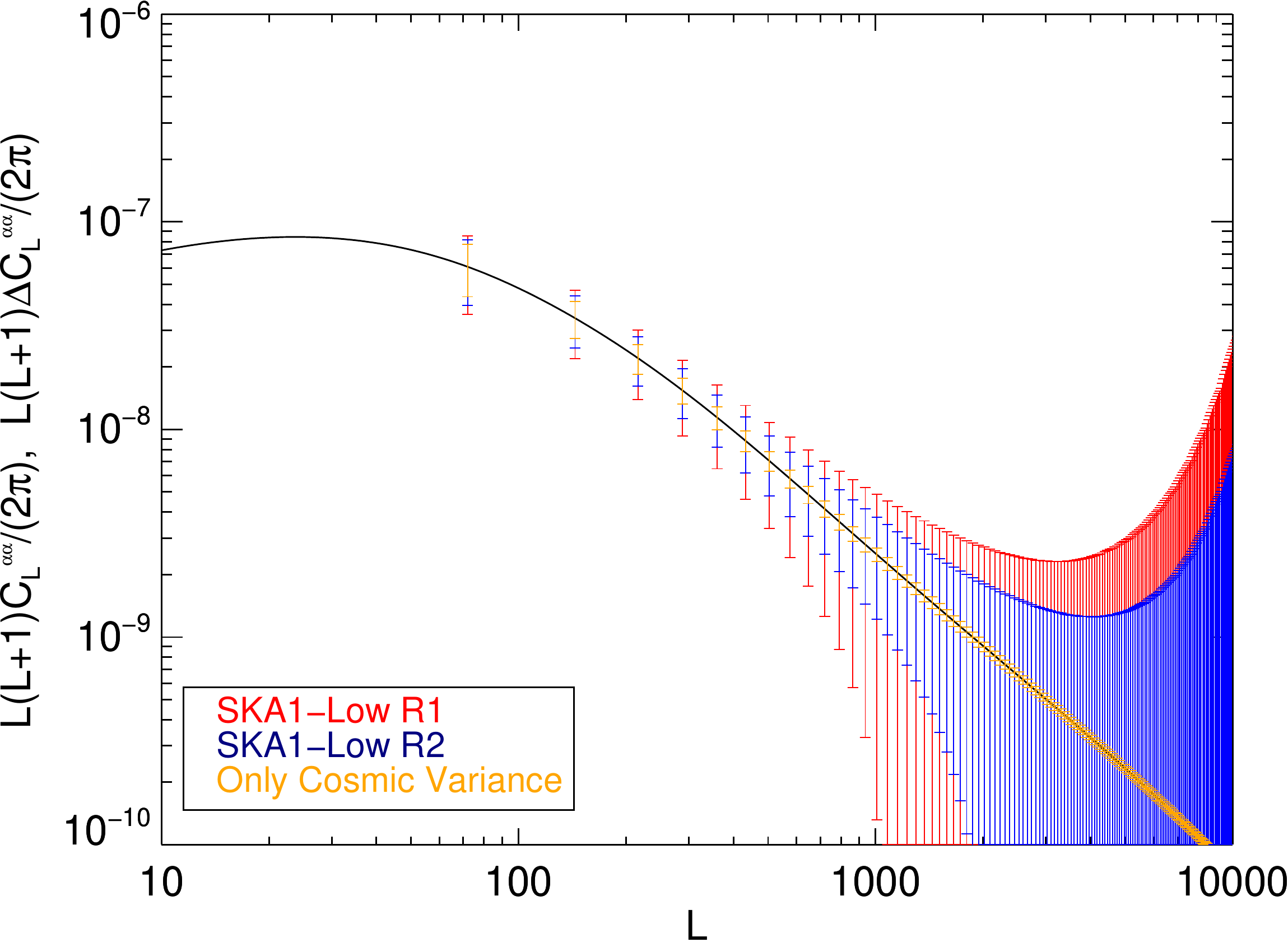}} \,
\subfigure[Multi-band Errors SKA2-Low R1/R2]{\label{fig:DC_MB_SKA2R1R2}\includegraphics[scale=0.245]{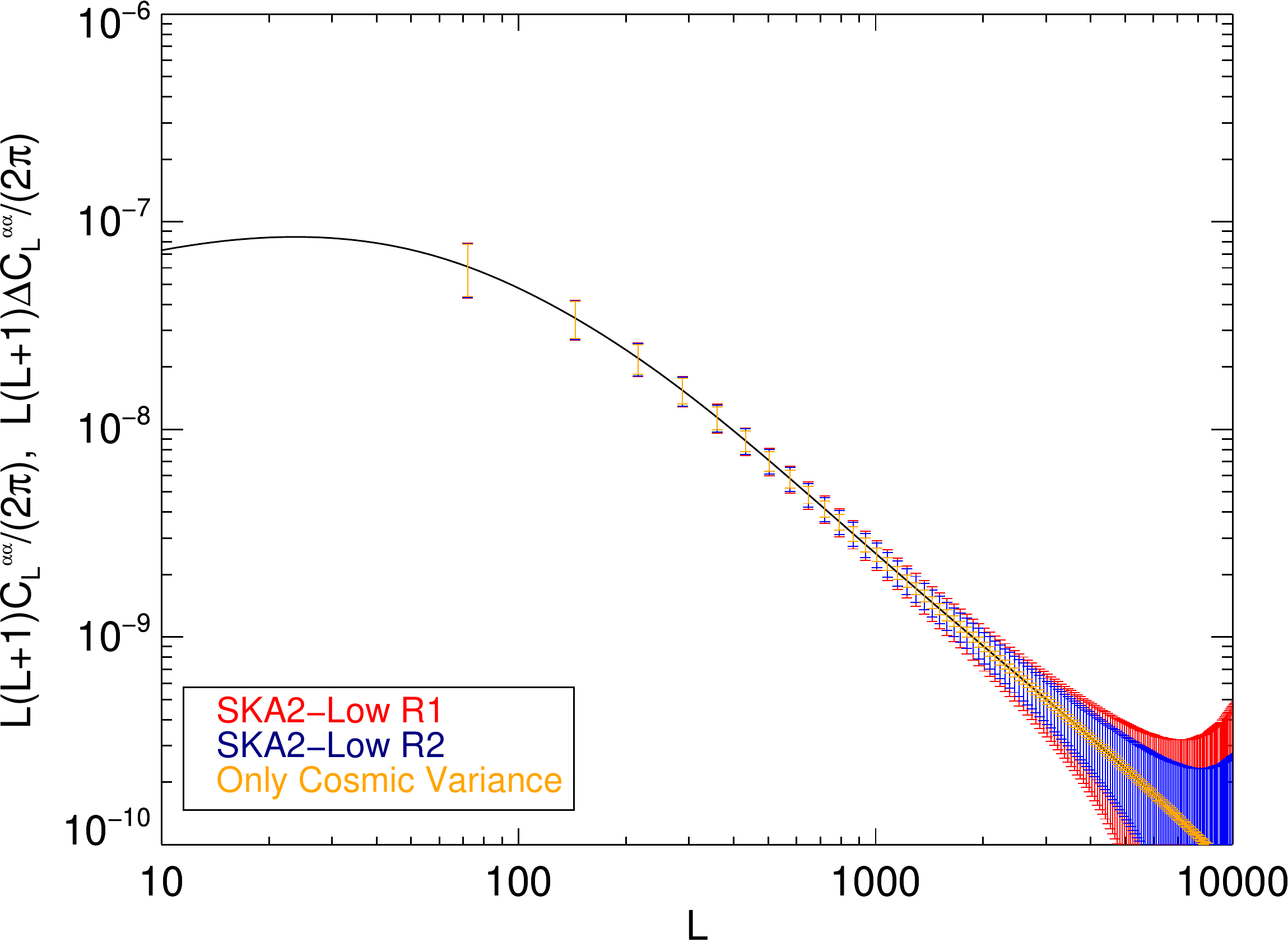}} 
\caption{\textit{First row panels}: on the left the deflection field power spectrum and measurement error bars for a SKA2-Low R2 model, considering a single band detection at $z=8$, with $f_{\rm sky} = 3.14\times 10^{-4}$, $\Delta L = 100$, and $k_p^{\rm min} = 3$. On the center we see the fractional error for the same experiment (blue), compared to the sample variance limit (red). On the right the deflection field power spectrum and measurement error bars for a SKA1-Low (red) and a SKA2-Low R2 (blue) models, considering a multiple frequency band detection in the range $z=7-11.6$, with $f_{\rm sky} = 6\times 10^{-4}$, $\Delta L = 72$, and $k_p^{\rm min} = 3$. \textit{Second row panels}: on the left we see the fractional error for the same experiments, compared to the sample variance limit (orange). On the center the deflection field power spectrum and measurement error bars for a SKA1-Low R1 (red) R2 (blue) models. On the right the same thing for SKA2-Low R1 (red) R2 (blue) models. The cosmic variance limit result is everywhere represented by the orange bars and lines.}
\label{fig:ErrorPSMeas}
\end{figure*}

\subsection{Lensing Power Spectrum Measurement} 
\label{sec:PSerror}
Giving analytic estimates of the recovered power spectrum or providing forecasts on cosmological parameters is beyond the scope of this work, but it would be interesting to understand how accurately the power spectrum can be measured with a $\Omega_s = 5\degree \times 5\degree$ survey like the one considered for a multi-band detection simulated in Section~\ref{sec:MultiBand} or with a $\Omega_s = 3.6\degree\times 3.6\degree$ survey for a single-band detection simulated in Section~\ref{sec:Imaging}.

A 21~cm lensing survey covering a large enough fraction of the sky would be able to measure the two-point statistics of the underlying lensing field. The statistical error in the deflection field power spectrum given by Eq.~(\ref{eqn:DefField}) is
\begin{equation}
\label{eqn:PSerror}
\Delta C^{\alpha\alpha}_L = \sqrt{\frac{2}{\left(2L+1\right)f_{sky}\Delta L}}\left(C_L^{\alpha\alpha} + \mathcal{N}_L^{\hat{\alpha}}\right),
\end{equation}
where $\Delta L$ is the multipole binning, $f_{\rm sky} = \Omega_s [\mbox{sr}] / 4\pi[\mbox{sr}] = \Omega_s [\mbox{deg}]^2 / 41253[\mbox{deg}]^2$ is the observed fraction of the sky, and $\mathcal{N}_L^{\hat{\alpha}}$ is the discrete estimator reconstruction noise related to Eq.~(\ref{eqn:BeamEstNoise}) via $\mathcal{N}_L^{\hat{\alpha}} = L^2\mathcal{N}_L^{\hat{\Phi}}$. When multi-band measurements are considered, the total band estimator reconstruction noise Eq.~(\ref{eqn:totBandEstN}) is used in Eq.~(\ref{eqn:PSerror}). From Eq.~(\ref{eqn:PSerror}) it can be noted that if the sky fraction $f_{\rm sky}$ is too low, the errors will be sample variance dominated. Moreover, a high reconstruction noise would increase these errors at all scales (especially at smallest ones), compromising the measurement of the lensing power spectrum. As stated also by \citet{PourtsidouMetcalf14}, the larger observed fraction of the sky planned for SKA-Mid will greatly improve these measurements, since the error is $\propto f_{sky}^{-1/2}$ and the signal is detected with a much higher signal-to-noise with respect to SKA1 and SKA2-Low phases. This will allow for a competitive estimate of cosmological parameters from such high-fidelity images. 

\subsubsection{Single-Band Constraints}
Let us first discuss the single-band results shown in Figure~\ref{fig:DC_SB}, in which the deflection power spectrum measurement errors are plotted for the SKA2-Low R2 model with a $5$ MHz bandwidth centered around a redshift of $z=8$. The SKA1-Low cases are not considered because the noise is well above the signal, as seen from Figure~\ref{fig:EstNvarSKA-New}. In this case $f_{\rm sky} = 3.14\times 10^{-4}$ and the multipole resolution is $\Delta L = 100$. We assume that the foreground cleaning makes the first 3 $k_p$ modes unusable. The result is obtained by applying Eq.~(\ref{eqn:PSerror}), and this is compared to the errors given by the sample variance limit for $\mathcal{N}^{\hat\alpha}_L \rightarrow 0$. 

To understand how accurate the power spectrum measurement is, we can consider the fractional error from Eq.~(\ref{eqn:PSerror}), namely
\begin{equation}
\label{eqn:fracDC0}
\frac{C_L^{\alpha\alpha}}{\Delta C_L^{\alpha\alpha}} \approx \sqrt{\frac{2L+1}{2}f_{\rm sky} \Delta L}\left(1 + \frac{\mathcal{N}^{\hat\alpha}_L}{C_L^{\alpha\alpha}}\right)^{-1},
\end{equation}
where the ratio of the power spectra is the inverse of the signal-to-noise ratio. For negligible estimator reconstruction noise we obtain the sample variance fractional error limit
\begin{equation}
\label{eqn:fracDC}
\frac{C_L^{\alpha\alpha}}{\Delta C_L^{\alpha\alpha}} \approx \sqrt{\frac{2L+1}{2}f_{\rm sky} \Delta L}.
\end{equation}
These are displayed in Figure~\ref{fig:DCR_SB}, where we show the fractional error ratio Eq.~(\ref{eqn:fracDC0}) compared to the sample variance fractional error Eq.~(\ref{eqn:fracDC}). It can be noticed that a good fidelity image does not correspond to an accurate measurement of the power spectrum even in the region where the reconstruction noise is small compared to the deflection field signal, being quite far from the sample variance limit.

\subsubsection{Multi-Band Constraints}
For a multi-band measurement the results are quite different. In Section~\ref{sec:MultiBand} we have seen that a lower level for the estimator reconstruction noise can be achieved when multiple frequency bands are stacked up and used simultaneously, even excluding some $k_p$ modes because of foreground subtraction. Figure~\ref{fig:DC_MB_SKAR1} shows the measurement error bars obtained for our most conservative survey strategy R1, for both SKA1-Low (red) and SKA2-Low (blue) telescope models. As described in Section~\ref{sec:4.3.4}, we used 14 bands in the redshift range $z=7-11.6$,  with $f_{\rm sky} = 6\times 10^{-4}$, $\Delta L = 72$, and $k_p^{\rm min} = 3$. We can notice that even if in Figure~\ref{fig:FidelMultiNewNoise} the images fidelities were more or less comparable, here SKA2-Low shows far better results with respect to the SKA1-Low model in measuring the power spectrum. This can be better appreciated in Figure~\ref{fig:DCR_MB_SKAR1}, that shows the fractional error ratio of the above mentioned models, compared to the sample variance limit result (orange). We can see that a SKA2-Low survey could measure the power spectrum with an accuracy comparable to the sample variance one for $L\lesssim 1000$. Phase 2 of SKA-Low considerably improves the accuracy with respect to Phase 1 also in the estimator reconstruction noise limited regime at high $L$.

Changing the survey strategy by doubling the observational time from model R1 to model R2 we do not see significant benefits. In fact, from Figure~\ref{fig:DC_MB_SKA1R1R2} for the SKA1-Low model and from Figure~\ref{fig:DC_MB_SKA2R1R2} for the SKA2-Low model we see that the improvement is minimal. SKA1-Low is still far away from the sample variance limit, while SKA2-Low gets a bit closer to it, but the total accuracy is only slightly improved.

As already stated at the beginning of this section, the situation can be improved by considering larger surveyed areas of the sky (like the ones explored by SKA-Mid) in order to have a larger $f_{\rm sky}$. Other possibilities consist in mosaicking different patches of the sky in order to increase the FoV, or detecting the signal for different patches in the sky. This latter observation method can increase the statistics of a given mode range by lowering the sample variance error, especially in the intermediate $L$-range $200\lesssim L \lesssim 1000$. Such a measurement can be done in a reasonable amount of time even with the SKA1-Low aperture array, and we plan to compute the constraints coming from this survey strategy in a future work.

Finally, further improvements in measuring the lensing power spectrum can be achieved by considering the detected convergence field in cross-correlation with CMB measurements, or, if lower redshifts are observed with the HI or galaxy density fields. \citet{PourtsidouBacon15} found that this last case considerably improves the 21~cm lensing detection prospects and excellent results can be achieved within frequencies observed by SKA-Mid and MeerKAT. In general, cross-correlating the galaxy and HI densities at post-EoR redshifts can be particularly useful for constraining HI and cosmological parameters and alleviate issues arising from systematic effects that are relevant for one type of survey but not the other \citep{Masui:2012zc,Wolz:2015ckn,Wolz:2015lwa,Pourtsidou:2016dzn}.

\subsection{Cluster Detection}
Another application of our code for 21-cm lensing concerns the possible detection of a galaxy cluster signal. To investigate this we generated a deflection field using the GLAMER\footnote{http://glenco.github.io/glamer/} library \citep{PetkovaMetcalfGiocoli14, MetcalfPetkova14}. This is a C++ library for performing gravitational lensing simulations using the output of cosmological simulations or analytic lens models or combinations of them. We generated a NFW halo profile with density 
\begin{equation}
\rho (r) = \frac{\rho_s}{r/r_s\left(1+r/r_s\right)^2}
\end{equation}
\citep{NFW97}, where the scale density $\rho_s$ is the normalisation of this profile and $r_s$ is a scale radius. These quantities are often described in terms of the concentration parameter $c = r_{200}/r_s$, with $r_{200}$ being the radius of the sphere in which the average density is 200 times the critical density and the enclosed mass is $M_{200}$. Its value is
\begin{equation}
r_{200} = 1.63\times 10^{-2}\left(\frac{M_{200}}{h^{-1}M_\odot}\right)^{1/3}\left[\frac{\Omega_0}{\Omega(z)}\right]^{-1/3} (1+z)^{-1}h^{-1} \mbox{ Kpc}.
\end{equation}
The mass of the cluster is linked to the concentration parameter via $M = 4\pi r^3_s\rho_s \left[\ln(1+c) - c/(1+c)\right]$. The lensing potential produced by the NFW profile is
\begin{equation}
\Phi_{NFW}(\theta) = 4\rho_s r_s \Sigma_{cr}^{-1} g(\theta/\theta_s),
\end{equation}
where $\Sigma_{cr}$ is the critical surface density, $\theta = r/\mathcal{D}(z)$ and $\theta_s = r_s/\mathcal{D}(z)$. The function $g(x = \theta/\theta_s)$ is defined as
\begin{equation}
g(x) = \frac{1}{2}\ln^2\frac{x}{2}+\left\{
\begin{array}{rl}
2 \arctan^2 \sqrt{\frac{x-1}{x+1}}, & (x>1) \\
-2\mbox{ arctanh}^2\sqrt{\frac{1-x}{1+x}}, & (x<1) \\
0, & (x=1)
\end{array}
\right. .
\end{equation}

Following previous works like \citet{GiocoliMeneghetti14} and \citet{SerenoGiocoli14}, we have simulated the deflection field produced by a plausible galaxy cluster placed in the centre of our lensed 21 cm radiation map with mass $M = 10^{15} M_\odot$ and concentration $c=7$. This cluster is placed at $z=0.5$, while the source is at $z_s=8$. 

This lensing source has been used to deflect our simulated 21 cm intensity maps, as discussed in Section~\ref{sec:LensMaps}. Following the procedure described in Section~\ref{sec:Imaging} and modeling a SKA2-Low R2 experiment, we applied these deflected maps to the estimator Eq. (\ref{eqn:BeamEst-2}). We find that the NFW cluster under consideration (with a few arcseconds Einstein radius) is basically undetectable because the recovered signal is totally consistent with the estimator reconstruction noise. Analysing the input deflection field power spectrum, we found that, even for a multi-band detection constructed by stacking bands from $z_c=6.5$ to $z_c=12$, this is well below the estimator reconstruction noise level by four orders of magnitude. This result agrees with the one obtained by \citet{KovetzKamionkowski13} for a lower redshift $(z=7)$.
Perhaps observations beyond SKA at lower frequency and/or higher resolution might make detecting clusters possible.

It would be interesting to study this detection at a lower redshift such as $z \sim 1-3$, where $x_H\neq 1$ and point source signal represents an important contribution to 21~cm source. For this reason we will need to take into account non-negligible Poissonian source terms in our estimator. We could indeed place in random positions more realistic clusters in our simulated map, in order to detect the total signal coming from them, but big improvements are not expected, since at those redshifts the reconstruction signal-to-noise is lower for each mode. This case will be studied in future work.

\section{Conclusions and future outlook}
\label{sec:Conclusions}
In this work we have seen how 21~cm lensing can be a leading cosmological probe during the next decade. Using the forthcoming observations from the SKA and other radio telescopes, a huge amount of cosmological information can potentially be extracted over a wide range of redshifts, in order to constrain the standard $\Lambda$CDM paradigm, and help us understand the nature of the dark sector of our Universe. The innovative technique of Intensity Mapping treats the 21~cm brightness temperature fluctuations as a continuous three-dimensional field, opening up the possibility of using alternative analysis methods similar to those successfully applied to the CMB, and, given the narrow channel frequency resolution, measuring redshifts with excellent precision.

We investigated the potential offered by the weak gravitational lensing of the 21~cm brightness temperature fluctuation field, focusing at a typical EoR redshift ($z=8$) in which the modelled HI is fully ionized. For this purpose, we implemented a simulation pipeline capable of dealing with issues that can not be treated analytically, like the simulation of a full telescope beam, the non-uniform visibility space coverage, the non-linearity of the lensing source field, and the discreteness of visibility measurements. 

With the simulation code built in this work it will be possible to simulate the weak gravitational lensing of 21~cm field at post-EoR redshifts, by taking into account the discreteness of the point sources and including it as an additive discrete Poisson noise to a clustering Gaussian three-dimensional signal. This allows for an improvement of the reconstruction signal-to-noise.

Moreover, in the theoretical and numerical framework established in this work, it is possible to include and investigate other complicated issues regarding our ignorance about the reionization process history, like the non Gaussianity of the 21~cm source in the considered EoR redshift range. In fact, it is very likely that EoR was a non-homogeneous process expanded over a considerable redshift range, and the detected signal strongly depends on the number density of ionized regions which are causing inhomogeneities in 21 cm temperature signal that is not possible to investigate analytically. Another important non-analytic issue our code is designed to handle concerns foreground subtraction techniques. With the pipeline developed in this work we can implement foreground contamination and study how foreground removal techniques can affect the accuracy of our results. These methods would indeed produce residual noises and cause cross-correlations among different frequencies, and their influence can be treated only numerically.

By taking advantage of the 21~cm source signal division into multiple statistically independent maps along the frequency direction, we have demonstrated how the lensing mass distribution can be reconstructed with high fidelity using a three dimensional optimal quadratic lensing estimator in Fourier space. This would provide a great opportunity to correlate mass with visible objects and test the dark matter paradigm. 

Considering the current SKA plans, we studied the performance of the quadratic estimator for detections aimed to observe EoR redshifts, for different observational strategies and using a thermal noise model which takes into account a realistic SKA-Low density distribution of the stations. These noise models have been added to simulated lensed 21~cm brightness temperature fluctuation maps, produced by interpolating on the grid the lensed positions of the temperature maps. To accomplish this task we followed the weak lensing assumption widely used in the CMB case, which is valid for the 21~cm field as well at the scales considered in this study. 

We successfully implemented the 3D Fourier space quadratic estimator in our simulation code, taking into account the smoothing effect caused by the beam of the telescope (set by the baseline maximum dimensions) and the discreteness of visibility measurements, paving the way for future numerical studies aimed to investigate more realistic issues. We showed that the discrete 21~cm estimator can be employed by using a single frequency band or by combining multiple frequency band measurements as well. 

We found that Phase 1 of the SKA-Low interferometer could obtain high-fidelity images of the underlying mass distribution only if several bands are stacked together, covering a redshift range from $z=7$ to $z=11.6$ and with a total resolution of $1.6$ arcmin. We also implemented a simple de-noising procedure in order to filter out the small-scale noise which is likely to strongly contaminate the estimated signal. 
Phase 2 of SKA-Low, modelled in order to improve the sensitivity of the instrument by at least an order of magnitude, should be capable of providing reconstructed images with good quality even when the signal is detected within a single frequency band. In this case the reconstructed image has a resolution of  $1.15$ arcmin at $z=8$, within a field of view of $13 \mbox{ deg}^2$.

Considering the serious effect that foregrounds could have on these detections (by making the first few $k_p$ modes unusable), we discussed the limits of these results as well as the possibility of measuring an accurate lensing power spectrum. In the case of multi-band detection of the lensed 21~cm signal made with an SKA2-Low telescope model we found constraints close to the sample variance ones in the range $L<1000$, even for a small field of view such as a $25 \mbox{ deg}^2$ survey area. Good constraints have also been found for SKA1-Low in multi-band detection, and for SKA2-Low in single band detection.

We finally explored the possibility to detect a cluster lensing signal coming from redshift $z=0.5$ with a mass of $M = 10^{15} M_\odot$, but we found its signal to be overwhelmed by the estimator reconstruction noise by several orders of magnitude, going well below the saturation limit of the noise imposed by sample variance also for multi-band analysis. These results could be improved if different patches of the sky are observed to decrease the sample variance of the larger-scale modes, or if the data coming from the lensing power spectrum measurements are cross-correlated with the ones coming from CMB. If lower redshifts are observed, the cross-correlation with galaxy surveys data can be used to better constrain the power spectrum accuracy.

\vspace{0.3cm} 
\leftline{\bf Acknowledgments} 
This research is part of the project GLENCO, funded under the Seventh Framework Programme, Ideas, Grant Agreement n. 259349. AP acknowledges support by a Dennis Sciama Fellowship at the University of Portsmouth. The authors would like to thank Ethan Anderes, Carlo Giocoli and Andrea Negri for useful discussions.

\bibliographystyle{mn2e}
\bibliography{biblio_paper}

\appendix
\onecolumn
\section{Fast quadratic estimator derivation}
\label{sec:AppA}
In this appendix we will provide an explicit derivation\footnote{Interested readers are advised to consult \citep{Anderes13} if they want to see how this procedure can be demonstrated for generic estimators in 2D case, like polarization-polarization or polarization-temperature estimators.} for Eq.(\ref{eqn:BeamEst-2}), found by applying to beamed 3D 21 cm case what can be found on other works already cited on this paper. Our starting point is the quadratic estimator expression, namely
\begin{equation}
\label{eqn:App1}
\hat{\phi}_{\bm{L}} = \frac{\mathcal{N}^{\hat{\phi}}_{L}}{2\Omega_s}\sum_{\bm{l},k_p}{\left\{\frac{W_lW^\star_{l-L}\left[\bm{L}\cdot\bm{l}\,C_{l,k_p} + \bm{L}\cdot\left(\bm{L}-\bm{l}\right)C_{l-L,k_p}\right]}{\mathcal{C}^{T}_{l,k_p}\mathcal{C}^T_{l-L,k_p}}\right\}\, \tilde{\mathcal{T}}_{\bm{l},k_p}\tilde{\mathcal{T}}^\star_{\bm{l}-\bm{L},k_p}}
\end{equation}
with $\mathcal{C}^T_{l,k_p} = |W_l|^2\left( C_{l,k_p} + N_{l,k_p}^{Sky} \right) + N^{Rcv}_{l,k_p}$ and the estimator reconstruction noise $\mathcal{N}_L^{\hat{\phi}}$ given in Eq.(\ref{eqn:BeamEstNoise}). In order to recover a beamed expression for the faster estimator of Section~\ref{sec:FastEst}, we can rewrite Eq.(\ref{eqn:App1}) in a smarter way:
\begin{equation}
\hat{\phi}_{\bm{L}} = -\frac{\mathcal{N}^{\hat{\phi}}_{L}}{2\Omega_s}\, \left(i\bm{L}\right)\cdot\sum_{k_p} \sum_{\bm{l}}\left[\frac{i\bm{l}W_lC_{l,k_p}\tilde{\mathcal{T}}_{\bm{l},k_p}}{\mathcal{C}^T_{l,k_p}}\, \frac{W^\star_{l-L}\tilde{\mathcal{T}}^\star_{\bm{l}-\bm{L},k_p}}{\mathcal{C}^T_{l-L,k_p}}-\frac{i\left(\bm{l}-\bm{L}\right)W^\star_{l-L}C_{l-L,k_p}\tilde{\mathcal{T}}^\star_{\bm{l}-\bm{L},k_p}}{\mathcal{C}^T_{l-L,k_p}}\, \frac{W_l\tilde{\mathcal{T}}_{\bm{l},k_p}}{\mathcal{C}^T_{l,k_p}} \right]. 
\end{equation}
Now, if we define the following small-scale filtered fields
\begin{equation}
\mathcal{F}_{\bm{l},k_p} = \frac{W_l\tilde{\mathcal{T}}_{\bm{l},k_p}}{\mathcal{C}^T_{l,k_p}}, \qquad \mathcal{G}_{\bm{l},k_p} = \frac{W_lC_{l,k_p}\tilde{\mathcal{T}}_{\bm{l},k_p}}{\mathcal{C}^T_{l,k_p}},
\end{equation}
our estimator will be
\begin{equation}
\label{eqn:App4}
\hat{\phi}_{\bm{L}} = -\frac{\mathcal{N}^{\hat{\phi}}_{L}}{2\Omega_s}\, \left(i\bm{L}\right)\cdot\sum_{k_p} \sum_{\bm{l}}\left\{ i\bm{l}\mathcal{G}_{\bm{l},k_p}\mathcal{F}^\star_{\bm{l}-\bm{L},k_p} + \left[i\left(\bm{l}-\bm{L}\right)\mathcal{G}_{\bm{l}-\bm{L},k_p}\right]^\star\mathcal{F}_{\bm{l},k_p} \right\}. 
\end{equation}
Let us consider the first $\bm{l}$-sum for a given $k_p$ mode. We can see that it is equivalent to a convolution in Fourier space and, for the convolution theorem this can be written as a product of two dual-complex space functions when the non-hermitian filtered fields are transformed into the dual complex space:
\begin{equation}
\sum_{\bm{l}}i\bm{l}\mathcal{G}_{\bm{l},k_p}\mathcal{F}^\star_{\bm{l}-\bm{L},k_p} = \sum_{\bm{l}}i\bm{l}\mathcal{G}_{\bm{l}, k_p} \left[\sum_{\bm{m}} e^{-i\left(\bm{l}-\bm{L}\right)\cdot\bm{m}}\mathcal{F}_{\bm{m}, k_p}\right]^\star = \sum_{\bm{m}} e^{-i\bm{L}\cdot\bm{m}}\mathcal{F}_{\bm{m},k_p}^\star \sum_{\bm{l}} e^{i\bm{l}\cdot\bm{m}} i\bm{l}\mathcal{G}_{\bm{l}, k_p} = \sum_{\bm{m}} e^{-i\bm{L}\cdot\bm{m}}\mathcal{F}^\star_{\bm{m},k_p}\left(\bm{\nabla}_{\bm{m}}\mathcal{G}_{\bm{m},k_p}\right).
\end{equation}
Analogously proceeding, one can show that the second sum in \ref{eqn:App4} can be analogously treated:
\begin{equation}
\left[i\left(\bm{l}-\bm{L}\right)\mathcal{G}_{\bm{l}-\bm{L},k_p}\right]^\star\mathcal{F}_{\bm{l},k_p}  = \sum_{\bm{l}}\mathcal{F}_{\bm{l}, k_p} \left[\sum_{\bm{m}} e^{-i\left(\bm{l}-\bm{L}\right)\cdot\bm{m}}\bm{\nabla}_{\bm{m}}\mathcal{G}_{\bm{m}, k_p}\right]^\star = \sum_{\bm{m}} e^{-i\bm{L}\cdot\bm{m}}\left(\bm{\nabla}_{\bm{m}}\mathcal{G}_{\bm{m}, k_p}\right)^\star\sum_{\bm{l}} e^{i\bm{l}\cdot\bm{m}} \mathcal{F}_{\bm{l}, k_p} = \sum_{\bm{m}} e^{-i\bm{L}\cdot\bm{m}}\mathcal{F}_{\bm{m},k_p}\left(\bm{\nabla}_{\bm{m}}\mathcal{G}_{\bm{m},k_p}\right)^\star.
\end{equation}
So in the end, we find for every independent $k_p$ mode 
\begin{eqnarray}
\hat{\phi}_{\bm{L}} &=& -\frac{\mathcal{N}^{\hat{\phi}}_{L}}{2\Omega_s}\, \left(i\bm{L}\right)\cdot\sum_{k_p}\sum_{\bm{m}} e^{-i\bm{L}\cdot\bm{m}}\left[\mathcal{F}^\star_{\bm{m},k_p}\left(\bm{\nabla}_{\bm{m}}\mathcal{G}_{\bm{m},k_p}\right)+\mathcal{F}_{\bm{m},k_p}\left(\bm{\nabla}_{\bm{m}}\mathcal{G}_{\bm{m},k_p}\right)^\star\right] =  -\frac{\mathcal{N}^{\hat{\phi}}_{L}}{2\Omega_s}\, \left(i\bm{L}\right)\cdot\sum_{k_p}\sum_{\bm{m}} e^{-i\bm{L}\cdot\bm{m}}\left(\bm{\mathcal{H}}_{\bm{m},k_p}+\bm{\mathcal{H}}^\star_{\bm{m},k_p}\right) \nonumber \\
&=& -\frac{\mathcal{N}^{\hat{\phi}}_{L}}{2\Omega_s}\, \left(i\bm{L}\right)\cdot\sum_{k_p}\sum_{\bm{m}} e^{-i\bm{L}\cdot\bm{m}} 2\mathcal{R}\left(\bm{\mathcal{H}}_{\bm{m},k_p}\right),
\end{eqnarray}
where the subscript $k_p$ means that every complex-space Fast Fourier Transform involving these filtered fields has to be computed for a fixed $k_p$ contribution. These FFTs hence produce a real vectorial field $\mathcal{H}_{\bm{m},k_p} = \mathcal{H}_{\bm{\theta},k_p}$, so we recover our final form for the beamed quadratic estimator presented on Eq.(\ref{eqn:BeamEst-2}):
\begin{equation}
\hat{\phi}_{\bm{L}} = - \frac{\mathcal{N}^{\hat{\Phi}}_{L}}{\Omega_s} \left(i\bm{L}\right) \cdot \sum_{k_p}\bm{\mathcal{H}}_{\bm{L},k_p} .
\end{equation}
Hence, this Fourier space estimator is basically the sum of the FFTs performed over $k_p$ modes of the multiplication between the small-scale filtered field $\mathcal{F}_{\bm{m},k_p}$ and the gradient of the other filtered field $\mathcal{G}_{\bm{m},k_p}$ added to its conjugate. The estimator appears to be real by construction, since one can easily see that $\hat{\phi}^\star_{\bm{L}} = \hat{\phi}_{-\bm{L}}$. Thus, the optimal estimator $\phi_{\bm{L}}$ is the divergence of the Hermitian vectorial field $\bm{\mathcal{H}}_{\bm{L}}$, summed for all the $k_p$ modes and normalized by its Fourier space variance, as one can notice from the presence of the operator $i\bm{L}$ in Eq.(\ref{eqn:BeamEst-2}). 
This result may depend on the way the derivatives are implemented in the code. Generally spectral derivative are accurate enough, but fourth-order finite differences methods can provide slightly different results. Moreover one needs to include a $\sqrt{2}$ constant factor in order to take into account the different normalization between FFTs and classic DFTs.

\end{document}